\documentclass[10pt,twocolumn]{IEEEtran}
\usepackage{soul}
\usepackage{graphicx}
\usepackage{mathtools}
\usepackage{amsmath}
\usepackage{amsfonts}
\usepackage{array}
\newcommand{\PreserveBackslash}[1]{\let\temp=\\#1\let\\=\temp}
\newcolumntype{C}[1]{>{\PreserveBackslash\centering}p{#1}}
\newcolumntype{R}[1]{>{\PreserveBackslash\raggedleft}p{#1}}
\newcolumntype{L}[1]{>{\PreserveBackslash\raggedright}p{#1}}
\usepackage[usenames]{color}
\usepackage{colortbl,booktabs}
\usepackage{tabularx}
\usepackage{multicol}
\usepackage{multirow}
\usepackage{booktabs}
\usepackage[Symbol]{upgreek}
\usepackage{subfigure}
\usepackage{bm}
\usepackage[hyphens]{url}
\usepackage[usenames,dvipsnames,svgnames,table]{xcolor}
\usepackage{threeparttable}
\usepackage{tikz}
\newcommand*\circled[1]{\tikz[baseline=(char.base)]{
            \node[shape=circle,draw,inner sep=2pt] (char) {#1};}}
\usepackage{diagbox}						
\usepackage{cite}
\usepackage{balance}
\usepackage{times}

\usepackage{stfloats}
\usepackage{balance}

\usepackage[nonumberlist,acronym,shortcuts]{glossaries}
\makeglossaries
\makeindex
\newacronym{A2A}{A2A}{Air-to-Air}
\newacronym{A2G}{A2G}{Air-to-Ground}
\newacronym{A2S}{A2S}{Air-to-Satellite}
\newacronym{AAC}{AAC}{Airline Administrative Control}
\newacronym{AANET}{AANET}{Aeronautical Ad-hoc Network}
\newacronym{ACARS}{ACARS}{Aircraft Communication Addressing and Reporting System}
\newacronym{ACAS}{ACAS}{Airborne Collision Avoidance System}
\newacronym{ACI}{ACI}{Airports Council International}
\newacronym{ACO-OFDM}{ACO-OFDM}{Asymmetrically Clipped Optical OFDM}
\newacronym{ADS}{ADS}{Automatic Dependent Surveillance}
\newacronym{ADS-B}{ADS-B}{Automatic Dependent Surveillance-Broadcast}
\newacronym{ADS-C}{ADS-C}{Automatic Dependent Surveillance-Contract}
\newacronym{AeroMACS}{AeroMACS}{Aeronautical Mobile Airport Communication System}
\newacronym{AMACS}{AMACS}{All-purpose Multi-channel Aviation Communication System}
\newacronym{AOC}{AOC}{Aeronautical Operational Control}
\newacronym{ASAS}{ASAS}{Airborne Separation Assurance System}
\newacronym{A-STAR}{A-STAR}{Anchor based Street and Traffic Aware Routing}
\newacronym{ATC}{ATC}{Air Traffic Control}
\newacronym{ATG}{ATG}{Air-To-Ground}
\newacronym{ATCT}{ATCT}{Air Traffic Control Tower}
\newacronym{ATM}{ATM}{Air Traffic Management}
\newacronym{B-AMC}{B-AMC}{Broadband Aeronautical Multi-carrier Communications}
\newacronym{BBR}{BBR}{Border node Based Routing}
\newacronym{BER}{BER}{Bit Error Rate}
\newacronym{BROADCOMM}{BROADCOMM}{BROADcast COMMunication based on geographical routing}
\newacronym{BS}{BS}{Base Station}
\newacronym{CCDF}{CCDF}{Complementary Cumulative Distribution Function}
\newacronym{CDMA}{CDMA}{Code Division Multiple Access}
\newacronym{COIN}{COIN}{Clustering for Open  Inter-Vehicle Communication Network}
\newacronym{COMETS}{COMETS}{Communications and Broadcasting Engineering Test Satellites}
\newacronym{CPFSK}{CPFSK}{Continuous-Phase Frequency-Shift Keying}
\newacronym{CRL}{CRL}{Communications Research Laboratory}
\newacronym{CRT}{CRT}{Cognitive Radio Technology}
\newacronym{CSMA}{CSMA}{Carrier Sense Multiple Access}
\newacronym{DCO-OFDM}{DCO-OFDM}{Direct-Current offset Optical OFDM}
\newacronym{DHT}{DHT}{Distributed Hash Table}
\newacronym{DIR}{DIR}{Diagonal-Intersection-based Routing}
\newacronym{DL}{DL}{DownLink}
\newacronym{UL}{UL}{UpLink}
\newacronym{DME}{DME}{Distance Measuring Equipment}
\newacronym{DRG}{DRG}{Distributed Robust Geocast}
\newacronym{DTSG}{DTSG}{Dynamic Time-Stable Geocast Routing}
\newacronym{EAN}{EAN}{European Aviation Network}
\newacronym{EM}{EM}{Electromagnetic}
\newacronym{FAA}{FAA}{Federal Aviation Administration}
\newacronym{FANET}{FANET}{Flying Ad-hoc Network}
\newacronym{FCI}{FCI}{Future Communications Infrastructure}
\newacronym{FDD}{FDD}{Frequency-Duplex Division}
\newacronym{FM}{FM}{Frequency Modulation}
\newacronym{FSO}{FSO}{Free-Space Optical}
\newacronym{GMSK}{GMSK}{Gaussian Minimum-Shift Keying}
\newacronym{GPCR}{GPCR}{Greedy Perimeter Coordinator Routing}
\newacronym{GPS}{GPS}{Global Positioning System}
\newacronym{GS}{GS}{Ground Station}
\newacronym{GSM}{GSM}{Global System for Mobile communications}
\newacronym{GSR}{GSR}{Geographic Source Routing}
\newacronym{GTO}{GTO}{Ground To Orbit}
\newacronym{GvGrid}{GvGrid}{Geographic Source Routing}
\newacronym{HAP}{HAP}{High-Altitude Platform}
\newacronym{HF}{HF}{High Frequency}
\newacronym{HDTV}{HDTV}{High-Definition TeleVision}
\newacronym{HIP}{HIP}{Host Identity Protocol}
\newacronym{HV-TRADE}{HV-TRADE}{History-enhanced V-TRADE}
\newacronym{ICAO}{ICAO}{International Civil Aviation Organization}
\newacronym{ICN}{ICN}{Information Centric Networking}
\newacronym{ICMANET}{ICMANET}{Information Centric MANET}
\newacronym{IFE}{IFE}{In-Flight Entertainment}
\newacronym{IP}{IP}{Internet Protocol}
\newacronym{ISP}{ISP}{Internet Service Provider}
\newacronym{IVG}{IVG}{Inter-Vehicles Geocast protocol}
\newacronym{JAUS}{JAUS}{Joint Architecture for Unmanned Systems}
\newacronym{km/h}{km/h}{kilometers per hour}
\newacronym{LAP}{LAP}{Low-Altitude Platforms}
\newacronym{LD}{LD}{Light Diode}
\newacronym{L-DACS}{L-DACS}{L-band Digital Aeronautical Communication System}
\newacronym{LED}{LED}{Light Emitting Diode}
\newacronym{LEO}{LEO}{Low-Earth Orbit}
\newacronym{LORA-CBF}{LORA-CBF}{Location Routing Algorithm with Cluster Based Flooding}
\newacronym{LOS}{LOS}{Line-Of-Sight}
\newacronym{LOUVRE}{LOUVRE}{Landmark Overlays for Urban Vehicular Routing Environments}
\newacronym{LTE}{LTE}{Long-Term Evolution}
\newacronym{MAC}{MAC}{Media Access Control}
\newacronym{MANET}{MANET}{Mobile Ad-hoc Network}
\newacronym{MDDV}{MDDV}{Mobility-centric Data Dissemination algorithm designed for Vehicular networks}
\newacronym{MCA services}{MCA services}{Mobile Communication services on Aircraft}
\newacronym{MCRT}{MCRT}{Monte Carlo Ray-Tracing}
\newacronym{MHVB}{MHVB}{Multi-Hop Vehicular Broadcast}
\newacronym{MHz}{MHz}{MegaHertz}
\newacronym{MIMO}{MIMO}{Multiple-Input Multiple-Output}

\newacronym{MSAS}{MSAS}{MTSAT Satellite-based Augmentation System}

\newacronym{MTSAT}{MTSAT}{Multi-functional Transport SATellite}
\newacronym{NASDA}{NASDA}{National Space Development Agency}
\newacronym{NextGen}{NextGen}{Next Generation air transportation}
\newacronym{NM}{NM}{Nautical Mile}
\newacronym{OFDM}{OFDM}{Orthogonal Frequency-Division Multiplexing}

\newacronym{PHY}{PHY}{PHYsical}
\newacronym{PPM}{PPM}{Pulse-Position Modulation}
\newacronym{PSR}{PSR}{Primary Surveillance Radar}
\newacronym{PKI}{PKI}{Public Key Infrastructure}
\newacronym{QAM}{QAM}{Quadrature Amplitude Modulation}
\newacronym{QoE}{QoE}{Quality-of-Experience}
\newacronym{QoS}{QoS}{Quality-of-Service}
\newacronym{RA}{RA}{Resolution Advisorie}
\newacronym{RF}{RF}{Radio Frequency}
\newacronym{SATCOM}{SATCOM}{SATellite COMmunication}
\newacronym{SELCAL}{SELCAL}{SELective CALling}
\newacronym{SESAR}{SESAR}{Single European Sky Air Traffic Management Research}
\newacronym{SHF}{SHF}{Super High Frequency}
\newacronym{SNR}{SNR}{Signal-to-Noise Ratio}
\newacronym{SSR}{SSR}{Secondary Surveillance Radar}
\newacronym{SSB}{SSB}{Single Side Band}
\newacronym{STDMA}{STDMA}{Self-organized Time Division Multiple Access}
\newacronym{STBC}{STBC}{Space-Time Block Coding}

\newacronym{TA}{TA}{Traffic Advisorie}
\newacronym{TCAS}{TCAS}{Traffic Collision Avoidance System}
\newacronym{TDD}{TDD}{Time-Division Duplex}
\newacronym{TDMA}{TDMA}{Time Division Multiple Access}
\newacronym{TRACON}{TRACON}{Terminal Radar Approach CONtrol}
\newacronym{UAT}{UAT}{Universal Access Transceiver}
\newacronym{UAV}{UAV}{Unmanned Aerial Vehicle}
\newacronym{UHF}{UHF}{Ultra High Frequency}
\newacronym{UMB}{UMB}{Urban Multi-Hop Broadcast protocol}
\newacronym{US}{US}{United States}
\newacronym{VANET}{VANET}{Vehicular Ad-hoc Network}
\newacronym{VDL}{VDL}{VHF Data Link}
\newacronym{VHF}{VHF}{Very High Frequency}
\newacronym{VOR}{VOR}{VHF Omnidirectional Range}
\newacronym{VoIP}{VoIP}{Voice over IP}
\newacronym{VPKI}{VPKI}{Vehicular Public Key Infrastructure}
\newacronym{V-TRADE}{V-TRADE}{Vector-based TRAcking DEtection}
\newacronym{WAM}{WAM}{Wide-Area Multilateration}
\newacronym{WiFi}{WiFi}{Wireless Fidelity}
\newacronym{WiMAX}{WiMAX}{Worldwide interoperability for Microwave Access}

\usepackage[ulem=normalem]{changes}
\begin{document}
{\linespread{2.0}
\title{Aeronautical Ad-Hoc Networking for the Internet-Above-The-Clouds}
\author{Jiankang~Zhang,~\IEEEmembership{Senior~Member,~IEEE,}
        Taihai~Chen,
        Shida~Zhong,
		Jingjing~Wang,
	    Wenbo~Zhang,
	    Xin~Zuo,
        Robert~G.~Maunder,~\IEEEmembership{Senior~Member,~IEEE},
				Lajos Hanzo~\IEEEmembership{Fellow,~IEEE}
\thanks{J.~Zhang, R.~G.~Maunder and L.~Hanzo are with Electronics and Computer Science, University of Southampton, U.K. (E-mails: jz09v@ecs.soton.ac.uk, rm@ecs.soton.ac.uk, lh@ecs.soton.ac.uk)}
\thanks{T.~Chen is with AccelerComm Ltd., U.K. (E-mail: taihai.chen@accelercomm.com)}
\thanks{S. Zhong is with College of Information Engineering, Shenzhen University, China (E-mail: shida.zhong@szu.edu.cn)}
\thanks{J.~Wang is with the Department of Electronic Engineering, Tsinghua University, Beijing, 100084, China. (E-mail: chinaeephd@gmail.com)}
\thanks{W. Zhang is with College of Information and Communication Engineering, Beijing University of Technology, China. (E-mail: wenbozhang@bjut.edu.cn)}
\thanks{X.~Zuo is with Huawei Technologies Co., Ltd. Shenzhen, China (E-mail: xinzuo$\_$cn@163.com)}
\thanks{The financial support of the EPSRC projects EP/Noo4558/1, EP/PO34284/1 as well as of the European Research Council's Advanced Fellow Grant QuantCom is gratefully acknowledged.}
}
}								
 
\maketitle
{\color{black}
\begin{abstract}
The engineering vision of relying on the ``smart sky" for supporting air traffic and the ``Internet above the clouds" for in-flight entertainment has become imperative for the future aircraft industry. Aeronautical {\it{ad hoc}}
Networking (AANET) constitutes a compelling concept for providing broadband communications above clouds by extending the coverage of Air-to-Ground (A2G) networks to oceanic and remote airspace via autonomous and self-configured wireless networking amongst commercial passenger airplanes. The AANET concept may be viewed as a new member of the family of Mobile {\it{ad hoc}} Networks (MANETs) in action above the clouds. However, AANETs have more dynamic topologies, larger and more variable geographical network size, stricter security requirements and more hostile transmission conditions. These specific characteristics lead to more grave challenges in aircraft mobility modeling, aeronautical channel modeling and interference mitigation as well as in network scheduling and routing. This paper provides an overview of AANET solutions by characterizing the associated scenarios,  requirements and challenges. Explicitly, the research addressing the key techniques of AANETs, such as their mobility models, network scheduling and routing, security and interference are reviewed. Furthermore, we also identify the remaining challenges associated with developing AANETs and present their prospective solutions as well as open issues. The design framework of AANETs and the key technical issues are investigated along with some recent research results. Furthermore, a range of performance metrics optimized in designing AANETs and a number of representative multi-objective optimization algorithms are outlined.
\end{abstract}
}
\IEEEpeerreviewmaketitle
 
\section{Introduction}\label{S1}
\begin{figure*}[bp!]
\begin{center}
\subfigure[Physical topology]{\includegraphics[width=0.77\textwidth]{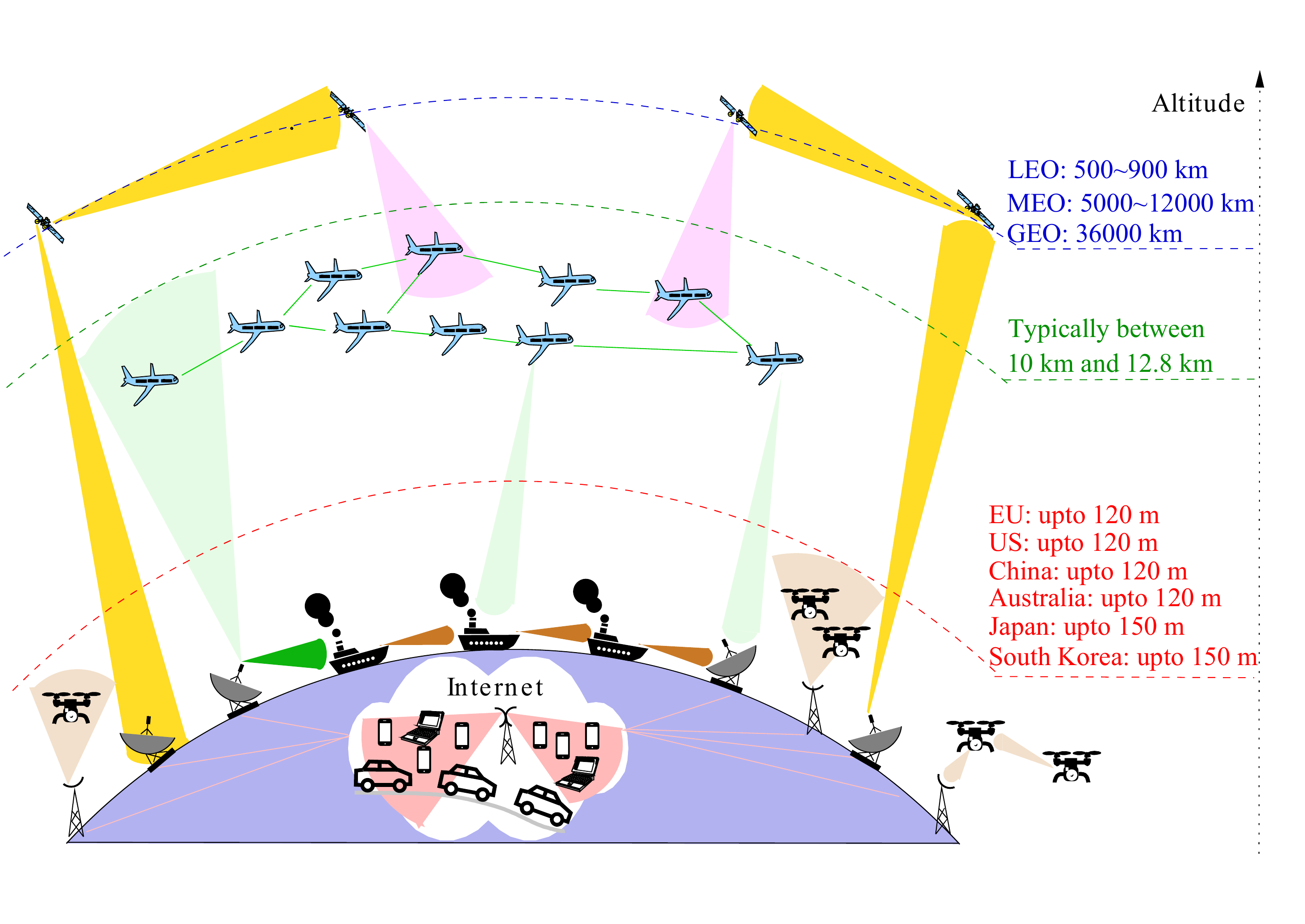} 
\label{FIG2a}}
\subfigure[Logical topology]{\includegraphics[width=0.21\textwidth]{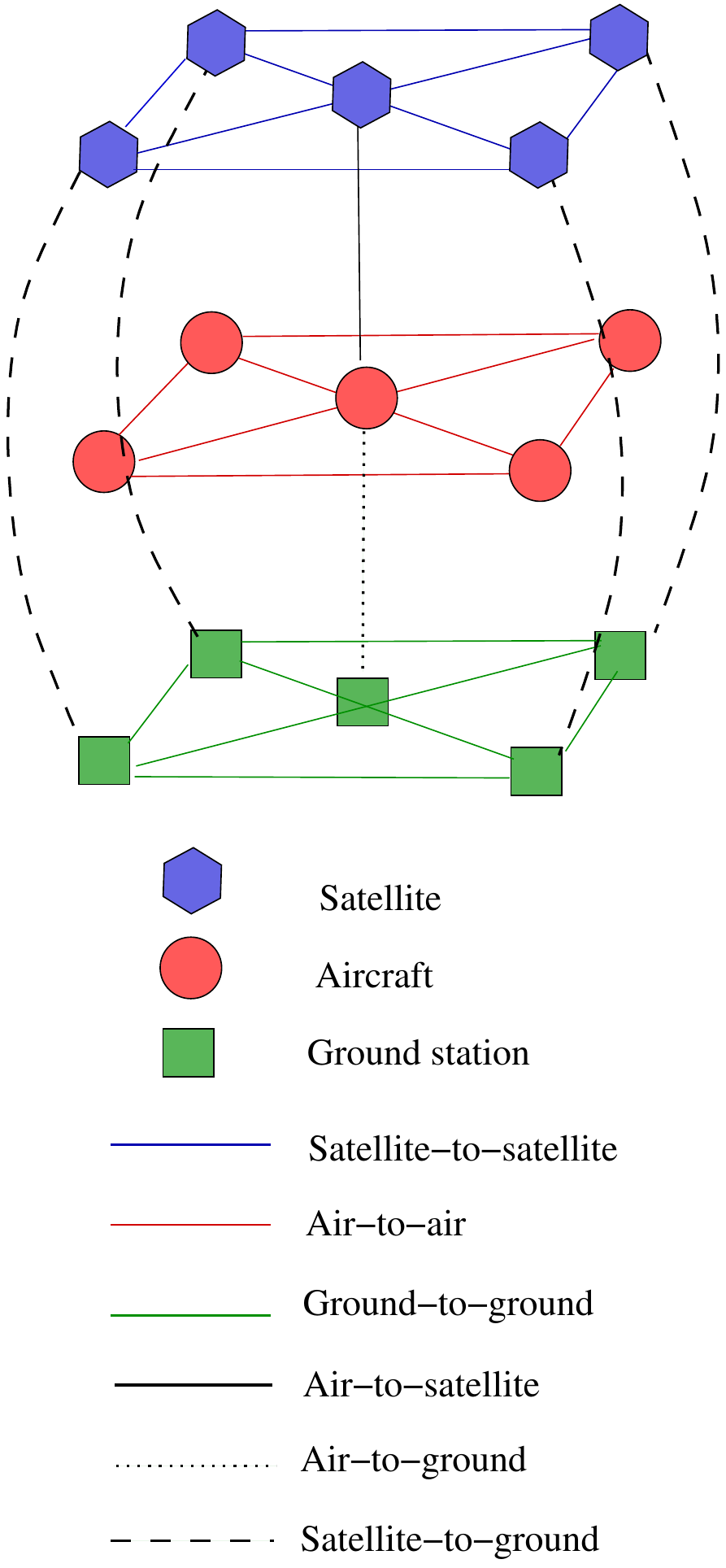}
\label{FIG2b}}
\end{center}
\caption{The AANET topology and the corresponding logical topology.}
\label{FIG1:AANET_topology}
\vspace{-2mm}
\end{figure*}

Trans-continental travel and transport of goods has become part of the
economic and social fabric of the globe. Therefore, the number of
domestic and international passenger flights is expected to grow
for years to come. For Europe as an example, it is
predicted that there will be 14.4 million flights in 2035, which
corresponds to a $1.8\%$ average annual growth compared to the flights
in 2012~\cite{eurocontrol2013Challenges}. However, passenger aircraft
remain one of the few places where ubiquitous data connectivity cannot
be offered at high throughput, low latency and low cost.  A survey by Honeywell \cite{Honeywell2016} revealed that
nearly $75\%$ of airline passengers are ready to switch airlines to secure
access to a faster and more reliable Internet connection on-board and
more than $20\%$ of passengers have already switched their airline for
the sake of better in-flight Internet access. Furthermore, in an
effort to provide potentially more efficient air traffic management
capabilities, ``free flight"~\cite{erzberger2001method} has been
developed as a new concept that gives pilots the ability to change
trajectory during flight, with the aid of \glspl{GS} and/or
\gls{ATC}. These demands have inspired both the
academic and industrial communities to further develop aeronautical communications. The joint European-American research activities were launched in 2004 for further developing the future communication infrastructure~\cite{neji2013survey}, led by the \gls{NextGen} in the US and by the \gls{SESAR} in Europe. However, they mainly focused their attention on improving  aeronautical communications for \gls{ATM} rather than on providing stable Internet access during cross-continental flights. Nonetheless, the ever-increasing interest in providing both Internet access and cellular connectivity in the passenger cabin has led to the emergence of in-flight \gls{WiFi} based both on satellite connectivity and on the Gogo \gls{A2G} network. However, they suffer from expensive subscription, limited coverage, limited capacity and high end-to-end delay. As a complement and/or design alternative, the \gls{AANET}~\cite{vey2014aeronautical,medina2011airborne} concept has been conceived as a large-scale
multi-hop wireless network formed by aircraft, which is capable of
exchanging information using multi-hop \gls{A2A} radio communication
links {\color{black}as well as integrating both the satellite networks and the ground networks.}, as shown in Fig.~\ref{FIG1:AANET_topology}. More explicitly, the
middle layer of objects is constituted by the aircraft of an 
\gls{AANET}, which are capable of exchanging information with the
satellite layer (top layer) and \gls{GS} layer (bottom layer) via
inter-layer links. Furthermore, \glspl{AANET} are also beneficial for automatic
node and route discovery as well as for route maintenance as aircraft fly within
the communications range of each other, hence allowing data to be
automatically routed between aircraft and to or from the
\gls{GS}.
The representative benefits of AANET are summarized as follows
\begin{itemize}
\item \emph{Extended Coverage}: AANETs extend the coverage of \gls{A2G} networks offshore
to oceanic or remote airspace by establishing an {\it{ad hoc}} network among
aircraft and \glspl{GS}. The \glspl{GS} may also communicate with each
other as part of an \gls{AANET} or they may act as a gateway for
connecting with the Internet via a fixed line. More specifically,
\glspl{AANET} are capable of substantially extending the coverage
range in the oceanic and remote airspace, without any additional
infrastructure and without relying on satellites.
\item \emph{Reduced Communication Cost}: Avoiding satellite
links directly reduces the airlines' cost of aeronautical
communication, since the cost of a satellite link is usually
significantly higher than that of an \gls{A2G} link~\cite{medina2011geographic}.
\item \emph{Reduced Latency}: Another potential benefit of \gls{AANET} is its reduced latency compared to geostationary satellite-based access, hence it is capable of supporting more delay-sensitive applications such as
    interactive voice and video conferencing.
\end{itemize}

The improvements that
\gls{AANET} offers to civil aviation applications may be much
appreciated by the passengers, operators, aircraft manufacturers and
\glspl{ATC}. More specifically, \gls{AANET} allows aircraft to
upload/download navigation data and passenger service/entertainment
data in a wireless live-streaming manner. The congestion of the
airspace at peak times will be mitigated by the more punctual
scheduling of takeoff/landing. Furthermore, \gls{AANET} allows direct
communication among aircraft for supporting formation flight or for
preventing disasters and terrorist attacks. It also grants access to
the Internet and facilitates telephone calls above the clouds, as well
as maintaining communications with airlines for various purposes, such
as engine performance or fuel consumption reports.

\subsection{Motivation}

As a new breed of networking, \gls{AANET} aims to establish an {\it{ad hoc}} network amongst aircraft for their direct communication in  high-velocity and high-dynamic scenarios, in order to handle the increasing flow of data generated by aircraft and to provide global coverage.
\glspl{AANET} have become an increasingly important research topic in recent years, and considerable progress has been made in conceiving the network structure~\cite{vey2014aeronautical,medina2011airborne} and network topology~\cite{sakhaee2007stable}. However, the significant remaining challenges must be overcome before they can be implemented in commercial systems. Airlines are demanding the connectivity offered by AANETs to provide on-board WiFi, while governments need AANETs for improving the operating capacity of the airspace. To motivate researchers both in the academic community and in the industrial community, as well as those who are concerned with the development of aeronautical communication, it is essential to understand the potential applications, requirements and challenges as well as the existing aeronautical communication systems/techniques. Despite these compelling inspirations, at the time of writing there is a paucity of detailed comparative surveys of aeronautical communication solutions designed for commercial aircraft taking into account their specific characteristics, scenarios, applications, requirements and challenges. Hence, we aim to fill this gap by conceiving this survey of AANETs. The objective of this survey is to offer an insight for inquisitive readers into the current status and the future directions of AANETs. We aim for motivating engineers in the aviation industry and researchers in the academic community to contribute to the development of  AANETs.

{\color{black}
\subsection{Our Contributions}
More specifically, we compare the AANETs to the existing family members of wireless {\it{ad hoc}} networks by identifying the specific features of AANETs. Following this, we investigate different scenarios of AANETs, including airports as well as populated and unpopulated areas, which result in strict requirements and impose challenges on the design of AANET. Before we scrutinize the remaining challenges, we comprehensively review the field of aeronautical communications in terms of A2G communications, A2A communications, A2S communications, in-cabin communications and multi-hop communications. Their capabilities in meeting the requirements as well as in accommodating diverse fundamental and enhanced aeronautical applications with the aid of Table~IX.

Then, the challenges associated with designing AANETs are analyzed and the recent research progress in addressing these challenges is also discussed. To facilitate future research in investigating AANETs, we provide a general design framework for AANETs and highlight some key technical issues in designing AANETs as well as present some of our recent experimental results. Moreover, we outline a range of performance metrics in jointly optimizing the AANET design as well as a number of representative multi-objective optimization algorithms. Based on the lessons learned from prior research, we also suggest promising research directions for AANETs, as well as highlight the open issues to be solved for implementing AANETs in practice.
}

\begin{figure}
\begin{center}
 \includegraphics[width=1.2\columnwidth]{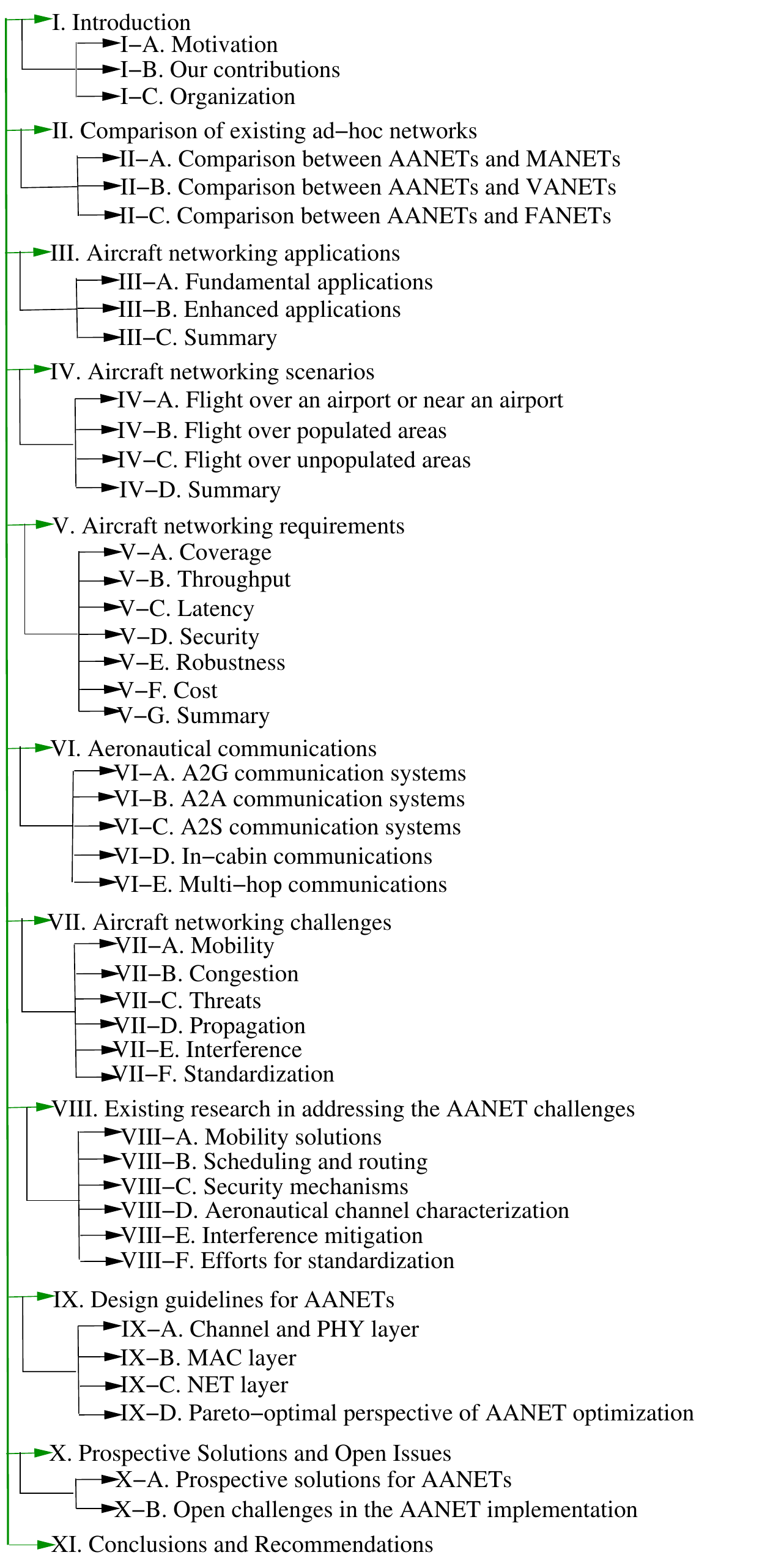}
\end{center}
\vspace{-1mm}
\caption{The skeleton structure of this paper}
\label{FIG3:contents}
\end{figure}

\subsection{Organization}

The rest of this paper is organized as follows. {\color{black}The comparison between the family members of wireless {\it{ad hoc}} networks is presented in Section~\ref{S2add}.} Section~\ref{S2} is devoted to the description of aircraft networking applications, including flight data delivery at airports, air traffic control, aircraft tracking, formation flight, free flight and passenger entertainment. The typical aircraft networking scenarios of airports, populated areas and unpopulated areas will be discussed in Section~\ref{S3}. In order to reliably operate in miscellaneous scenarios and applications, \glspl{AANET} have to meet specific requirements, which are discussed in Section~\ref{S4}. In Section~\ref{S5}, we describe both existing and emerging aircraft communications systems and discuss their suitability to \glspl{AANET}. We will also demonstrate that there are still some challenges to be addressed, as discussed in Section~\ref{S6}. In Section~\ref{S7}, we review the research efforts invested in addressing the open \gls{AANET} challenges. {\color{black} The relevant design guidelines and the key technologies are illustrated in Section~\ref{S8aa1}.} The prospective solutions and open issues of \glspl{AANET} are discussed in Section~\ref{S8}. Finally, our conclusions  are offered in Section~\ref{S9}.
The organization of this paper is shown at a glance in Fig.~\ref{FIG3:contents}.

\section{Comparison of existing {\it{ad hoc}} networks}\label{S2add}
{\color{black}The Mobile {\it{ad hoc}} Network (MANET) paradigm has been developed for providing direct network connections among wireless devices. During the earliest stage evolution the nodes were mobile users, later the nodes evolved to vehicles and then the nodes evolved to unmanned aerial vehicles (UAVs) and in this treatise the nodes evolve to airplanes. The acronym MANET constitutes a gereralised terminology, which includes {\it{ad hoc}} networks mobile users, vehicles and UAVs as well as airplanes. However, the specific terminologies of Vehicular {\it{ad hoc}} Networks (VANETs), Flying {\it{ad hoc}} Networks (FANETs) and AANETs refer to the networks constituted by vehicles, UAVs and airplanes, respectively. In this section, we will compare the existing MANETs, VANETs and FANETs with the AANETs in terms of their fundamental characteristics.} 

\subsection{Comparison between AANETs and MANETs}

The \gls{MANET} concept was conceived in the
1990s to enable nearby users to directly communicate with each other
without the need for any central network infrastructure. This is achieved by
exploiting the user devices' wireless interfaces in an {\it{ad hoc}} manner,
in order to exchange data or relay traffic~\cite{conti2014Mobile}. Given their high-velocity node mobility,
dynamic topology, decentralised architecture and limited transmission
range, legacy Internet routing and transport protocols do not work
properly in \glspl{MANET}, hence the recent research efforts have been
focused on these areas. More particularly, Abid {\it{et al}}.~\cite{abid2015survey} present a survey of the distributed hash table
based routing protocols that are capable of enhancing \glspl{MANET},
whilst identifying their features, strengths and weaknesses as well as
the corresponding research challenges. With the advent of the future
Internet architecture, relying on the information centric network concept, the Internet
is moving from the conventional host-centric design to the
content-centric philosophy. A new form of \glspl{MANET} termed as the
information centric MANET has also emerged. In~\cite{liu2017information}, the
authors interpret the formation of information centric MANET and define the conceptual
model of its content routing. Three types of schemes, namely
proactive, reactive and opportunistic arrangements, are also described
to exemplify the content routing procedure. Apart from routing,
surveys on important aspects such as security and neighbour discovery
have also been produced by the \gls{MANET} research community, which
shows that the routing information and encryption defeating approaches~\cite{dorri2015security} are the most effective \gls{MANET} security
solutions. Dorri {\it{et al}}.~\cite{dorri2015security} have also
characterized a range of energy-efficient neighbour discovery
protocols~\cite{sun2014energy} in order to emphasize the need for
connectivity maintenance and context awareness.

\glspl{AANET} constitute a new member of the \glspl{MANET} family, since they inherit some of the general mobility features of the nodes, as well as the dynamic nature of the network topology, and the self-organizing traits of the network.  However, they also differ quite significantly in a range of specific aspects. In terms of mobility, a mobile device of \glspl{MANET} typically travels at human walking speed in random directions and exhibits varying node density. By contrast, the aircraft of \glspl{AANET} typically travel at high speed along planned flight trajectories whilst exhibiting a much lower node density while en-route over the ocean. Thus, the mobility models for emulating the node behaviours are completely different for \glspl{MANET} and \glspl{AANET}. Hence the random walk model designed for \glspl{MANET}~\cite{nayak2015analysis} is not applicable to \glspl{AANET}, whereas the smooth semi-Markov model designed for \glspl{AANET} is not directly usable for \glspl{MANET}. Additionally, the power consumption is also rather different for \glspl{MANET} and  \glspl{AANET}. The typically battery-powered nodes of \glspl{MANET} have to utilise energy-efficient methods in order to extend the network lifetime. On the other hand,  aircraft are powered by jet engines, which can provide ample power for communication systems. Hence, the power consumption of \glspl{AANET} does not impose challenges.  Furthermore, radio propagation characteristics are also different. \glspl{MANET} operate on the ground and they often suffer from Rayleigh fading. By contrast, a \gls{LOS} path propagation exists for a pair of \gls{A2A} communicating aircraft.  The above differences may also result in notable routing and forwarding differences at the network layer. Explicitly, the routing and forwarding methods  assist in avoiding  congestion by aiming for the maximum throughput per aircraft, while maintaining the shortest possible end-to-end delay from one continent to another. This is because both the throughput and the delay constitute key requirements for passengers  accessing the Internet. On the other hand, energy-aware routing as well as proactive and reactive routing are favoured in  \glspl{MANET}, since they maximise the network lifetime and mitigate the effects of its unpredictable dynamic nature.

\subsection{Comparison between AANETs and VANETs}

\gls{AANET} relies on {\it{ad hoc}} networking, which can disseminate information using multi-hop communication without a central  infrastructure. {\it{ad hoc}} networks have already been developed for ground-based \gls{VANET}~\cite{raya2007securing}, which constitute a subclass of \gls{MANET}~\cite{chlamtac2003mobile,wang2018vehicular}. For example, \glspl{VANET} have been successfully applied in collision warning~\cite{sou2013modeling}, in road trains for allowing vehicles to be driven in formation~\cite{coelingh2012all},  as well as for providing Internet connectivity~\cite{gramaglia2011overhearing,wang2018internet} in vehicles. Although both \gls{AANET} and \gls{VANET} are forms of {\it{ad hoc}} networking, the transceivers, receivers and routers in \gls{AANET} are carried by aircraft. These systems must be designed for exploiting all aircraft assets, in order to connect with satellite- and ground-networks for the sake of constructing a seamless communications platform across the air, space and terrestrial domains.

Furthermore, \glspl{AANET} have many different characteristics compared to conventional \glspl{VANET}. First of all, the speed of nodes in \glspl{VANET} is much lower, staying within the range of a few \gls{km/h} to tens of \gls{km/h} for the higher-speed \glspl{VANET}~\cite{sakhaee2007stable}. But the nodes in \glspl{AANET}, namely aircraft, fly at velocities of 800 km/h to 1000 km/h. This very high velocity leads to serious Doppler shift and highly dynamic mobility, which results in the frequent setup and breakup of communication links between aircraft. Secondly, aircraft may fly over a very large-scale range, spanning across oceans, deserts and continents. In unpopulated areas, such as the North Atlantic, the aircraft density is relatively
low, but the aircraft routes are planned and updated twice daily by taking into account both the jet stream and the principal
traffic flow, leading to a very sparse but relatively stable network topology. However, the route can be adjusted according to an aircraft's own specific circumstances, in order to reach the destination airport faster
or to aim for a particular landing slot. Furthermore, in continental airspace or near airports, aircraft move at random angles to each other, especially in the high-density European airspace~\cite{medina2008topology}. Thirdly, the antenna employed by aircraft have the rigorous requirements of good isolation, high efficiency and ease of integration with the aircraft~\cite{hsu2009dual}. The suitable antenna installation sites on aircraft are typically limited to the wings, tail units or pertaining control flaps~\cite{mehltretter2003structural}. In addition, the implementation and deployment of  communication systems may be overlooked in some research work, but it is of vital significance in aircraft communication systems. The lack of understanding of the economic value, the security mechanisms and its added value for users may make the deployment of \gls{AANET} appear risky to manufacturers, airlines and regulatory organizations. Even once \glspl{AANET} have been implemented and deployed in aircraft, they also face the challenges of meeting different regulations for aircraft communication and Internet access in different countries. Therefore, further significant challenges are faced by \gls{AANET}.

\subsection{Comparison between AANETs and FANETs}

\begin{table*}[htp!]
\caption{Most recent surveys related with FANET and their main contributions} 
\label{tab:AANET-FANET}
\begin{tabular}{|C{0.7cm}|L{2.9cm}|L{9.7cm}|L{3.0cm}|}
\hline\hline
Year
&\multicolumn{1}{c}{Paper}&\multicolumn{1}{|c|}{Focus/Main Contributions}&\multicolumn{1}{c|}{Object}\\\hline\hline
2011 &Bauer and Zitterbart~\cite{bauer2011survey} & Protocols that support IP Mobility& Airplanes\\\hline
2013 &Neji {\it{et al}}~\cite{neji2013survey} & Development activities from 2004 to 2009, PHY layer and MAC layer for L-DACS& Airplanes\\\hline
2013 &Bekmezci {\it{et al}}~\cite{bekmezci2013flying} & Applications, design considerations, communication protocols& UAVs\\\hline
2015 &Saleem {\it{et al}}~\cite{saleem2015integration}&Cognitive radio technology& UAVs\\\hline
2016 &Gupta {\it{et al}}~\cite{gupta2016survey}&Routing protocols, handover schemes, energy conservation& UAVs\\\hline
2016 & Zafar and Khan~\cite{zafar2016flying}&Societal concerns& UAVs\\\hline
2016 & Hayat  {\it{et al}}~\cite{hayat2016survey}&Applications, network requirements& UAVs\\\hline
2017 & Sharma and Kumar~\cite{sharma2017cooperative}&Taxonomy of multi-UAVs, network simulators and test beds& UAVs\\\hline
2018 & Khuwaja {\it{et al}} ~\cite{khuwaja2018survey}&Propagation characteristics of UAV channels& UAVs\\\hline
2018 & Cao {\it{et al}} ~\cite{cao2018airborne}&LAPs, HAPs and integrated airborne communication networks& UAVs, airships, balloons\\\hline
{\color{black}2018} & {\color{black}Liu {\it{et al}} ~\cite{liu2018space}}&{\color{black}Integration of satellite systems, aerial networks, and terrestrial
communications}& {\color{black}Satellites, UAVs, airships, balloons, ground base station and mobile users}\\\hline
\multicolumn{2}{|c|}{\multirow{2}{*}{This paper}}  &Scenarios, applications, requirements, challenges, comprehensive survey of existing aeronautical systems& \multirow{2}{*}{{\color{black}Passenger} airplanes}\\\hline
\end{tabular}
\end{table*}

A relatively new research area of {\it{ad hoc}} networks has gained significance in the wireless research community, namely \glspl{FANET}. This is a type of {\it{ad hoc}} network that connects \glspl{UAV} allows them to autonomously conduct their tasks. No doubt that \glspl{AANET} and \glspl{FANET} share some common features in terms of their mobility and dynamic topology. But \glspl{UAV} are rather  different in terms of their flying speed, flying altitude, trajectory and geographic coverage. Furthermore, 
 they also differ in terms of their technical specifications, applications and requirements. This section identifies the contributions of this survey on \glspl{AANET} beyond some of the most recent surveys of \glspl{FANET}, accentuating the differences between the two.  

The insightful surveys~\cite{bekmezci2013flying,saleem2015integration,gupta2016survey,zafar2016flying,hayat2016survey,sharma2017cooperative,khuwaja2018survey,liu2018space} have covered a wide range of fundamental issues, such as channel modelling~\cite{khuwaja2018survey}, radio frequency aspects~\cite{saleem2015integration}, communication protocols~\cite{bekmezci2013flying,gupta2016survey}, simulators and testbeds~\cite{sharma2017cooperative}, application issues~\cite{bekmezci2013flying,hayat2016survey} as well as societal concerns~\cite{zafar2016flying}. In Table~\ref{tab:AANET-FANET}, we summarize the most representative survey papers on \glspl{FANET} from the past five years, which have focused on various subjects of research and challenges. More specifically, Bekmezci {\it{et al}}.~\cite{bekmezci2013flying} covers diverse application scenarios and design considerations for the physical layer as well as communication protocols up to the transport layer of FANETs. Gupta {\it{et al}}.~\cite{gupta2016survey} focuses on the three important issues in UAV communications networks, namely on existing and new routing protocols conceived for meeting various requirements, such as seamless handovers to allow flawless communication, as well as energy conservation across different communications layers. Zafar {\it{et al}}.~\cite{zafar2016flying} touches not only on the technical aspects of previous surveys, but also on the societal concerns in terms of privacy, safety, security and psychology, plus on the military aspects.  Meanwhile, Hayat {\it{et al}}.~\cite{hayat2016survey} primarily focuses on the communication demands of FANETs from two unique perspectives, namely identifying the qualitative communication demands as well as quantitative communication requirements in the context of  four main applications. By contrast,  Saleem {\it{et al}}.~\cite{saleem2015integration} stresses the need for and potential applications of cognitive radio technology designed for UAVs, as well as its integration issues and future challenges. Sharma {\it{et al}}.~\cite{sharma2017cooperative} focuses on the network architecture and, in particular, on the taxonomy of multi-UAVs, as well as on network simulators/test beds constructed for UAV network formation.
Recently, Khuwaja{\it{et al}}.~\cite{khuwaja2018survey} provided an extensive survey of the measurement methods
of UAV channel modeling and discussed various channel characterization for UAV communications. Cao {\it{et al}}.~\cite{cao2018airborne} presented an overall view on research efforts in the areas of \gls{LAP}-based communication networks, \gls{HAP}-based communication networks, and integrated airborne communication networks. {\color{black}Liu {\it{et al.}} ~\cite{liu2018space} comprehensively surveyed the integration of satellite systems, aerial networks and terrestrial communications, focusing on the aspects of cross layer  operation aided system design, mobility management, protocol design, performance analysis and optimization.}

Despite some similarities between AANETs and FANETs, however, there are also significant dissimilarities between them. In terms of mobility, AANETs have to cope with a significantly higher velocity than FANETs, since commercial aircraft travel at cruise speeds of 880 to 926~km/h, while most UAVs typically travel at speeds of 30 to 460~km/h~\cite{bekmezci2013flying}\footnote{Military \gls{UAV}s can reach the speed of commercial planes or even exceed it, as exemplified by the RQ-4 Global Hark UAV reaching 640~km/h and the X-47B unmanned combat air system reaching Mach 0.9 subsonic speed, i.e. roughly 1100~km/h. However, these UAVs are not designed for {\it{ad hoc}} networking and so they are not considered in our discussions.}.
Owing to this substantial difference in mobility patterns whilst flying, AANETs require mobility models characterizing high Doppler wireless channel fluctuations as well as more prompt topology changes than FANETs.

\begin{table*}[htp!]
\caption{Comparison between existing networks of AANET, MANET, VANET and FANET} 
\label{TAB2:Com_networks}
\begin{tabular}{|C{1.1cm}||C{1.1cm}|C{1.2cm}|C{1.5cm}|C{1.7cm}|C{1.2cm}|C{1.8cm}|C{2.7cm}|C{1.6cm}|}
\hline\hline
Networks
&Objects&Channel&Speed (m/s)&Altitude (m)
&Scale (km)&Power&Density&Security\\\hline\hline
MANET &Mobile phones&Rayleigh&$0$ $\sim~1.5$& $1\sim 250$&0.25&Constraint&Dense&Medium\\\hline
VANET &Vehicles&Rayleigh/ Rician&$4\sim36$& $0.5\sim 5$&1&Non-constraint&City: dense  Rural: sparse
&Life critical\\\hline
FANET &UAVs&Rayleigh/ Rician&$8\sim128$& Upto $122$&80&Constraint&Mission dependent&Medium\\\hline
AANET &Airplanes&Rician&$245\sim257$& $9100 \sim 13000$&740&Non-constraint&Populated area: dense  Unpopulated area: sparse&Life critical\\\hline
\end{tabular}
\end{table*}

Furthermore, because their size is of a completely different scale, their antenna design considerations have to be different. In FANETs, the \glspl{UAV} are generally not large, hence the type and structure of the antenna is of grave concern. By contrast, we have to avoid blocking the signal propagation path in AANETs, when installing antennas on aircraft. In addition to the antenna and aircraft size, the altitude also makes a difference.
For AANETs, the flying altitude is typically 10.68~km, while for FANETs the maximum flying altitude is generally regulated as 122~m (400~feet)~\cite{saleem2015integration}.
More specifically, because of the limitations of currently available sensors as well as the wind speed at higher altitudes and the weather conditions, UAVs are constrained in terms of their altitudes. Therefore, the communication range and communication structure of FANETs is different from that of AANETs.

One of the most substantial differences between AANETs and FANETs is their throughput requirement. Because of the massive throughput demand of the 'Internet above the clouds', the throughput requirements of AANETs have to satisfy the passengers' needs in the aircraft. The recently developed GoGo@2Ku is capable of delivering 70~Mbps peak transmission rate for each aircraft and its next-generation version aims for achieving a 200+~Mbps peak transmission rate~\cite{gogo20162ku}. On the FANET side, although the throughput requirements vary between applications, the one that requires the highest throughput is visual tracking and surveillance, as exemplified by a rate of 2~Mbps for video streaming~\cite{hayat2016survey}. The total throughput requirement of the most demanding applications in FANETs is still at least one or even two order(s) of magnitude lower than that of AANETs. This huge difference leads to significant design and implementational differences in their architecture.

In contrast to the previous FANET surveys concentrating on UAVs, there are also two valuable contributions on airplanes, as summarized in Table~\ref{tab:AANET-FANET}. However, the aeronautical networks they considered were designed for \gls{ATM}. Explicitly, Bauer and Zitterbart~\cite{bauer2011survey} investigated the protocols that can be used to support IP Mobility for aeronautical communications amongst airplanes. Neji {\it{et al}}~\cite{neji2013survey} gave an overview of aeronautical communication infrastructure development activities spanning from 2004 to 2009 and focused their attention on the \gls{L-DACS} in terms of its \gls{PHY} characteristics and \gls{MAC} characteristics. However, given the more advanced solutions that we have a decade later, the time has come for us to focus our efforts on aeronautical communication designed for commercial airplanes. Specifically, we offer insights into AANETs, covering their networking scenarios, applications, networking requirements and real-life communication systems designed for commercial aircraft, as well as into the challenges to be addressed.

A birds eye perspective of \gls{AANET}, \gls{MANET}, \gls{VANET} and \gls{FANET} is illustrated in Table~\ref{TAB2:Com_networks}, where issues, such as the propagation channel, speed, altitude, network scale, power constraint, node density and security are considered. Although, the \gls{MANET} has initially been designed both for mobile phones and for vehicles, we have classified vehicles into \glspl{VANET}, which are specifically developed for connecting vehicles. \gls{AANET} distinguishes itself from \gls{MANET},  \gls{VANET} and \gls{FANET} in terms of its features, such as its flying speed,  network coverage and altitude, which directly result in new propagation characteristics and impose challenges both on the data link layer and network layer design.
\section{Aircraft Networking Applications}\label{S2}

\begin{figure}[tp!]
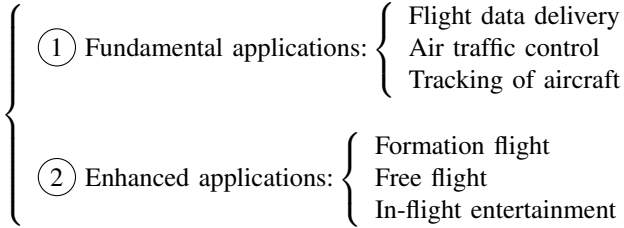

\begin{align}
\left\{\begin{array}{l}
\text{\circled{1} Fundamental applications:}
\left\{
\begin{array}{l}
\text{Flight data delivery}\\
\text{Air traffic control}\\
\text{Tracking of aircraft}
\end{array}
\right. \nonumber
\\[25pt]
\text{\circled{2} Enhanced applications:}
\left\{\begin{array}{l}
\text{Formation flight}\\
\text{Free flight}\\
\text{In-flight entertainment}
\end{array}
\right.\nonumber
\end{array}
\right.
\end{align}
\caption{Applications of AANET}
\label{FIG2:applications}
\end{figure}

\glspl{AANET} are capable of supporting various aviation applications and services for their passengers. More specifically, \glspl{AANET} are capable of enhancing the existing applications of wireless communication in aviation, such as flight data delivery at airports, air traffic control, aircraft tracking, satisfying the emerging vision of formation flight and free flight, as well as entertainment, as shown in Fig.~\ref{FIG2:applications}.

\subsection{Fundamental Applications}

\begin{itemize}[\setlength{\listparindent}{\parindent}]
\item \emph{Flight Data Delivery}: In preparation for an aircraft's next flight, compressed navigation data and passenger service as well as entertainment data are typically uploaded/downloaded to/from the aircraft at an airport. Traditionally, this information is stored and transported using a large-capacity physical disk. However, getting the right disk to the right aircraft at the right time requires significant effort and resources~\cite{wright2001wirelesspread}, which is expensive. 
    
    By contrast, \glspl{AANET} allow the upload/download of flight summary reports, raw flight data and passenger audio/video files to/from the aircraft, whenever the aircraft is within communication range of an airport's \glspl{GS}~\cite{wright2000wireless}. \gls{AANET} can also allow those databases to be maintained in real time, while the aircraft is in flight. In this way, time and expense  can be saved by uploading/downloading data not only while the aircraft is parked, but also while it is landing, takeoff and even enroute.

\item \emph{Air Traffic Control}: Maintaining safety and high efficiency are the main objectives of a communication system supporting air-traffic, as it will be discussed in Section~IV. The current system relies on ground-based radar solutions for centralized surveillance, which allows the air traffic service providers or airline operation centers to receive reports sent by aircraft. However, it is expensive to deploy radar systems, which rely on very large antenna structures and require costly regular maintenance~\cite{park2014hybrid}. Moreover, radar-based solutions fail to achieve the \gls{ATC}'s expectation of global coverage, since it is typically not possible to deploy radars in unpopulated areas, owing to the cost and challenge of maintaining them. Furthermore, they are also impractical to implement on a large scale, since they require information from all aircraft to be relayed to the central facility. This problem will be exacerbated as the traffic demand increases~\cite{park2014hybrid}. 
    
    As a solution to this, \glspl{AANET} can be used in both congested regions and in low-density regions. This may be achieved by using self-organization and multi-hop relaying for the aircraft to exchange their \gls{GPS} locations, instead of merely relying on \glspl{GS}~\cite{strohmeier2014realities}. Owing to this, \gls{AANET} is more capable of meeting the requirements of achieving long range, low latency, automated discovery/healing/control, strong security and robustness. As an extra benefit,  \glspl{AANET} may allow a minimum safety separation of 5 \glspl{NM} in unpopulated areas, instead of the current safety separation requirement of 50 \glspl{NM}.
    
\item \emph{Tracking of Aircraft}: Almost 900 people lost their lives in the aircraft disappearances and aircraft accidents that happened in 2014~\cite{yu2015aftermath}. This motivates an \gls{AANET}-based live-streaming solution for maintaining connection between aircraft and the rest of the world, especially in unpopulated areas. Moreover, in the event of an emergency or terrorist attack, the \gls{ATC} may desire to take control of the aircraft and take whatever action is necessary to maintain safety~\cite{nelson1998flight}, such as engaging the autopilot~\cite{kainrath2016communication} and locking out any unauthorized access to the aircraft controls. When the aircraft is flying over a populated area, this could be achieved using \glspl{GS}. However, when the aircraft is flying over an unpopulated area, such as an ocean or desert, the only option today is to use a satellite link~. However, the round-trip latency of satellite links can be as long as 250 ms~\cite{medina2012geographic}, which may be considered unsuitable for the delivery of emergency control signals. Moreover, during disasters and accidents, organizing, coordinating, and commanding an aircraft are significant technical and operational challenges, which require timely collection, processing and distribution of accurate information from disparate systems and platforms. 
    
    These issues impose great challenges in the design of a robust solution for tracking aircraft. However, by exploiting direct \gls{A2A} communication and multi-hop relays among aircraft, \gls{AANET} may provide an efficient and robust network that is capable of meeting these challenges with low latency and high reliability.
\end{itemize}

\begin{figure*}
\begin{center}
 \includegraphics[width=1.0\textwidth,angle=0]{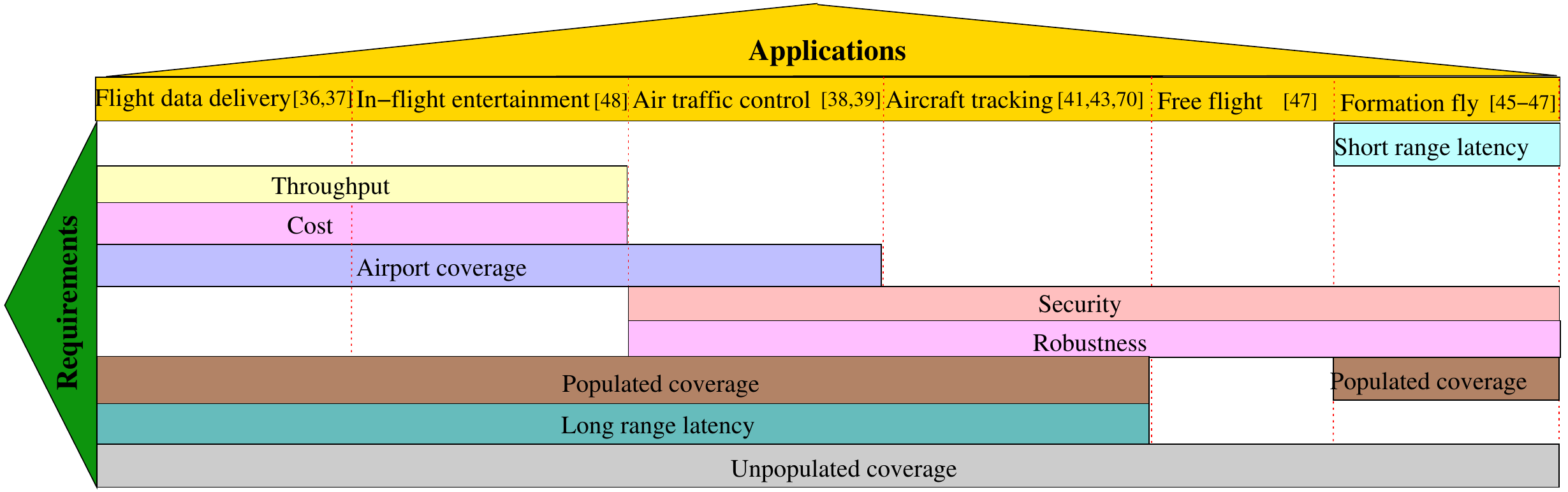}
\end{center}
\vspace*{-1mm}
\caption{Requirements imposed by the potential applications}
\label{FIG3:app_requirement}
\end{figure*}

\subsection{Enhanced Applications}

\begin{itemize}[\setlength{\listparindent}{\parindent}]
\item \emph{Formation Flight}: As stated by the European Union's 2020 Vision, the carbon dioxide footprint of aviation
should decrease by $50\%$ by the year 2050 compared to 2005. To achieve this goal, one of the promising solutions is formation flight. This approach is inspired by the formation flight of birds, where a $71\%$ increase in flying range can be observed~\cite{rayner1979vortex}. In principle, aircraft
could also save fuel by taking advantage of formation flight during the enroute flight phase, thus leading to further
reduction of $\text{CO}_{2}$ emissions and costs. A case study of an aircraft design specifically optimized for
formation flight was reported in~\cite{Bos2010formation}, where an average of $54\%$ fuel savings were observed
over the most fuel-efficient long-haul state-of-the-art Boeing 787-8 available in 2011. 

One of the key concerns in formation flight is collision avoidance, which has rigorous requirements in terms of latency, security and robustness. At the time of writing the Global Navigation
Satellite System (GNSS), together with the Inertial Navigation System (INS), is used for ensuring accurate and safe spacing between aircraft~\cite{dijkers2011integrated,Bos2010formation}. However, satellite communications tend to be unreliable and of high latency. Motivated by this, the Airbus 2050 vision for ``Smarter Skies"~\cite{Airbus2015} suggests that aircraft should be able to communicate with each other
for the purpose of collision avoidance, which will enable aircraft to autonomously maintain the most beneficial separation during formation flight. This requires low-latency communication to enable the
real-time autonomous reaction of an aircraft to the movement of a neighbour aircraft during
turbulence. This {\it{ad hoc}} inter-aircraft communication is one of the long-term applications of \glspl{AANET}.

\item \emph{Free Flight}: Pilots now have to ask for permission from \gls{ATC} for any deviation from the original flight path, as required for example by poor weather conditions. \gls{ATC} responds to the pilots' request according to its perception of the air traffic conditions in the vicinity of the aircraft. However, the \gls{ATC} may not be able to accurately assess the surrounding conditions of the requesting aircraft, since it relies on off-site monitoring by a radar system, for example. The recent aircraft crash of AirAsia flight QZ8501 had asked for permission to climb in order to avoid a storm cluster. However, its request was deferred by \glspl{ATC} owing to heavy air traffic, since there were seven other aircraft in the vicinity. If aircraft in this situation were able to form an \gls{AANET} then they would be able to adjust their heading, altitude and speed based on the information shared by the other aircraft in the \gls{AANET}.
    
    Free flight~\cite{Airbus2015} has become a concept recommended by the \gls{ICAO} for future air traffic management. Although the concept of free flight was first proposed about two decades ago, it currently has no feasible technical solution, when relying on the traditional centralized technologies, which are unavailable in unpopulated areas. However, free flight may indeed be achieved by exploiting the self-discovering, self-organizing and self-healing, as well as the distributed control of \glspl{AANET}, which is capable of providing robust \gls{A2A} communication at a low latency.

\item \emph{In-Flight Entertainment}: Design studies carried out by airlines and market surveys of in-flight network providers demonstrate the necessity of high-data-rate communications services for airliners, with an obvious trend toward in-flight entertainment, Internet applications and personal communications~\cite{jahn2003evolution} regardless of whether the aircraft is in a populated or unpopulated area. 
    
    However, the provision of a global airborne Internet service is only possible at the time of writing via satellite links, which are costly and suffer from long round trip delays of up to several seconds~\cite{medina2010routing}. This leads to one of the bottlenecks for the future expansion of the aeronautic industry. \glspl{AANET} would facilitate multi-hop communications between aircraft and the ground, by establishing an {\it{ad hoc}} network among aircraft within reliable communications range of each other. In this way, each aircraft will transmit or relay data packets to the next aircraft or \gls{GS}, potentially offering lower latency, lower cost and higher throughput than satellite-aided relaying.
\end{itemize}

\begin{figure}
\begin{center}
\subfigure[Heathrow airport]{\includegraphics[width=0.47\textwidth]{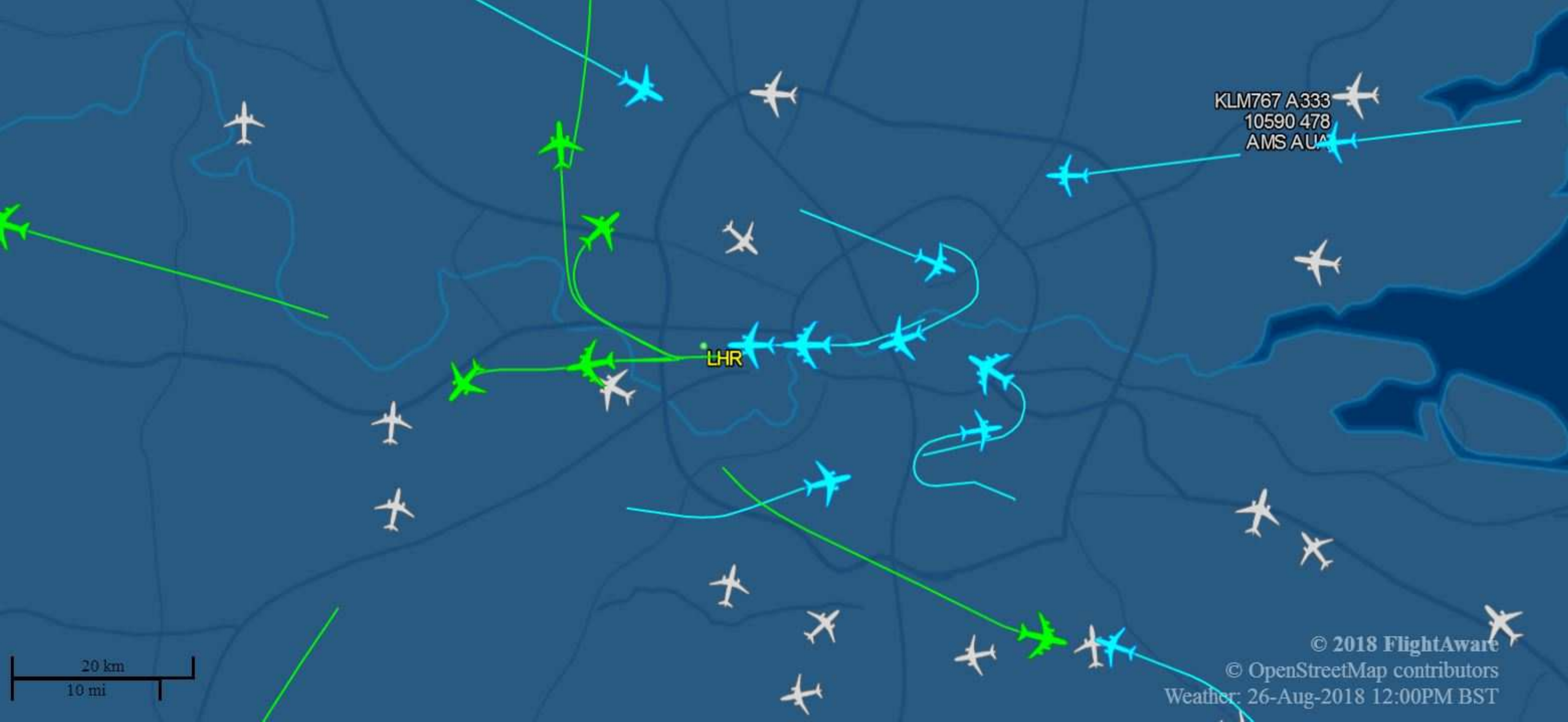}
\label{FIG4a}}
\subfigure[European airspace]{\includegraphics[width=0.47\textwidth]{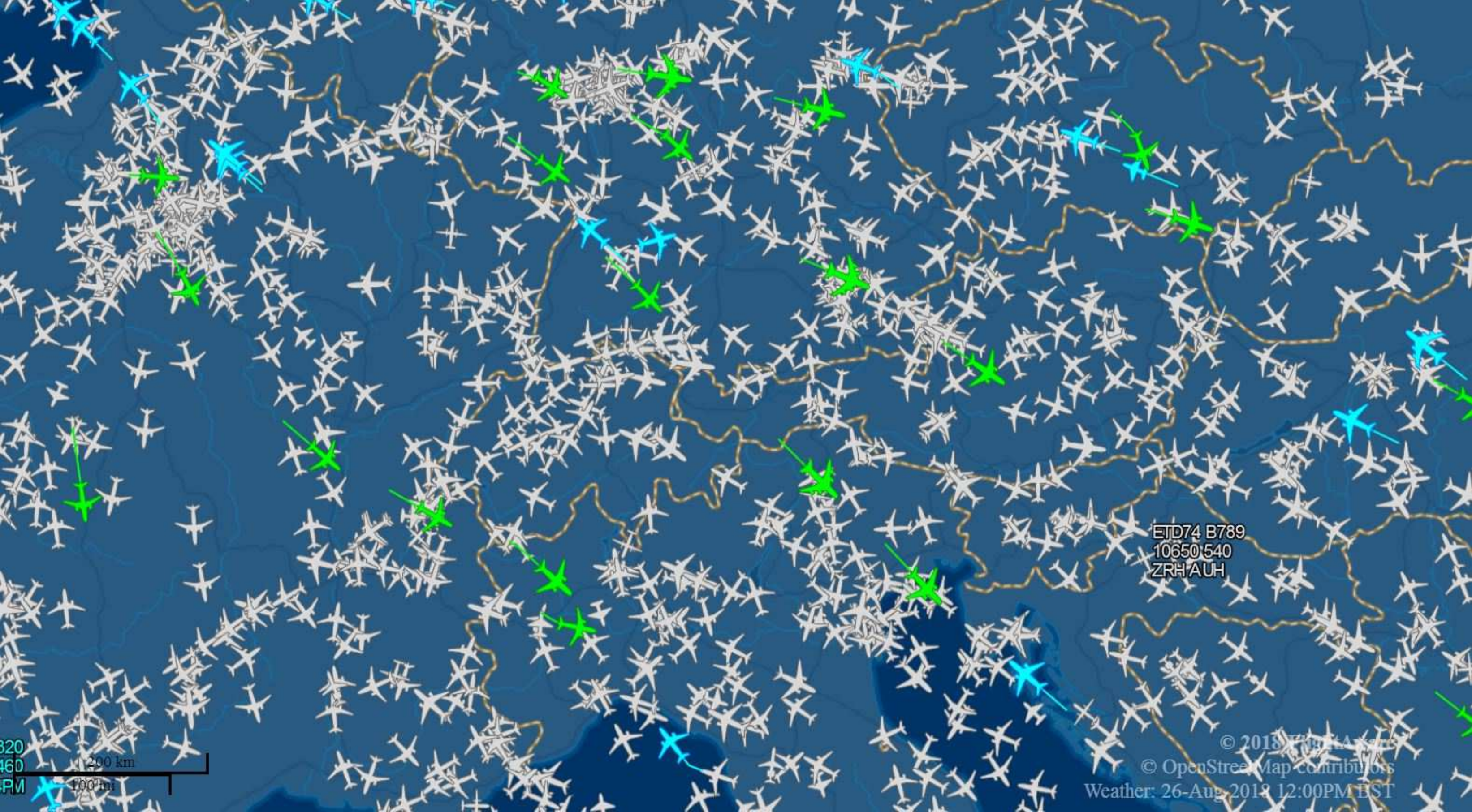}
\label{FIG4b}}
\subfigure[North Atlantic]{\includegraphics[width=0.47\textwidth]{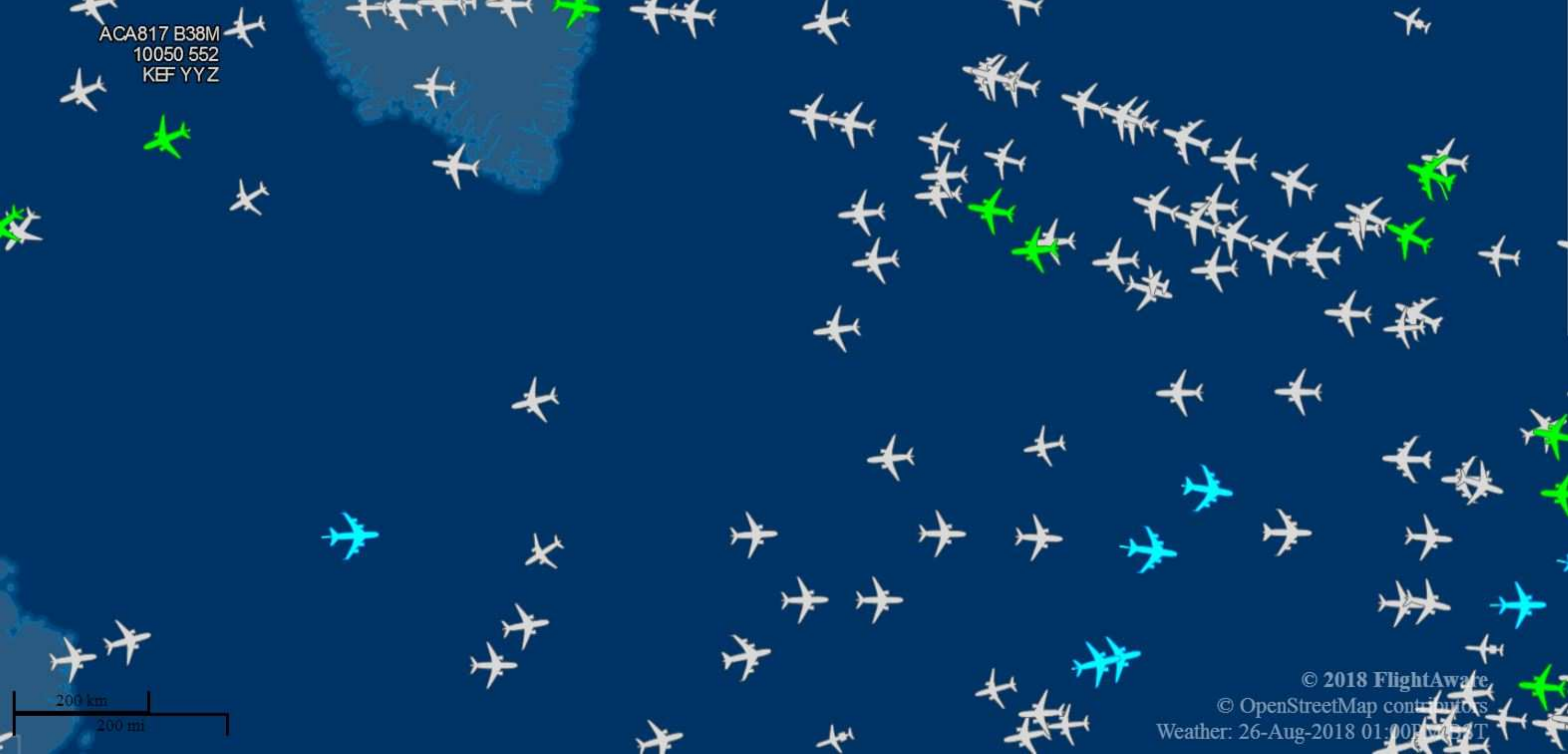}
\label{FIG4c}}
\end{center}
\caption{Aircraft mobility patterns.}
\label{FIG4:Mobility_pattern}
\end{figure}

Finally, as a speculative but high-impact research topic, it is worth investigating, whether the myriads of planes in the air might be able to enhance the terrestrial coverage as mobile \glspl{BS} for users on the ground.

\subsection{Summary}

Each application may impose particular requirements, as seen in Fig.~\ref{FIG3:app_requirement}. Explicitly, online flight data uploading/downloading requires the network to have a global coverage, a high throughput and acceptable latency in the face of the multi-hop links invoked for delivering the information. The cost has to be low, since there is a requirement for abundant information to be delivered frequently. Air traffic control is safety-related, hence it has strict requirements in terms of security and robustness. Meanwhile, air traffic control also requires both global coverage and low latency of the multi-hop links. Furthermore, the networks have to have the capability of self discovery/healing/control. Aircraft tracking is generally required in unpopulated airspace, and again, low latency is required. Aircraft heading in the same direction for a long trans-Atlantic journey for example may consider formation flight, which will significantly reduce the fuel consumption~\cite{Bos2010formation}. Thus, the network should cover both populated and unpopulated areas. The short range latency must be very low for internal communications within the formation, and the requirements of discovery/healing/control, security as well as robustness are also fundamental to secure formation flying. Free flight may be considered in the airspace over unpopulated areas, which again imposes appropriate requirements on the latency and self discovery/healing/control. Meanwhile, the security and robustness of the network are also crucial for guaranteeing flight safety. Passenger entertainment appeals to a large potential market for the aeronautical industry, but it requires a high throughput at low cost. Furthermore, the passengers tend to expect that the connection is seamless over their entire journey, regardless of whether they are departing from the airport, flying over populated areas or   unpopulated areas. The requirements imposed by the potential applications will be detailed in Section~\ref{S4}.
\section{Aircraft Networking Scenarios}\label{S3}
The potential applications supported by \glspl{AANET} are strictly coupled to the different phases of flight, such as landing/takeoff, taxiing, parking, holding pattern and being en-route. Hence, characterizing \glspl{AANET} with the aid of a uniform model may not always be possible for different aircraft scenarios, since the air traffic density, the mobility pattern as well as the propagation environment varies significantly both geographically and throughout the day. \glspl{AANET} are often categorized into three different geographical scenarios, namely operation in the vicinity of an airport area, a populated area and an unpopulated area. In Fig.~\ref{FIG4:Mobility_pattern} we capture the aircraft pattern of the above-mentioned three representative scenarios of airport, populated area, and unpopulated area, such as London's Heathrow airport, the European airspace, and the North Atlantic airspace, respectively. It can be observed that the aircraft density and mobility patterns are distinctly different. More specifically, in the rest of this section we will characterize the aircraft networking scenarios of airports, populated areas and unpopulated areas. The implications of these characterizations will be further discussed as requirements in Section~\ref{S4}.

\begin{figure*}
\begin{center}
 \includegraphics[width=1.00\textwidth,angle=0]{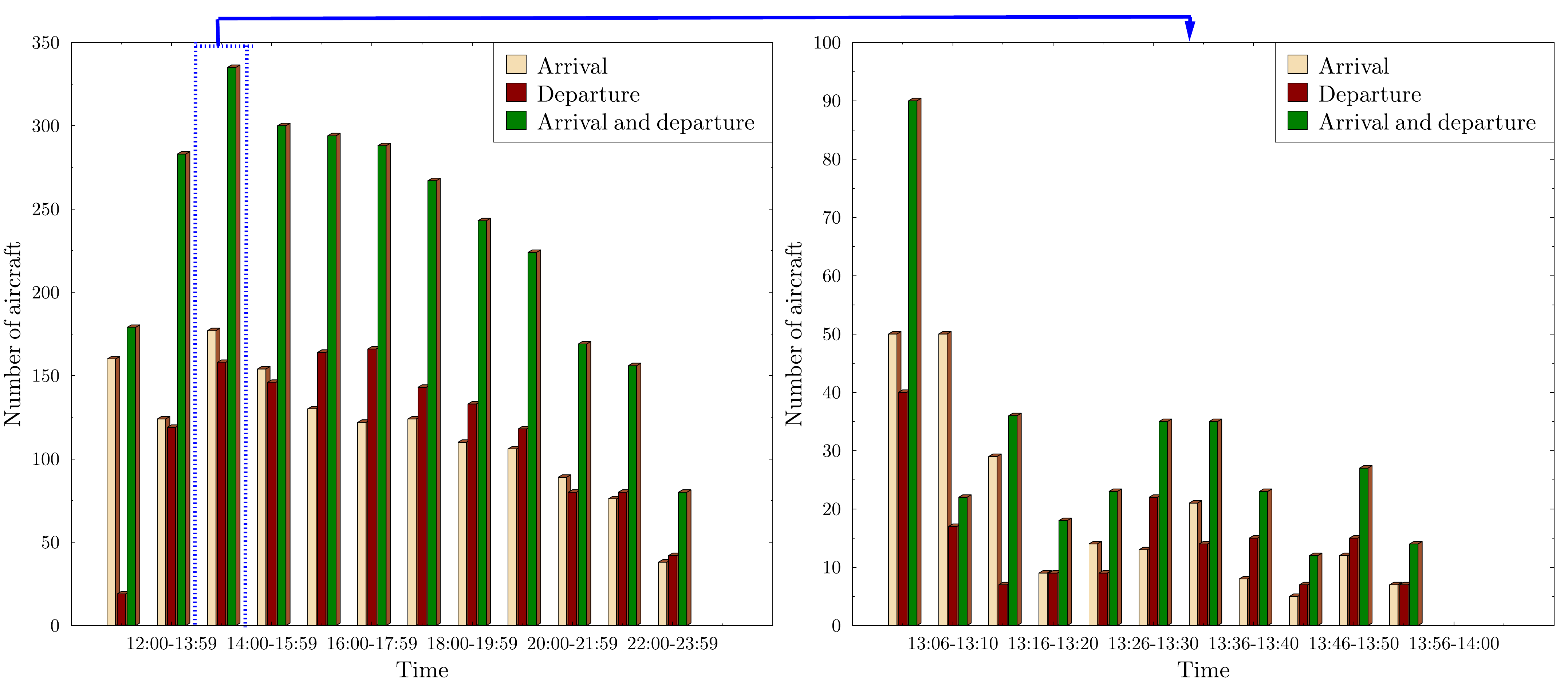}
\end{center}
\vspace{-1mm}
\caption{The number of aircraft arrival/departure events at London's Heathrow airport observed on August 26th, 2018}
\label{FIG5:aircraft_number}
\end{figure*}

\subsection{Flight Over An Airport or Near An Airport}\label{S2-1}
\begin{figure*}
\begin{center}
 \includegraphics[width=1.0\textwidth,angle=0]{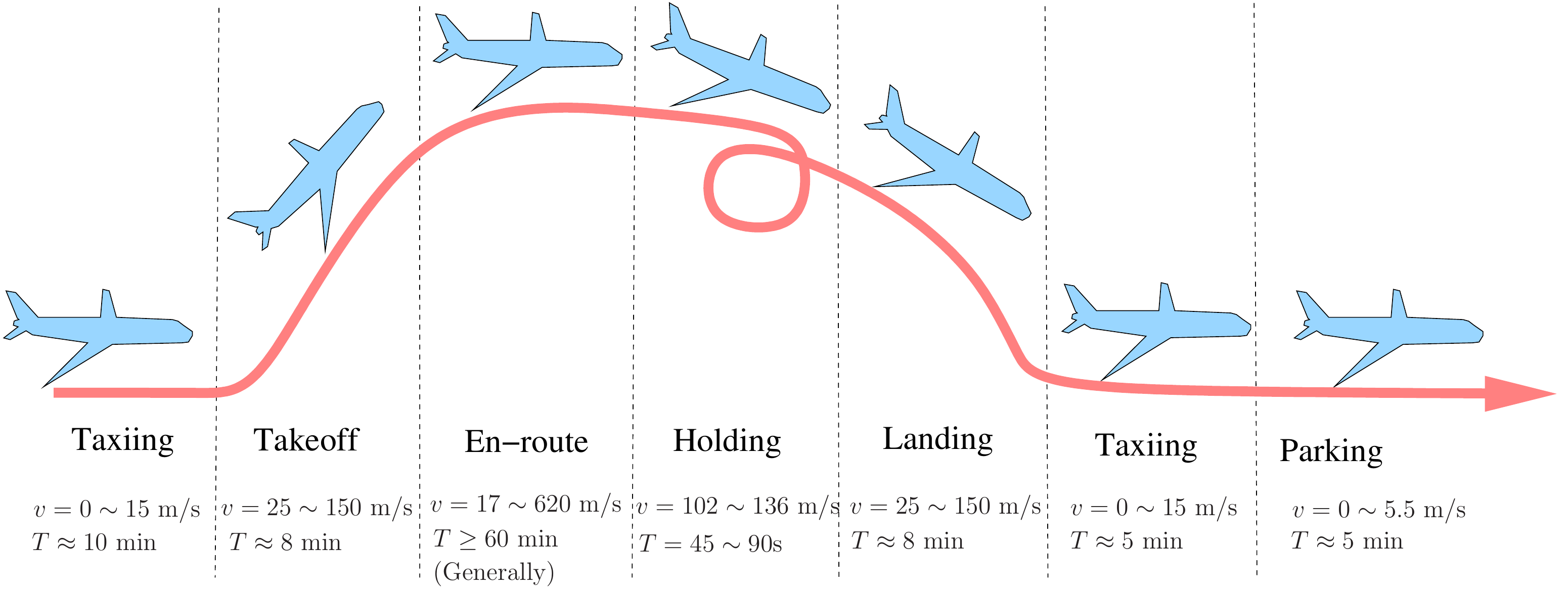}
\end{center}
\vspace*{-1mm}
\caption{Phases of aircraft flight}
\label{FIG6:scenarios}
\end{figure*}

A high-density area of aircraft is typical in the vicinity of airports, especially near the world's busiest airports. For example, there are about 2544, 1623 and 1378 aircraft takeoffs/landings per day on average at Chicago O'Hare International Airport, Beijing Capital International Airport and London Heathrow Airport respectively, as reported by the \gls{ACI} in August, 2014~\cite{Aircraft2014}. It was also reported that at Heathrow  there is typically a flight takeoff or landing every 45 seconds~\cite{Heathrow}. More specifically, the number of landing/takeoff aircraft at Heathrow airport during twelve hours of August 26th, 2018 is illustrated in Fig.~\ref{FIG5:aircraft_number}. It can be seen from Fig.~\ref{FIG5:aircraft_number} that there were 335 aircraft takeoff/landing events during the peak hours between 13:00 and 14:00. Furthermore, during our observed period, there were up to 90 aircraft takeoff/landing events in a period of 5 minutes. The exhausted airspace capacity\footnotemark\footnotetext{The airspace capacity represents the capability of accommodating aircraft within a given airspace in line with current safety separation.} problems could potentially be overcome with the aid of efficient \glspl{AANET}, which would enable aircraft to directly communicate with each other for self-organizing takeoff/landing. This presents an opportunity for establishing an {\it{ad hoc}} network amongst aircraft, but also imposes challenges in terms of both scheduling and interference management. In \glspl{AANET}, \glspl{GS} may be deployed at and around the airport, which allow the aircraft to directly communicate with \gls{ATC} or for messages to be relayed by other aircraft, for the sake of arranging their landing/takeoff, taxiing and holding patterns~\cite{haas2002aeronautical,haque2011ofdm}, as depicted in Fig.~\ref{FIG6:scenarios}.

\begin{itemize}[\setlength{\listparindent}{\parindent}]
\item \emph{Holding Pattern}: The holding pattern phase is encountered when the aircraft approaches the airport, but has no clearance to land yet, which can be seen both from the flight tracks of Fig.~\ref{FIG4:Mobility_pattern} and from the flight phases of Fig.~\ref{FIG6:scenarios}. Holding pattern procedures are designated to absorb any flight delays that may occur along an airway, during airport arrival and on missed approach. This phase results in a Rician distributed wireless communication channel, as shown in Table~\ref{TAB1:channel}, since a strong \gls{LOS} communication path can always be expected due to the proximity to the \gls{GS} at the airport, together with multi-path reflections from the ground and the various airport buildings. Frequency selective fading also takes place in the vicinity of airport owing to the delay spread of these multi-path reflections. Note that a high Doppler shift may be expected for the \gls{LOS} path owing to the speed of the aircraft~\cite{haas2002aeronautical}, leading to time selective fading. In particular, while performing waiting rounds, it is very likely that the aircraft will fly over the \gls{GS}, causing a rapid change of the Doppler shifts due to the Doppler rate, which results in rapidly changing Rician fading channels~\cite{haas2002aeronautical}.

\item \emph{Landing and Takeoff}: In the takeoff phase an aircraft becomes airborne and commences climbing under instruction from \gls{ATC}, whilst increasing its speed. By contrast, in the arrival phase, an aircraft reduces its cruising speed and altitude, descending towards landing. These two phases result in a Rician distributed wireless communication channel, due to the high likelihood of having a strong \gls{LOS} communication path. As in the holding pattern scenario, frequency selective fading may be expected owing to multi-path reflections. Furthermore, a high Doppler shift may be expected because of  the speed of the aircraft~\cite{haas2002aeronautical}.

\item \emph{Taxiing}: Upon touchdown, the aircraft leaves the runway via one of the available turn-off ramps toward the terminal, and vice versa for takeoff. The maximum Doppler and the maximum delay spread have been decreased compared to the landing phase and the takeoff phase. However, the \gls{LOS} path can be expected to be weaker than the reflections from the ground and from surrounding buildings, which results in a reduction of the Rician $K$ factor, leading to more severe fading of the signal~\cite{haas2002aeronautical}.

\item \emph{Parking}: In the parking phase, the aircraft is on the ground and traveling at a slow speed close to the terminal, before parking at the terminal. In this phase, the \gls{LOS} path is often blocked, therefore the information contained in the received signal has to be reconstructed from the echo paths alone. The fading will be Rayleigh distributed and frequency-selective, depending on the bandwidth of the signal~\cite{haas2002aeronautical}.
\end{itemize}

\begin{table*}
\caption{Aeronautical channel characteristics for different aircraft flight phases~\cite{haas2002aeronautical}} 
\label{TAB1:channel} 
\begin{small}
\begin{center} 
\begin{threeparttable}
\begin{tabular}{|C{2.3cm}|C{1.6cm}|C{1.6cm}|C{2.3cm}|C{2.3cm}|C{2.6cm}|C{2.4cm}|} 
\hline\hline 
Parameters & Parking  & Taxiing & Holding & Landing/Takeoff & En-Route (\gls{A2A}) & En-Route (\gls{A2G}) \\ [0.5ex] 
\hline 
Aircraft velocity (m/s) & $0-5.5$&$0-15$&$102-136$ &$25-150$ & $245-257$ & $245-257$ \\
\hline 
Maximum delay (s) & $7 \times 10^{-6}$& $0.7 \times 10^{-6}$& $66 \times 10^{-6}$ & $7 \times 10^{-6}$& $66 \times 10^{-6}-200 \times 10^{-6}$ &  $33 \times 10^{-6}-200 \times 10^{-6}$  \\
\hline 
Number of echo path & $20$& $20$& $20$& $20$& $20$ &  $20$  \\
\hline 
Rice factor (dB) & /& $6.9$& $9-20$ (mean $15$)& $9-20$ (mean $15$)& $2-20$ (mean $15$)&  $2-20$ (mean $15$)  \\
\hline 
$f_{D_{LOS}}/f_{D_{max}}$ & /& $0.7$& $1.0$& $1.0$& $1.0$ &  $1.0$  \\
\hline 
Lowest angle of beam ($^{\circ}$) & $0$ & $0$& $-90$& $-90$& $178.25$ &  $178.25$  \\
\hline 
Highest angle of beam ($^{\circ}$) & $360$ & $360$& $181.75$& $90$& $181.75$ &  $181.75$  \\
\hline 
Delay power spectrum& Exponential & Exponential&Exponential / Two-ray& Exponential&Two-ray&Two-ray  \\
\hline 
Slope time$^{1}$& $1.0 \times 10^{-6}$ & $1.09 \times 10^{-7}$&$1.0 \times 10^{-6}$&$1.0 \times 10^{-6}$&/&/  \\
\hline 
\end{tabular}
\begin{tablenotes}\footnotesize
\item [1] `Slope time' is the rate of decay in the exponential function that describes the distribution function of delay.
\end{tablenotes}
\end{threeparttable}
\end{center}
\end{small}
\end{table*}

\subsection{Flight Over Populated Areas}\label{S2-2}
In populated areas, aircraft occupy the international airspace very heterogeneously. Some regions experience dense air traffic, with the aircraft directions being largely uncorrelated, as exemplified by the particular instantiation of the European airspace shown in Fig.~\ref{FIG4:Mobility_pattern}. Other regions remain only  sparsely populated, with aircraft typically flying parallel to each other. Moreover, the number of airborne aircraft in a given region changes significantly throughout the day~\cite{medina2008topology}. In order to ensure the aircraft are adequately separated when en-route, a minimal aircraft separation of 5 \glspl{NM} is required for continental flights, which is managed by several ground based surveillance radars. \glspl{GS} are typically deployed in populated areas, and aircraft fly over \glspl{GS} when en-route. Thus, satellites may constitute a less attractive solution for relaying information between the \glspl{GS} and aircraft. However, they may be suitable for relaying information from aircraft to aircraft. In this scenario, the aircraft always engage in \gls{A2G} communications or may engage in \gls{A2A} communications, when they communicate with each other and act as relays.
\begin{itemize}
\item \emph{\gls{A2A}}: In the en-route phase, the aircraft is typically navigating a flight-planned route at `optimum' altitudes for its specific weight and engine configuration. This phase will result in rapidly fluctuating frequency-selective fading~\cite{haas2002aeronautical}. The channel-induced dispersion is more severe when compared to terrestrial channels due to the high velocity of the aircraft~\cite{dovis2002small}, particularly when two aircraft are flying in opposite directions. In this case, the \gls{LOS} path is dominant, resulting in Rician distributed fading.
\item \emph{\gls{A2G}}: If an aircraft passes over a \gls{GS} during the flight, the polarity of the associated Doppler shift will change. However, the Doppler shift does not change its polarity abruptly, but rather it decreases gradually upon decreasing the projected distance of an aircraft, when flying over the \gls{GS} at altitude. Additionally, aircraft turns will also cause a Doppler shift, but in general lead to lower values of the frequency change~\cite{haas2002aeronautical}.
\end{itemize}

\begin{figure}
\begin{center}
 \includegraphics[width=1.0\columnwidth,angle=0]{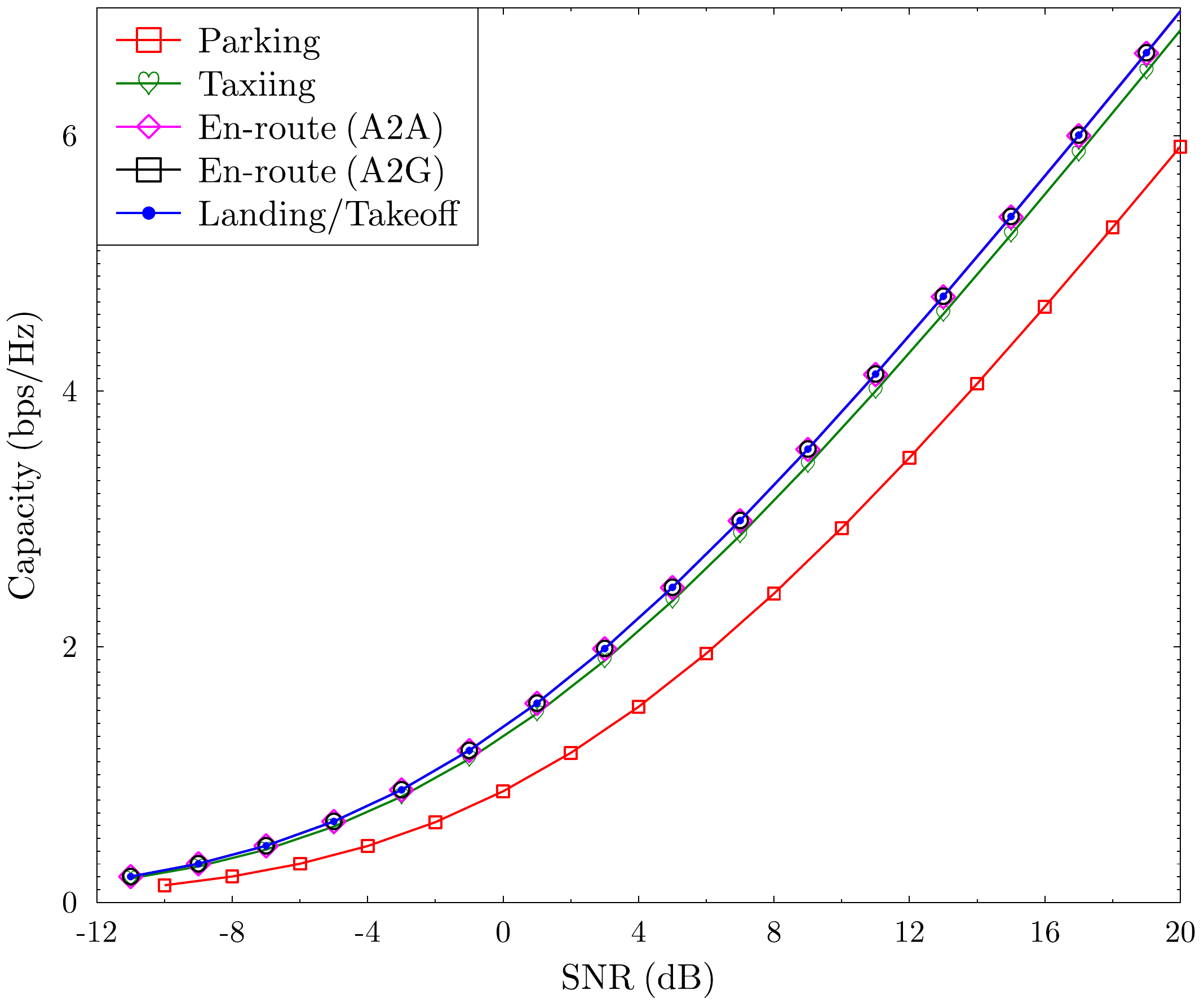}
\end{center}
\vspace{-0mm}
\caption{The capacity of aeronautical channel in different flight phases. The Rice factor is set as 15 for landing/takeoff, en-route (A2A), en-route (A2G), while the Rice factor is set as 6.9 for the taxiing phase}
\label{FIG7:channel-capacity}
\end{figure}

The dynamic nature of aircraft motion will generate different phases associated with diverse channel characteristics. These phases will be characterized by different types of fading, Doppler shifts and delays endured by the system. The \gls{A2G} or \gls{A2A} channel links will experience time-varying conditions, depending on the specific state of the aircraft journey~\cite{haque2011ofdm}. Table~\ref{TAB1:channel} shows the channel characteristics of the above-mentioned phases of landing, takeoff, taxiing, parking, holding pattern and en-route. The capacities of the aeronautical channel in different phases are shown in Fig.~\ref{FIG7:channel-capacity}. The achievable capacities of different flight-phases are simulated according to the parameters of Table~\ref{TAB1:channel}. Explicitly, the Rician $K$-factor is set as 15 for landing/takeoff, en-route (A2A) and en-route (A2G), while it is set to $K =  6.9$ for the taxiing phase. Observe from Fig.~\ref{FIG7:channel-capacity} that the parking phase has the lowest capacity, because it suffers from Rayleigh fading. However, the capacities of the taxiing, landing/takeoff, en-route (A2A) and en-route (A2G) phases are higher than that experienced during parking, as a benefit of \gls{LOS} propagation leading to Rician fading. More specifically, the capacities of landing/takeoff, en-route (A2A) and en-route (A2G) are almost the same, although they have different scattering  characteristics, dependent on the presence or absence of \gls{LOS}. The capacity of the taxiing phase becomes a little lower, since its $K$-factor is lower than that of the landing/takeoff and en-route phases.

\begin{figure}
\begin{center}
 \includegraphics[width=1.0\columnwidth,angle=0]{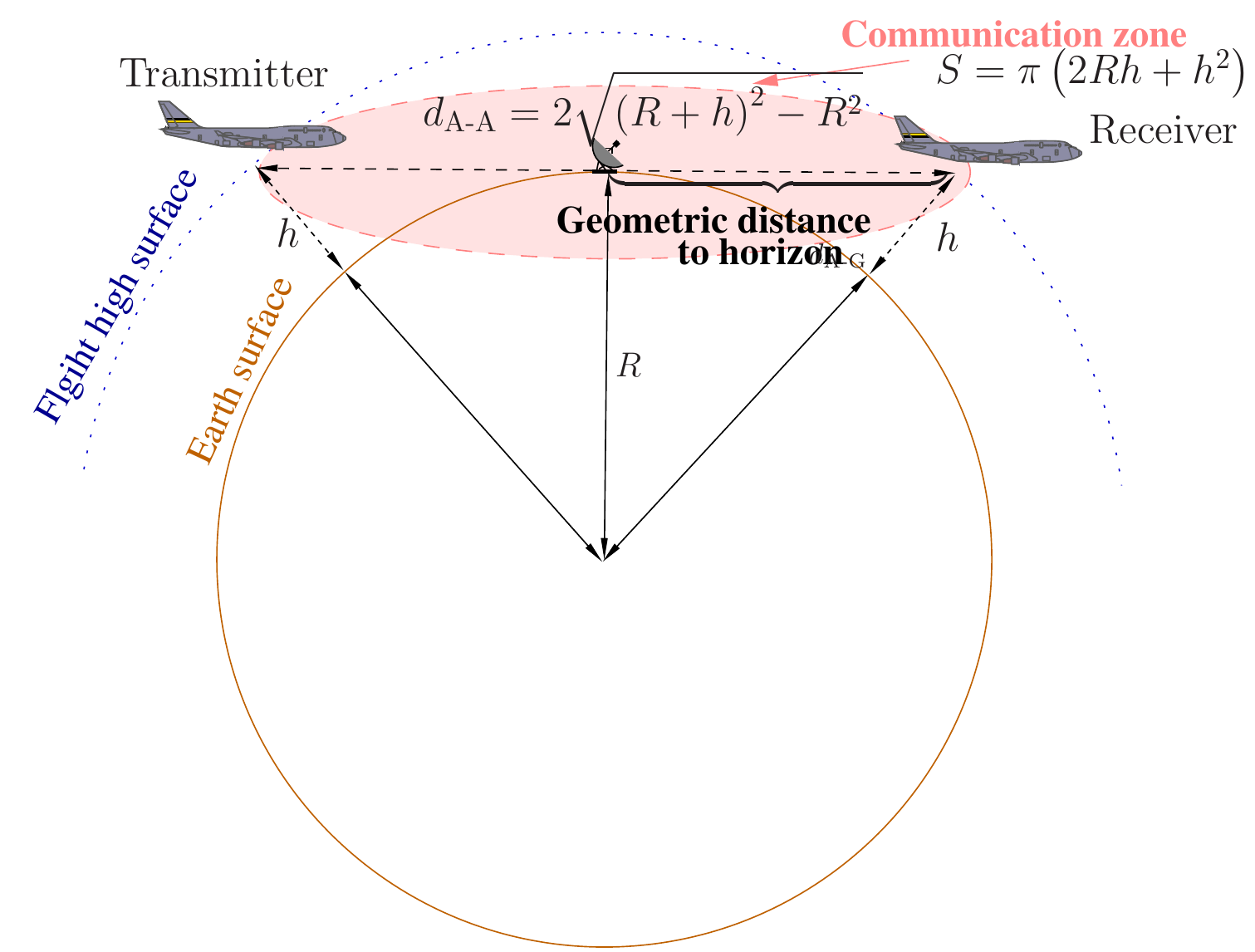}
\end{center}
\vspace{-0mm}
\caption{Geometrical area of the LOS communication zone $S$ at flying altitude $h$}
\label{FIG8:comm_zone}
\end{figure}

The communications range of aircraft is affected by the aircraft altitude, antenna height of the \gls{GS}, receiver sensitivity, transmitter power, antenna type, coax type and length as well as the terrain details, especially in the presence of hills, mountains, etc. Note that the aircraft type will directly affect both the antenna installment and the communications device deployment, which may also affect the communication range.  However, the two most crucial factors in determining the communications range are the aircraft altitude and the terrain characteristics. Since \gls{VHF} radio signals propagate along the \gls{LOS} path, aircraft that are behind hills or beyond the radio horizon (due to the Earth's curvature) cannot communicate with \glspl{GS}, even if they experience favorable channel conditions. Considering the Earth's curvature, the geometric distance of horizon (\gls{A2G} communication range) $d_{\textrm{A-G}}$ is given by {$d_{\textrm{A-G}} = \sqrt{\left(R + h\right)^{2} - R^{2}}$}, where $R$ is the radius of the Earth and $h$ is the altitude of the aircraft. More specifically, given a typical cruising altitude of $h = 10.68km$, the maximum \gls{A2G} communication range is approximately $d_{\textrm{A-G}} \approx 200$ \gls{NM}, as shown in Fig.~\ref{FIG8:comm_zone}. Furthermore, the approximate geometrical area of the communication zone~\cite{sakhaee2006global} covered by a \gls{GS} can be defined as the circle having the radius of the geometric distance $d_{\textrm{A-G}}$ to the horizon, as shown in Fig.~\ref{FIG8:comm_zone}. Thus the achievable \gls{A2G} communication zone is given by $S_{\textrm{A-G}} = \pi d_{\textrm{A-G}}^{2}= \pi\left(2Rh + h^{2}\right)$. The investigation shows that for lower altitudes, the communication range is predominately limited by the geometric distance of the horizon, while at higher altitudes it is limited by the signal strength~\cite{medina2011airborne}.

\subsection{Flight Over Unpopulated Areas}\label{S3-3}
With the development of a global economy, the number of international flights has increased considerably. Most international flights pass through the North Pacific Ocean or the North Atlantic Ocean, which are areas where it is not possible to build \glspl{GS}. One of the solutions to this problem is to arrange aircraft flying above unpopulated areas to use satellites as relays, albeit this solution is costly and suffers from long round-trip delays~\cite{medina2010routing}. It can be seen from Fig.~\ref{FIG4:Mobility_pattern} that a string of aircraft are heading towards the destination continent following a similar route, which may be identified as a pseudo-linear mobile entity~\cite{sakhaee2007stable}. The group mobility feature of aircraft flying over unpopulated airspace can be exploited for setting up a large-scale mobile {\it{ad hoc}} network by establishing multi-hop \gls{A2A} links amongst the aircraft~\cite{tu2009mobile}, where any aircraft can be viewed as a node communicating with its neighbor aircraft for data routing.

The aircraft trajectories above an unpopulated area follow a limited set of predefined routes, and the aircraft density is low when compared to populated airspace~\cite{vey2014aeronautical}. It may be expected that many \glspl{GS} and radars can be deployed in continental areas. However, these stations and radars are rare in unpopulated areas and they are totally absent in oceanic areas~\cite{tu2009proposal}. Thus, in the oceanic areas that are outside of radar range, the safety interval is required to be longer than the above-mentioned 5 \glspl{NM}. Specifically, they have to be as high as 50 \glspl{NM}~\cite{tu2009mobile}, for the sake of avoiding mid-air collisions.

\begin{figure}
\begin{center}
 \includegraphics[width=1.0\columnwidth,angle=0]{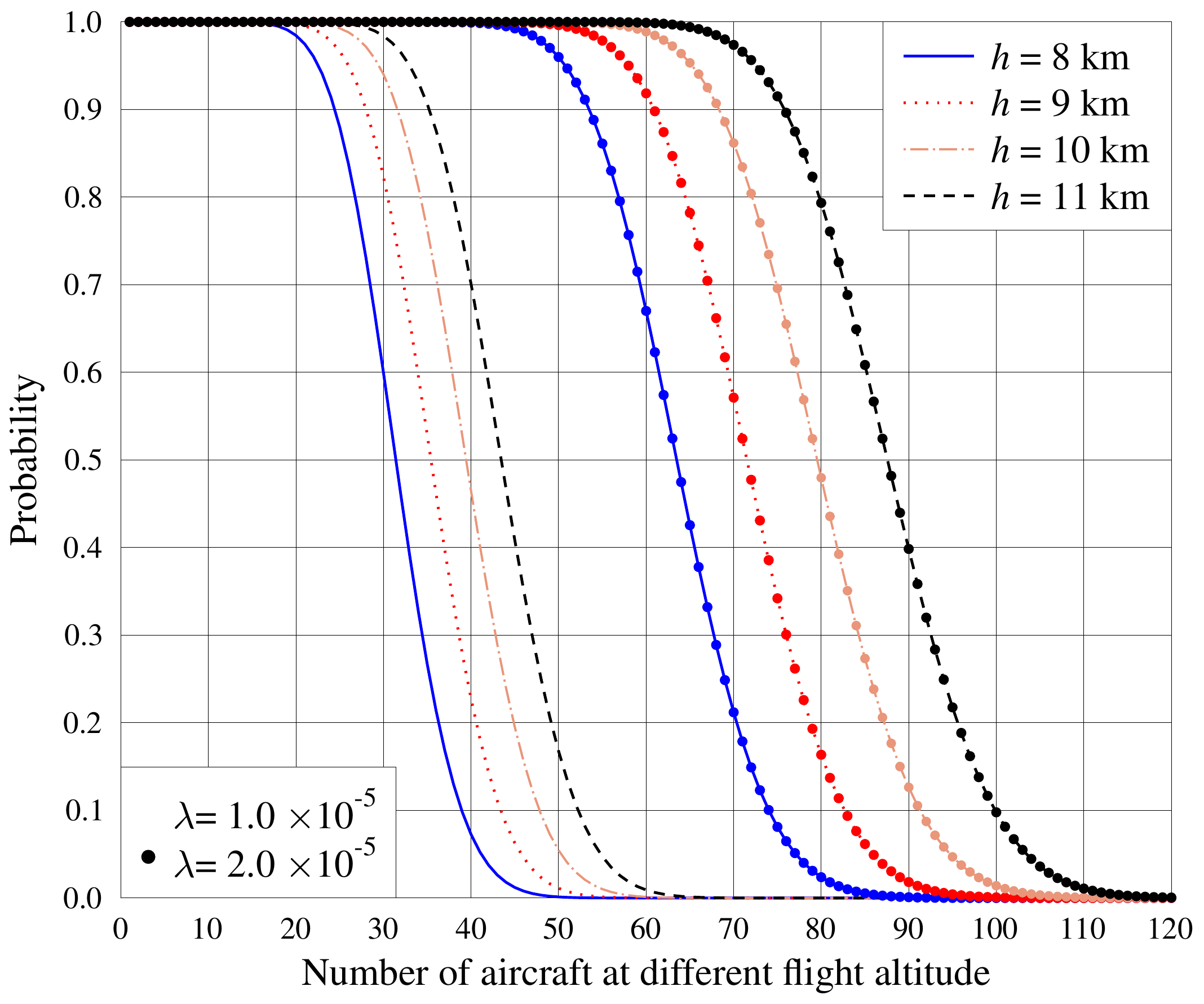}
\end{center}
\vspace{-0mm}
\caption{Probability of at least $n$ aircraft appearing in the communication zone $S$ illustrated in Fig.~8}
\label{FIG9:CDF}
\end{figure}

In order to create an \gls{AANET}, at least the following two criteria must be met~\cite{sakhaee2006global}. Firstly, there has to be an adequate number of aircraft in the airspace at any given time instant in order for {\it{ad hoc}} networking among aircraft to become possible. Secondly, an aircraft must be within the communication range of at least one other aircraft, in order for a link to be established and for multi-hop routing to become practical. The maximum geometrical communication range of two aircraft (\gls{A2A} communication range) at altitude $h$ is defined as {$d_{\textrm{A-A}} = 2d_{\textrm{A-G}} = 2\sqrt{\left(R + h\right)^{2} - R^{2}}$}. Specifically, considering a cruising altitude of $h = 10.68km$, the \gls{A2A} communication range is limited to a maximum of $d_{\text{A-A}} \approx 400$ \gls{NM}. The nominal communication range is likely to be smaller than the maximum communication range $d_{\text{A-A}}$, depending on the transmit power, the specific characteristics of the antenna, the particular modulation used for transmission and the target \gls{BER}.
Note that different types of aircraft will have different antenna and communications device deployment, which may also affect the attainable communication range.
 
Here we provide an example for the probability of having $n$ aircraft in a given \gls{A2A} communication zone of area  $S_{\textrm{A-A}} = \pi d_{\textrm{A-A}}^{2}= 4\pi\left(2Rh + h^{2}\right)$, which is over the oceanic airspace not covered by the \gls{GS} service.
The probability of the number of aircraft $n$ being within the given region $S_{\textrm{A-A}}$ may be estimated by~\cite{sakhaee2006global} {\color{black}$p\left(n,S_{\textrm{A-A}}\right) = \frac{(\lambda S_{\textrm{A-A}})^{n}}{n!}e^{-\lambda S_{\textrm{A-A}}}$}. Given the practical aircraft densities of 900 aircraft and 1800 aircraft over a $S_{\textrm{A-A}} = 9,000,000 \text{km}^{2}$ area of the Atlantic ocean at any given time, the average aircraft densities are $\lambda = 1 \times 10^{-5}$ aircraft/km$^2$ and $\lambda = 2 \times 10^{-5}$ aircraft/km$^2$, respectively.  The estimated number of aircraft at altitudes of $h = 8$ km, $h = 9$ km, $h = 10$ km and $h = 11$ km within $S_{\textrm{A-A}} = 9 000 000 \text{km}^{2}$ are illustrated in Fig.~\ref{FIG9:CDF}. It can be observed from Fig.~\ref{FIG9:CDF} that the probability of finding at least two or even as many as up to 52 aircraft within $S_{\textrm{A-A}}$ is close to $100\%$, when we have $h = 9$km and $\lambda = 2 \times 10^{-5}$ aircraft/km$^2$. Typically, dozens of aircraft may be expected to be within communication range at any given time. Therefore, the \gls{A2G} communication zone will be extended by exploiting both multi-hop as well as direct \gls{A2A} communication via AANET. More particularly, the communication zone of direct \gls{A2A} communication will be a circle having a radius of $d_{\textrm{A-A}}$, which may be capable of achieving global coverage with the aid of multi-hop {\it{ad hoc}} networking amongst aircraft.

\subsection{Summary}

In summary, we have compared the scenarios of flight over airport areas, flight over populated areas and flight over unpopulated areas in terms of the associated flight phases, fading type of the channel propagation, aircraft density, related data links and potential applications. In Table~\ref{TAB3:com_scenarios}, we summarise some of the salient distinguishing features of the three scenarios discussed in this section.  Explicitly, the flight phases of holding, takeoff/landing, taxiing and parking are always performed at an airport or near an airport. The channel propagation obeys Rician fading for holding, takeoff/landing, taxiing, while the parking phase suffers from Rayleigh fading, since the LOS path may be blocked by other parked aircraft and airport buildings. Intuitively, the density of aircraft is very high at an airport or near an airport, and the communications are mainly \gls{A2G}. During the flight phases of holding and takeoff/landing, the air traffic control messages are the most important ones, while during the phases of taxiing and parking, flight data may be delivered. In the populated continental airspace, the flight phase is en-route and there is typically always LOS propagation for all the \gls{A2G}, \gls{A2A} and \gls{A2S} data links. The continental air traffic is generally busy associated with a high aircraft density heading in different directions. When the aircraft fly over the continental airspace, offering on-board entertainment is attractive. In the unpopulated airspace, the flight phase is en-route, but the data links are restricted to LOS \gls{A2A} and \gls{A2S} propagation, owing to the absence of ground stations. In the unpopulated airspace,  formation flight and free flight may be applicable due to the relative simplicity of their flight lanes in this airspace. Moreover, on-board entertainment is routinely expected during these long periods of flight. In unpopulated areas, aircraft tracking is an important application which may prevent aircraft disappearance, such as that of the Malaysian Airline plane MH370.

\begin{table*}
\caption{Comparison of the different scenarios} 
\label{TAB3:com_scenarios} 
\begin{center} 
\begin{tabular}{|C{2.3cm}||C{1.9cm}|C{1.3cm}|C{1.0cm}|C{1.8cm}|C{6.5cm}|}\hline
Scenarios
&{Phase}&{Fading}&{Density}
&{Data links}&{Applications}\\ \hline\hline
\multirow{4}{*}{Airport/near airport}  &Holding&Rician&High&\gls{A2G} & Air traffic control\\
\cline{2-6}
 &takeoff/Landing&Rician&High&\gls{A2G} & Air traffic control\\
\cline{2-6}
  &Taxiing&Rician&High&\gls{A2G} & Upload/Download\\
	\cline{2-6}
 &Parking& Rayleigh&High&\gls{A2G} & Upload/Download\\\hline
Populated  &En-route&Rician&High&\gls{A2G}/\gls{A2A}/\gls{A2S} & Entertainment\\\hline
Unpopulated &En-route&Rician&Low&\gls{A2A}/\gls{A2S} &Formation fly/Free flight/Entertainment/Aircraft tracking\\\hline
\end{tabular}
\end{center} 
\end{table*}
\section{Aircraft Networking Requirements}\label{S4}
As discussed in Section~\ref{S2} and Section~\ref{S3}, \gls{AANET} is associated with specific characteristics and applications, which lead to particular requirements, as shown in Fig.~\ref{FIG3:app_requirement} and Table~\ref{TAB5:relationships-requirements-challenges}. This means that the system should treat various types of data differently, depending on the corresponding applications. For example, latency and robustness must be prioritized for safety-critical data. By contrast, in-flight entertainment requires a high throughput in order to service a large number of passengers. In this section, we focus our attention on some representative \gls{AANET} requirements, which are crucial for designing a feasible and reliable {\it{ad hoc}} network for linking up aircraft.

\subsection{Coverage}

\gls{AANET} requires global coverage, since aircraft traverse both continents and oceans. However, the  achievable \gls{AANET} coverage is affected by the aircraft mobility pattern, transmission power~\cite{shirazipourazad2011connectivity} and propagation environment. Note that the environmental factors also include the diverse effects of the topography, of the physical obstructions, of the atmosphere and of the weather. These effects potentially introduce propagation losses and delays, which have to be carefully considered, when designing an \gls{AANET}. Nevertheless, it is still possible to achieve global coverage by falling back upon satellite-aided relaying or upon hybrid satellite/\gls{AANET} solutions. Furthermore, some additional factors, such as clutter models and obstruction densities, need to be taken into account when considering the coverage in airports. More particularly, the clutter models represent the density of obstructions in the deployment area. An airport surface may have relatively open runways and taxi areas, but congested terminal areas may require the assistance of more fixed infrastructure \glspl{GS}. A plethora of parameters have to be considered for meeting the coverage requirements of \glspl{AANET}, such as the \gls{GS} and aircraft transmit/receive power, antenna gains, feeder losses and aircraft altitude.

\begin{table*}
\caption{Throughput for services of voice, data and video~\cite{LSTelcom2003Radio}.}
				\label{TAB2:throughput}
				\begin{center}
        \begin{tabular}{|c|c|c|c|c|}
          \hline
          Services & \acrshort{DL}/\acrshort{UL} & Throughput in kBit/s & Fixed channel rate in kBit/s & Required number of channels \\ \hline
          \multirow{2}{*}{Voice}             & \acrshort{DL} & 870.41 & 20 & 44\\
					\cline{2-5}
                            & \acrshort{UL} & 870.41 & 20 & 44\\ \hline
          \multirow{2}{*}{Data+Security}     & \acrshort{DL} & 809.17 & 64 & 13 \\
					\cline{2-5}
                            & \acrshort{UL} & 3.14 & 64 & 1 \\ \hline
          \multirow{2}{*}{Commercial Data}   & \acrshort{DL} & 197.60 & 64 & 4 \\
					\cline{2-5}
                            & \acrshort{UL} & 197.60 & 64 & 4 \\ \hline
          \multirow{2}{*}{Video}             & \acrshort{DL} & 768.00 & 384 & 2 \\
					\cline{2-5}
                            & \acrshort{UL} & 0 & 384 & 0 \\
          \hline
        \end{tabular}
    \end{center}
\end{table*}

\subsection{Throughput}

In general, the throughput requirement depends on that of any other networked aircraft owing to the provision of relaying service. Beside the traditional communication systems used for civil aviation, \glspl{AANET} have to offer comparable or even higher data rates for providing various services, for example, voice+audio, data+security, commercial data and video. The total bandwidth required depends on the number of passengers that are simultaneously using each service. Therefore, a sufficiently high throughput is required in order to provide services that meet the users' expectations.
The authors of~\cite{vey2014aeronautical} assessed the available throughput of an \gls{AANET} assuming that the capacity of each relaying node was 1~Mbps. Their assessment indicated that the achievable throughput of \glspl{AANET} in the oceanic airspace was better than that attained in the continental airspace, namely 68.2 kbps versus 38.3~kbps, respectively.
Wang  {\it{et al}}.~\cite{wang2015throughput} derived the upper bound of the throughput and the closed-form  average delay expression of a two-hop aeronautical communication network.
Furthermore, Schutz and Schmidt~\cite{LSTelcom2003Radio}  provided the calculation of the required  bandwidth 
in their final report on the radio frequency spectrum requirements for future aeronautical mobile systems. Their prediction was calculated by sharing the overall achievable throughput amongst 105 aircraft, which is predicted to be the Peak Instantaneous Aircraft Count (PIAC) for the airspace sector in the year 2029. Furthermore, a fixed channel rate was selected for the different services in order to calculate the number of channels required. Explicitly, the throughput achieved by each aircraft for the services of voice, data and video are summarized in Table~\ref{TAB2:throughput}.

Table~\ref{TAB2:throughput} estimates the throughput required by a single aircraft. However, only two video channels are supported in this analysis, allowing only two passengers to perform video streaming simultaneously. Due to the ever increasing demand of business/entertainment applications, this is insufficient. In particular, flawless quality video conferencing of a video call requires a throughput of 900~kpbs, given that the video call encodes high-quality voice alone at a throughput of around 40~kbps~\cite{xu2012video,zhang2012profiling}. Therefore, much more channels and higher throughput may be required in \glspl{AANET}.

In airport scenarios, the highest throughput requirements for \glspl{AANET} emanate from airlines and port authorities, which include communications with ground maintenance crews and airport security. According to a series of studies conducted for the \gls{FAA}, the company MITRE CAASD estimated the potential bandwidth requirements to the year 2020 and beyond for aeronautical communications~\cite{gheorghisor2006preliminary}. The highest total aggregate data capacity requirements for fixed and mobile applications are based on large airports relying on a \gls{TRACON} \gls{ATC} facility, which is not collocated with an \gls{ATCT}. The estimated aggregate data rate requirement of mobile applications is close to 20~Mbps~\cite{gheorghisor2006preliminary}, where the aeronautical operational control data services account for more than half of the 20 Mbps throughput. The estimated aggregate data rate requirement for these fixed applications is over 52~Mbps~\cite{gheorghisor2006preliminary}. The combination of video surveillance and sensory information, as well as that of the associated TRACON-to-ATCT data communications account for about $80\%$ of the total.

\subsection{Latency}

The applications of \glspl{AANET} having the highest sensitivity to latency are real-time interactive telephony and safety/control applications. For voice telephony, the maximum acceptable latency in a \gls{VoIP} network is 250 milliseconds~\cite{SANSInstitute}. Although a user may tolerate seconds of buffering delay in broadcast video streaming, such a long delay would seriously impair the user's \gls{QoE} in interactive video conferencing~\cite{xu2012video,jansen2011enabling}. However, the current latency  specification of aircraft \gls{VDL} is limited to delays below 3.5~seconds for $95\%$ of the time for data packets~\cite{stacey2008aeronautical}. Naturally, this is not suitable for real-time interactive services. Thus, \glspl{AANET} must be able to offer an order of magnitude improvement in latency, while overcoming the challenges associated with multi-hop scenarios. More specifically, EUROCONTROL/FAA has defined the \gls{QoS} in form of the distribution of one-way transmission latency designated by the acronym ``TT95-1 way''~\cite{EUROCONTROL2007Future} in aeronautical terminology. This terminology becomes explicit in Table VI. For example, observe in the first row of Table~VI that the one-way delay has to be lower than 1.4~s in $99.96\%$ of the scenarios at the airport.

{
\begin{table*}[ht]
\caption{Quality of service requirements in terms of connectivity$^{1}$ and the $95\%$ percentile of the one-way transmission latency (TT95-1 way)~\cite{EUROCONTROL2007Future}.} 
\label{TAB3:QoS-Latency} 
\centering 
\begin{threeparttable}
\begin{tabular}[t]{L{4.0cm}C{2.5cm}C{2.5cm}C{2.5cm}C{2.5cm}|} 
\cline{2-5}
  & \multicolumn{2}{ |c }{Without automatic execution service$^{2}$} & \multicolumn{2}{ |c| }{With automatic execution service}\\ [0.5ex] 
\hline
\multicolumn{1}{|C{4.0cm}}{Scenarios} & \multicolumn{1}{|C{2.5cm}}{TT95-1 way (s)} & \multicolumn{1}{|C{2.5cm}}{Connectivity  ($\%$)} & \multicolumn{1}{|C{2.5cm}}{TT95-1 way (s)} & \multicolumn{1}{|C{2.5cm}|}{Connectivity ($\%$)} \\
\hline
\multicolumn{1}{|L{4.0cm}}{Airport} & \multicolumn{1}{|C{2.5cm}}{1.4} & \multicolumn{1}{|C{2.5cm}}{99.96} & \multicolumn{1}{|C{2.5cm}}{/} & \multicolumn{1}{|C{2.5cm}|}{99.99} \\
\hline
\multicolumn{1}{|L{4.0cm}}{Terminal Maneuvering Area} & \multicolumn{1}{|C{2.5cm}}{1.4} & \multicolumn{1}{|C{2.5cm}}{99.96} & \multicolumn{1}{|C{2.5cm}}{0.74} & \multicolumn{1}{|C{2.5cm}|}{99.99} \\
\hline
\multicolumn{1}{|L{4.0cm}}{En Route} & \multicolumn{1}{|C{2.5cm}}{1.4} & \multicolumn{1}{|C{2.5cm}}{99.96} & \multicolumn{1}{|C{2.5cm}}{0.74} & \multicolumn{1}{|C{2.5cm}|}{99.99} \\
\hline
\multicolumn{1}{|L{4.0cm}}{Oceanic/Remote/Polar} & \multicolumn{1}{|C{2.5cm}}{5.9} & \multicolumn{1}{|C{2.5cm}}{99.96} & \multicolumn{1}{|C{2.5cm}}{/} & \multicolumn{1}{|C{2.5cm}|}{99.99} \\
\hline
\multicolumn{1}{|L{4.0cm}}{Autonomous Operations Area} & \multicolumn{1}{|C{2.5cm}}{1.4} & \multicolumn{1}{|C{2.5cm}}{99.96} & \multicolumn{1}{|C{2.5cm}}{/} & \multicolumn{1}{|C{2.5cm}|}{99.99} \\
\hline
\end{tabular}
\begin{tablenotes}\footnotesize
\item[1] The lose terminology of `connectivity' was introduced to indicate that a specific service request was indeed satisfied according to the specification discussed in~\cite{EUROCONTROL2007Future}.
\item[2] The automatic execution service is capable of automatically capturing situations where encounter-specific separation is being used and a non-conformance event occurs with minimal time remaining to solve the conflict.
\end{tablenotes}
\end{threeparttable}
\end{table*}
}

\subsection{Security}

An \gls{AANET} has to guarantee a high level of information security, so that the control network can maintain the safety of the flight, while the passenger network can protect the personal data of the passengers. Any networked system must address the following basic security requirements: confidentiality for ensuring the privacy of the end users and for protecting their data from spoofing; authentication for ensuring that only valid users have access to the network's resources; privacy for providing long-term anonymity and for preventing tracking of passengers; traceability and revocation for ensuring that malicious use of on-board units is traced and disabled in a timely manner; and integrity for ensuring that the data sent by the end user is not modified by any malicious element in the network~\cite{thanthry2005aviation}.

Thus, in addition to the above requirements, an aircraft network has to support additional security, for maintaining the separation amongst various network segments. Any security breach within the control network may result in serious consequences for flight safety. Hence, it is vital to maintain isolation between the control network and the passenger network~\cite{thanthry2005aviation,thanthry2006security}.

Table~\ref{tab:security-application} shows the security requirements of different AANET applications. In the control network, only the  authentication requirement is necessitated for the upload/download data-transmission services, while the security has to meet all of the requirements imposed by air traffic control, aircraft tracking, free flight and formation flight. In the passenger network, the extreme integrity of data may not be required for passenger entertainment, but all the other requirements have to be met to protect user safety.

\linespread{1.0}
\begin{table*}[!htbp]
\caption{Security requirements of AANETs applications.}
\label{tab:security-application}
\begin{center} 
\begin{tabular}{|c|l|c|c|c|c|c|}
\cline{2-7}
    \multicolumn{1}{C{0.7cm}|}{}   & \backslashbox{Applications}{Requirements} & Confidentiality & Authentication & Privacy & Revocation & Integrity  \\ \hline
       \multirow{5}{*}{CN} & Flight data delivery at airport~\cite{wright2000wireless,wright2001wirelesspread} &  $\surd$ & $\surd$ &  &  &  \\ \cline{2-7}
                     & Air traffic control~\cite{park2014hybrid,strohmeier2014realities} & $\surd$ & $\surd$ & $\surd$ & $\surd$ & $\surd$  \\ \cline{2-7}
 & Aircraft tracking~\cite{nelson1998flight,medina2008feasibility,medina2012geographic} & $\surd$ &  & $\surd$ & $\surd$ & $\surd$  \\ \cline{2-7}
                     & Free flight~\cite{Airbus2015} & $\surd$ & $\surd$ & $\surd$ & $\surd$ & $\surd$  \\ \cline{2-7}
                     & Formation fly~\cite{Bos2010formation,dijkers2011integrated,Airbus2015} & $\surd$ & $\surd$ & $\surd$ & $\surd$ & $\surd$  \\ \hline
       \multicolumn{1}{|C{0.75cm}|}{PN}           & In-flight entertainment~\cite{jahn2003evolution} & $\surd$ & $\surd$ & $\surd$ & $\surd$ &  $\surd$  \\ \hline
\end{tabular}
\begin{tablenotes}\footnotesize
\item CN: Control Network.
\item PN: Passenger Network.
\item $\surd$: Required.
\end{tablenotes}
\end{center}
\end{table*}

\subsection{Robustness}

The dynamic nature of the \gls{AANET} topology has been considered in~\cite{vey2014aeronautical}, which may result in interrupted connectivity or even node drop-out in the network.
In order to maintain connectivity, \glspl{AANET} must employ specific disruption prevention mechanisms to make them robust. Hence  self-healing relying on distributed and adaptive techniques is desired by \glspl{AANET}.
However, the highly dynamic nature of AANETs results in a potentially intermittent connectivity~\cite{ghosh2016acpm}. To elaborate, the connectivity of the aircraft network is primarily a function of velocity, position, direction and communication range of the aircraft, which is highly dependent on their mobility. 
Therefore, a realistic mobility model is required for reflecting the typical movement of the aircraft. The transmitter and receiver of the \glspl{AANET} have to be robust against both interference and latency, hence requiring sophisticated coding and modulation schemes~\cite{schnell2014ldacs,neji2013survey}. The routing of the network has to use redundant multi-hop paths and be capable of prompt self-healing in order to guarantee sustained connectivity~\cite{vey2014aeronautical}. The aircraft in the system are envisioned to participate as intelligent nodes in a global network of aerial, satellite and ground systems  for ensuring that all information reliably reaches the right place at the right time for both processing and decision making~\cite{sampigethaya2011future}.

\subsection{Cost}

The cost of manufacturing the \glspl{AANET} hardware and deploying it within the aircraft will affect how widely it is adopted. Intuitively, an airline will prudently consider the social and economical implications of introducing the function of \glspl{AANET}. Furthermore, as discussed in Section~\ref{S3}, the first criterion for creating an {\it{ad hoc}} network among aircraft is that there has to be an adequate number of compatible aircraft in a given area. Thus, the adoption rate of the \glspl{AANET} will affect the total value of the network.

Overall, \glspl{AANET} are required for reliable operation under different channel conditions, diverse network topologies and various scenarios, both with and without the support of \glspl{GS} and satellites.
It may be impossible to simultaneously satisfy all the requirements and applications. For example, improving the safety may gravely erode the capacity and efficiency of the system. The conflicting requirements will impose challenges, which must be carefully considered for achieving tradeoffs among various criteria.

\subsection{Summary}
In this section, we discussed the requirements of the different applications of \glspl{AANET} shown in Fig.~\ref{FIG1:AANET_topology}. As shown in Fig.~\ref{FIG3:app_requirement} and Table~\ref{TAB5:relationships-requirements-challenges}, the requirements vary depending on the corresponding applications. In order to cater for various applications, the coverage of aeronautical communication has to be global, although this is challenging due to the high-velocity aircraft mobility, and the associated propagation environment, which introduces propagation losses and delays. Therefore, multi-hop AANETs are required for meeting the global coverage requirement. Table~\ref{TAB2:throughput} characterizes the different data throughput requirements of diverse services. Latency is a critical requirement of \glspl{AANET} because of their safety/control applications, but also for real-time interactive telephony. Table~\ref{TAB3:QoS-Latency} summarizes the associated latency requirements~\cite{EUROCONTROL2007Future}. \glspl{AANET} must provide resilient end-to-end delivery of data, requiring a network relying on self-discovery/healing and reliable control of the communications network. \glspl{AANET} are also required to decentralize the communication,  when the associated aircraft are ready to land and leave the network. Another essential requirement for \glspl{AANET} is information security. Table~\ref{tab:security-application} shows the security requirements of different \gls{AANET} applications in terms of their grade of confidentiality, authentication, privacy, revocation and integrity. Due to the highly dynamic nature of \glspl{AANET}, the communication must be robust to   delays or link failures. The data routing in \glspl{AANET} must be intelligent,  in order to ensure  that all information reaches the right place at the right time for both processing and decision making~\cite{sampigethaya2011future}. The last requirement identified  in this section for \glspl{AANET} is the manufacturing cost, since this will affect how widely \glspl{AANET} are adopted. It may be necessary  to strike tradeoffs between the cost and other quality criteria, such as the throughput.
\section{Aeronautical Communications}\label{S5}
\begin{figure*}
\begin{center}
 \includegraphics[width=0.95\textwidth,angle=0]{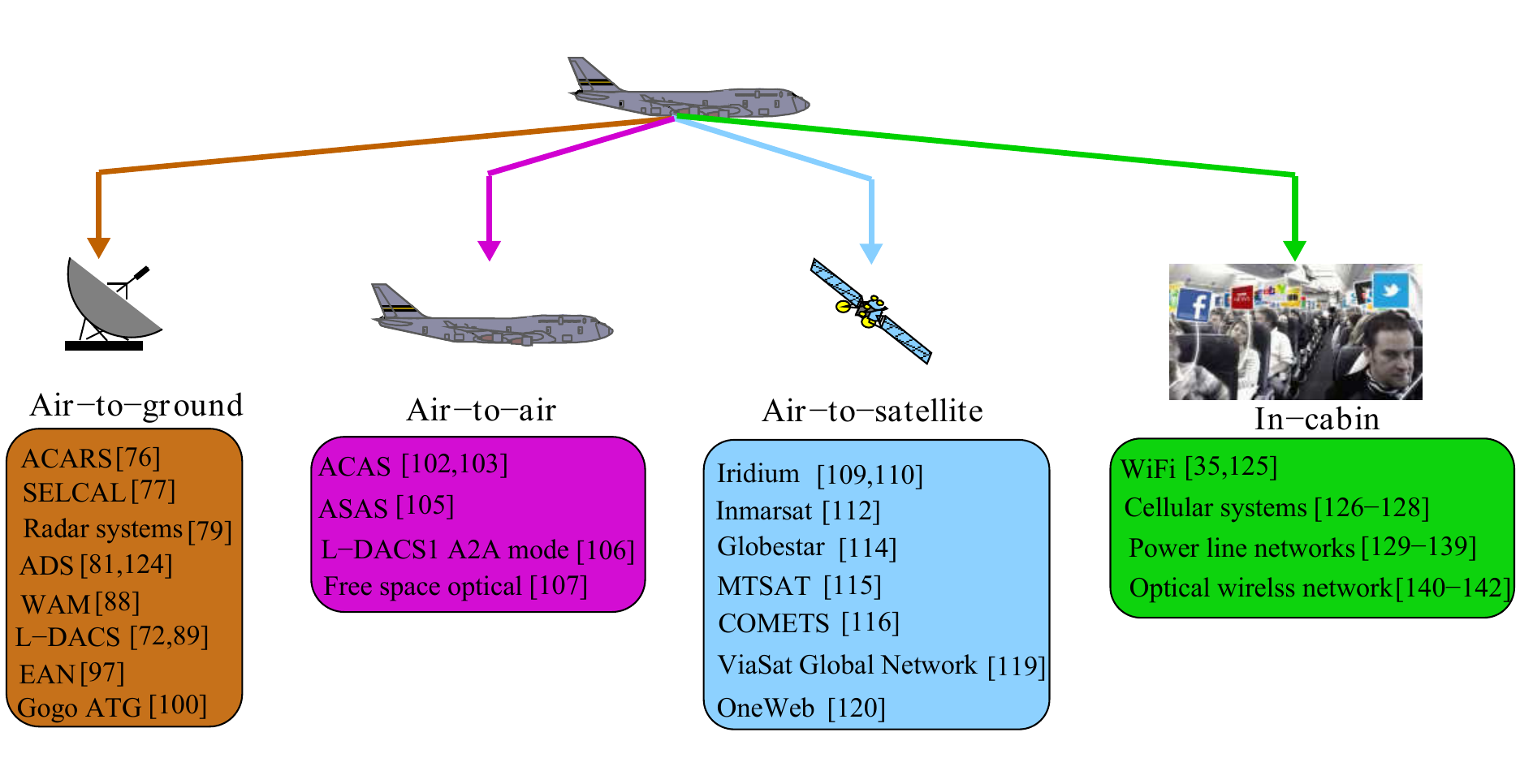}
\end{center}
\vspace*{-1mm}
\caption{Aeronautical communication systems and techniques.}
\label{FIG10:aircraft-communication-systems}
\end{figure*}

In this section, we will describe a range of existing aircraft communication systems and discuss future techniques~\cite{Aradhana2014Enabling,plass2014flight} considered for mitigating the increasing congestion and for meeting the future demands of sustainable air traffic worldwide~\cite{Aradhana2014Enabling}. As shown in Fig.~\ref{FIG10:aircraft-communication-systems}, the aircraft communication systems and technologies are categorized as \gls{A2G}, \gls{A2A} and \gls{A2S}. Meanwhile, existing and potential in-cabin communication techniques are also included in this section.

\subsection{A2G Communication Systems}
\gls{A2G} communication is the means by which people and systems on the ground, such as air traffic control or the aircraft operating agency, communicate with those in the aircraft, which may be outfitted with radio frequency, \gls{GPS}, Internet and video capabilities. Most commercial aircraft carry a device known as a transponder, which acts as an identification tool for the aircraft, allowing \gls{ATC} towers to immediately recognize the identity of each aircraft.

\begin{itemize}[\setlength{\listparindent}{\parindent}]
\item \emph{\gls{ACARS}}: \gls{ACARS}~\cite{EUROCONTROLACARS} defines a digital data link for transmitting messages between the aircraft and the \glspl{GS}, which has been in use since 1978. It was developed to reduce the pilot's workload by using computer based technology for exchanging routine reports, as well as \gls{ATC}, aeronautical operational control and airline administrative control information between the aircraft and the \glspl{GS}, which is now widely used near airports and in populated areas. \gls{ACARS} permits the secure, authenticated exchange of messages between the aircraft and the ground systems by using the security framework of \gls{ICAO} and safe public key infrastructure cryptographic algorithms. However, the \gls{ACARS} system only supports the transmission of short messages between the aircraft and the \glspl{GS} via \gls{HF}, \gls{VHF} or satellite links~\cite{tu2009mobile}, which is not sufficient for supporting applications requiring high throughput, low latencies as well as long-range multi-hop routing via a self-organizing mesh network.
    
\item \emph{\gls{SELCAL}}: The \gls{SELCAL} system~\cite{ASRI2013SELCAL} was introduced in civil aviation as early as 1957, which allows the operator of a \gls{GS} to alert the aircrew that the \gls{GS} wishes to communicate with them. This service is robust in the vicinity of airports and in populated areas where the \glspl{GS} 
have been well deployed. However, its service is poor in unpopulated areas.
The \gls{SELCAL} system is often employed for reducing the burden on the flight crew, owing to the intermittent nature of voice communication on long oceanic routes~\cite{wyatt2013aircraft}. However, \gls{SELCAL} is inadequate in latency-sensitive applications. This is because the aircraft receives and decodes the audio signal broadcast by a ground-based radio transmitter, where the transmission consists of a combination of four pre-selected audio tones whose transmission requires approximately two seconds~\cite{ASRI2013SELCAL}. The frequency employed is either in the \gls{HF} or \gls{VHF} range, where the \gls{GS} and the aircraft must be operated on the same frequency.

\item \emph{Radar Systems}: Radar systems were originally designed for military applications, but are now widely used also in civilian applications for the surveillance of aircraft. There are two types of radar systems used for the detection of aircraft: the so-called \gls{PSR} and \gls{SSR} systems~\cite{nolan2010fundamentals}. The \gls{PSR} system measures only the range and the bearing of targets by detecting   the radio signals that they reflect. In this way, the \gls{PSR} is capable of operating totally independently of the target aircraft, since no action is required from the aircraft for providing a response. However, the \gls{PSR} requires enormous amounts of power to be radiated, in order to receive a sufficiently high power from the target, which would be a health hazard if deployed in the vicinity of populated areas. In addition, the \gls{PSR} requires a significant effort and financial investment to install and maintain, owing to the mechanical nature of the rotating antenna.

    By contrast, the \gls{SSR} systems~\cite{nolan2010fundamentals} rely on targets equipped with a radar transponder, which replies to each interrogation signal by transmitting a response containing encoded data. There are three main advantages of \gls{SSR}. Firstly, since the reply is transmitted from the aircraft, it is much stronger when received at the \gls{GS}, hence providing a much higher range than the \gls{PSR}. Secondly, the transmission power required by the \gls{GS} for a given range is substantially reduced, together with the associated cost. Thirdly, since the signals are electronically coded, additional information can be transmitted between the aircraft and the \gls{GS}, such as the aircraft's position, heading direction and speed. However, the disadvantage of \gls{SSR} is that it requires the target aircraft to carry a compatible transponder. Hence, \gls{SSR} is a so-called `dependent' surveillance system.

    Passive radar is also a family member of radar systems, which is capable of opportunistically exploiting a variety of transmitters~\cite{howland2005editorial}, i.e. \gls{FM} radio, digital audio broadcast, digital video broadcast, global navigation satellite systems and cell-phone base-stations for object detection. Explicitly, the time difference of arrival between the signal arriving directly from the transmitter and the signal arriving via reflection from the object are measured for calculating the location, speed and even the bearing of the objects.

    Radar systems are capable of providing coverage in airports and in populated areas at a low latency, whilst ensuring robust operation. Owing to this, they constitute the main solutions invoked for discovering and monitoring  aircraft. However, radar systems cannot deliver data between the \glspl{GS} and aircraft, hence they are incapable of data upload/download and of the provision of passenger entertainment.

\item \emph{\gls{ADS}}: The \gls{ADS} system relies on a technique, in which the aircraft uses a data link for automatically providing data derived from onboard navigation and position-fixing systems, including aircraft identification, position and any additional information as appropriate. This system is automatic, because it requires no pilot or controller input for its operation (other than turning the equipment on and logging in to the system). Furthermore, it is referred to as being dependent, because it requires compatible airborne equipment, such as an \gls{SSR}. The original \gls{ADS} system is known as \gls{ADS-C}, because reports from the aircraft are generated in compliance with a contract set up with the ground system.

    Furthermore, in order to improve the performance of the \gls{ADS} system, \gls{ADS-B}~\cite{lester2007benefits} has been designed, which uses a combination of satellites to provide both flight crews and ground control personnel with very specific information about the location and speed of airplanes~\cite{richards2010new}. As an element of \gls{SESAR}~\cite{SESAR} in Europe and the \gls{NextGen}~\cite{NextGen} in the \gls{US}, \gls{ADS} is being rolled out in support of a variety of applications for oceanic, domestic en-route and terminal area flight operations, as well as to provide collision avoidance protection, when operating in-flight and on the airport ground~\cite{strohmeier2014realities}. A natural limitation of \gls{ADS-B} is that only the aircraft equipped with the same type of transceiver can exchange their position reports, and there are a variety of different kinds of transceivers,  which are designed for diverse aircraft types, regions, locations, etc. Furthermore, the broadcast aircraft remains unaware, as to whether its report has or has not been received, since there is no further broadcast from the receiving aircraft. Moreover, \gls{ADS-B} may be unsuited for delivering large amounts of data for passenger entertainment, since its data link throughput is only about 1 Mbps~\cite{costin2012ghost}. Explicitly, \gls{ADS-B} replies/broadcasts are encoded by a certain number of pulses with \gls{PPM}, each pulse being 1~$\mu$s long. Owing to this, \gls{ADS} is mainly used by \gls{ATC} for air traffic management. However, its security mechanisms are challenged by the lack of entity authentication, message signatures and message encryption~\cite{costin2012ghost}, which is requiring more research efforts in such issues. According to the OpenSky report 2016~\cite{schafer2016opensky}, about $70\%$ of all Mode S transponders in Europe and the \gls{US} have already benn upgraded with ADS-B capabilities. Moreover, ADS-B systems are to be deployed on most aircraft by 2020~\cite{strohmeier2015security}.

\item \emph{\gls{WAM}}: \gls{WAM}~\cite{james2007multilateration} constitutes a surveillance technique that exploits the various transmissions broadcast from the aircraft. The \gls{WAM} relies on a number of relatively simple \glspl{GS} deployed over the terrain for triangulating an aircraft's position, which is capable of providing accurate localization and robust tracking in its coverage. If a compatible transponder is installed on the aircraft, \gls{WAM} is capable of tracking various aircraft parameters,  such as identification, position, altitude, etc. This system has the advantage of requiring a much simpler and thus cheaper ground installation than the conventional \gls{SSR}, while not necessarily requiring the installation of expensive aircraft equipment, as required in \gls{ADS}. Although \gls{WAM} is capable of tracking and maintaining aircraft for the applications of \gls{ATC}, it cannot provide data-transmission services, such as passenger entertainment and upload/download. Furthermore, \gls{WAM} is not suited to latency-sensitive applications, such as formation flight and free flight, since it relies on \glspl{GS} and does not support \gls{A2A} communication.

\item \emph{\acrfull{L-DACS}}: The \gls{L-DACS}~\cite{EUROCONTROL2007actionplan,schnell2014ldacs} is one of the most important data links of the \gls{FCI}. It is designed for mitigating the saturation of the current continental \gls{A2G} aeronautical communication systems that operate in the \gls{VHF} band. Furthermore, \gls{L-DACS} is capable of providing secure and robust data services in populated areas. The \gls{ICAO} has recommended the further development and evaluation of two \gls{L-DACS} technology candidates, \gls{L-DACS}1~\cite{Eurocontrol2009dacs1s} and \gls{L-DACS}2~\cite{EUROCONTROL2009dacs2t}.

\begin{itemize}
\item \emph{{\gls{L-DACS}1}} is a combination of the P34 solution (TIA 902 standard)~\cite{TIA2002}, of the \gls{B-AMC} system~\cite{rokitansky2007b} and of the \gls{WiMAX}~\cite{etemad2008overview}, as illustrated in Table~\ref{TAB4:A2G}. The \gls{L-DACS}1 system is based on the classic \gls{FDD} technique, where the \gls{GS} and the airborne equipment transmit simultaneously using distinct frequency bands~\cite{neji2012coexistence}.
\item \emph{{\gls{L-DACS}2}} relies on a combination of the \gls{GSM}, of the \gls{UAT} and of the \gls{AMACS}. It uses the \gls{GSM} physical layer and the \gls{AMACS} \gls{MAC}, as shown in Table~\ref{TAB4:A2G}. The \gls{L-DACS}2 system is a narrowband single carrier system utilizing the \gls{TDD} technique, where both the \gls{GS} and the airborne equipment transmit using the same carrier frequency during distinct time intervals~\cite{neji2012coexistence,jain2011analysis}.
\end{itemize}

Overall, \gls{L-DACS}1 associated with \gls{OFDM} is more scalable, more spectrally-efficient and more flexible than \gls{L-DACS}2, which relies on single-carrier modulation. However, the \gls{TDD} structure of \gls{L-DACS}2 is more suitable for asymmetric data traffic, whilst the \gls{FDD} of \gls{L-DACS}1 is more suitable for symmetric voice traffic, but less suitable for data.

\item \emph{European Aviation Network}: Inmarsat and Deutsche Telekom are powering Europe's aviation connectivity via the \gls{EAN}~\cite{Inmarsat2017}, which consists of S-band satellite and \gls{LTE}-based \glspl{GS}. Explicitly, Inmarsat provides the satellite access service, while Deutsche Telekom build and manage approximately 300 \glspl{GS} across all the 28 European Union member states based on a 4G-\gls{LTE} mobile terrestrial network that seamlessly works together with Inmarsat's satellites. Note that the \gls{LTE}-based \glspl{GS} built for \gls{EAN} are different from the `standard' \gls{LTE} system designed for terrestrial networks, since they cater for speeds of up to 1200 km/h at cruising altitudes, requiring a cell diameter of up to 150 km~\cite{Nokia2017}. Furthermore, \gls{EAN} is capable of providing as high as 50Gbps total network capacity, which allows on-board passengers to enjoy broadband Internet access in the air just as well as on the ground. {\color{black} However, LTE-based A2G communications still suffer from the major challenges imposed by the uplink/downlink interference, high-Doppler mobility and hostile channel effects \cite{tadayon2016inflight}.}

\item \emph{Gogo ATG Network}: The US provider Gogo has built an \gls{ATG} network comprising about 200 \glspl{GS} in the continental area of USA, Alaska and Canada~\cite{gogo2017}. Explicitly, Gogo exploits the existing Airfone ATG phone relay stations and the newly built towers operating in the 850~MHz frequency band, in order to  provide 3.1~Mbps data rate for in-flight WiFi for on-board passengers. In order to meet the growing demand for bandwidth, Gogo developed its second-generation \gls{ATG} technology \gls{ATG}4, which exploits advanced multiple antenna technology on the aircraft. Explicitly, the aircraft is equipped with four omni-directional antennas for exchanging data information with the \glspl{GS}. ATG operates in the 800~MHz frequency band based on the CDMA2000 standard and it is capable of providing up to 9.8~Mbps data rate~\cite{gogo2017AGT4}.
\end{itemize}

As shown in Table~\ref{TAB4:A2G}, we compare the above-mentioned \gls{A2G} communication systems in terms of their duplexing  mode, modulation type, spectral efficiency, throughput and served domain as well as their applications.

{{\small
\linespread{1.0}
\begin{table*}
\caption{Comparison of different A2G communications systems.} 
\label{TAB4:A2G} 
{\footnotesize
\centering 
\begin{tabular}{|C{1.7cm}|C{0.7cm}|C{1.4cm}|C{1.7cm}|C{2.8cm}|C{2.0cm}|C{1.2cm}| C{1.6cm}| C{1.2cm}|} 
\hline\hline 
System  & Duplex & Combinations& Modulation / Mode Type & Spectrum & Throughput & Served Domain & Communication Distance & Applications \\ [0.5ex] 
\hline 
\acrshort{ACARS}~\cite{EUROCONTROLACARS}  & Full-duplex &  Telex System &Amplitude Modulation-Minimum Shift Keying (AM-MSK)  &3MHz---30MHz(HF), 129.15 MHz---136.90 MHz (\acrshort{VHF})  & 2400 bps & Airport/ Continent/ Oceanic & Upto 200NM& Data\\ %
\hline 
\acrshort{SELCAL}~\cite{ASRI2013SELCAL}     & FDD &  / &TDMA  &3MHz---30MHz(HF), 129.15 MHz---136.90 MHz (\acrshort{VHF})  & Up to 120 kbps & Airport/ Continent/ Oceanic & Upto 200NM & Voice\\ %
\hline 
\acrshort{PSR} ~\cite{nolan2010fundamentals}  & / & /&  / &2700 MHz---2900 MHz&/ &Airport/ Continent & Upto 220NM& Surveillance\\
\hline
\acrshort{SSR} ~\cite{nolan2010fundamentals}  &/ & / & Mode A, Mode C, Mode S & 1030 MHz---1090 MHz&\acrshort{UL} 0 bit/s/\acrshort{DL} 23 bps &Airport/ Continent& Upto 250NM& Surveillance\\ 
\hline
\gls{ADS-B}~\cite{lester2007benefits}  &  \gls{TDD} & \gls{UAT}/VDL4 & PPM & 960 MHz --- 1215 MHz (\acrshort{UAT})/117.975 MHz 137 MHz (VDL4)& 1 Mbps & Airport/ Continent/ Oceanic & Upto 250NM & Surveillance\\ 
\hline
\gls{WAM}~\cite{james2007multilateration}& /  & / &Mode A/C, Mode S, and Mode S ES (work with \acrshort{SSR})&Depend on the mode employed& /   &Airport & Determined by
the geometry of the GS & Surveillance\\[1ex] 
\hline 
\gls{L-DACS}1~\cite{Eurocontrol2009dacs1s} &  \gls{FDD} & P34, \gls{B-AMC} \& \gls{WiMAX}& \acrshort{OFDM}  & 960 MHz --- 1164 MHz & Upto 1373 kbps (forward link), upto 1038 kbps (reverse link)& Continent & Upto 20 NM & Data\\ 
\hline
\gls{L-DACS}2~\cite{EUROCONTROL2009dacs2t} &  \gls{TDD} & \gls{GSM}, \gls{UAT} \& \gls{AMACS}& \acrshort{CPFSK}/\acrshort{GMSK} & 960 MHz --- 1164 MHz &273 kbps (forward link/reverse link)& Continent & Upto 200NM & Data\\%
\hline
EAN ~\cite{Inmarsat2017} &  FDD & LTE &OFDM & 2 GHz --- 4 GHz &75 Mbps (peak rate)& Continent/ Oceanic& Upto 81NM & Data\\
\hline
GoGo ATG ~\cite{gogo2017} &  FDD & CDMA2000 &CDMA & 4 MHz of spectrum in the 850MHz band  & 9.8 Mbps per aircraft& Continent& Upto 81NM & Data\\
\hline 
\end{tabular}
}
\end{table*}
}}

\subsection{A2A Communication Systems}
Aircraft are routinely equipped with \gls{GPS} for navigation purposes and for \gls{A2A} communications between pilots. This provides a global time reference that can be exploited for synchronization among network nodes, for example, for scheduling contention-free transmissions~\cite{medina2011airborne}. Furthermore, \gls{A2A} communication is a key technology in future aeronautical communication systems, which aims for separation assurance and collision avoidance, as well as Internet surfing for passenger entertainment. In this section, we will discuss three main technologies used in \gls{A2A} and a potential technology for future \gls{A2A} communication, namely \gls{ACAS}, \gls{ASAS}, \gls{L-DACS}1 \gls{A2A} mode and \gls{FSO} communications.

\begin{itemize}[\setlength{\listparindent}{\parindent}]
\item \emph{\gls{ACAS}}: The \gls{ACAS}~\cite{williams2004airborne,kuchar2007traffic} operates independently of any ground-based equipment and air traffic control, which allows low-latency direct \gls{A2A} communication among aircraft. It is capable of warning pilots of approaching other aircraft that may present a threat of collision. Specifically, the only commercial version of \gls{ACAS}-II relies on \gls{SSR} transponder signals and it will generate a \gls{RA} to warn the pilot if a risk of collision is established by \gls{ACAS}-II. The \gls{ACAS} is a short-range system designed for preventing metal-on-metal collisions by providing secure and robust \gls{A2A} communication between pilots. There are three types of \gls{ACAS}~\cite{williams2004airborne}, namely \gls{ACAS}-I which gives \gls{TA} but does not recommend any maneuvers, \gls{ACAS}-II which gives \glspl{TA} and \glspl{RA} in the vertical direction and \gls{ACAS}-III which gives \glspl{TA} and \glspl{RA} in vertical and/or horizontal directions, respectively. More specifically, the implementation of \gls{ACAS}-II is referred to as the \gls{TCAS}-II version 7.0 and version 7.1~\cite{FAA2011TCAS}, but \gls{ACAS}-III has not as yet been rolled out. All the three types of \gls{ACAS} provide only emergency communication between pilots, without supporting data transmission for passenger applications.

\item \emph{\gls{ASAS}}: The \gls{ASAS} enables pilots to maintain separation from one or more other aircraft, while providing flight information concerning the surrounding traffic~\cite{EUROCONTROLASAS}. Against the background of the ``free flight'' concept, \gls{ASAS} was developed to assist pilots in self-separation, facilitating the flexible use of airspace along user-preferred trajectories, hence allowing direct routing. Similar to the \gls{ACAS}, \gls{ASAS} does not support any services for the passengers, since it only allows the exchange of flight information among aircraft at a low data rate. Explicitly, airborne surveillance and separation assurance processing equipment is used for processing surveillance reports from one or more sources, which has to assess the target data according to pre-defined criteria for assisting pilot-controlled self separation or self maneuver.

\item \emph{\gls{L-DACS}1 \gls{A2A} Mode}: \gls{L-DACS}1 \gls{A2A} mode has been designed for the periodic transmission of \gls{A2A} surveillance data, while supporting the transmission of a low volume of non-periodic \gls{A2A} messages~\cite{graupl2011dacs1}. The system relies on a self-adaptive slotted \gls{TDMA} protocol for providing \gls{A2A} data communication services, using the so-called  paired approach, self separation and \gls{ATC} surveillance. The maximum net user data rate is up to 273~kbps~\cite{graupl2011dacs1}. However, the delivery of passenger data is not supported in the current stage of \gls{L-DACS}, although this will be developed in future versions.

\item \emph{Free Space Optical}: \gls{FSO}~\cite{chan2006free} communication constitutes a promising technique that adopts \glspl{LD} as transmitters to communicate, for example between aircraft as well as between aircraft and a satellite, at high rates of up to 600~Mbps for \gls{MANET} applications~\cite{bilgi2010throughput}. Since \gls{FSO} signals are very directional and limited to a small diameter, it is virtually impossible to intercept \gls{FSO} signals from a non-desired destination. This feature can meet the very high security requirements of aeronautical communications. Establishing their applicability to \gls{A2G} communications requires further studies owing to eye-safety concerns. The directional and license-free features of \gls{FSO} are appealing in aeronautical communication, since the conventional radio-frequency communication is fundamentally band-limited. However, \gls{FSO} communications are vulnerable to mobility, because \gls{LOS} alignment must be maintained for high-integrity communication. In order to solve the associated problem of pointing and tracking accuracy, a feasible solution is to rely on the built-in \gls{GPS} system of the aircraft, along with the \gls{FSO} system's low transmission latency, which can also assist in formation-light. Furthermore, since there are no obstacles in the stratospheres, the main disadvantage of \gls{FSO} links in terms of requiring a \gls{LOS} channel becomes less of a problem. Thus, the \gls{FSO} communication links between aircraft have a promising potential in terms of constructing an \gls{AANET} for aircraft tracking and collision avoidance. Finally, since \gls{FSO} and \gls{RF} links exhibit complementary strengths and weaknesses, a hybrid \gls{FSO}/\gls{RF} link offers great promise in future aircraft communications.
\end{itemize}

\subsection{A2S Communication Systems}
Satellite systems are especially important for enabling communications to aircraft in oceanic and other unpopulated areas, since they are capable of providing global coverage, as well as global discovery and control for other communication systems. \gls{A2S} communication also complements \gls{A2G} communication where appropriate~\cite{schnell2014ldacs}, for example, for locating the position of aircraft. However, apart from having a high cost, aeronautical communication relying on satellites suffers from very long end-to-end propagation delays of approximately 250~ms~\cite{medina2012geographic}, which prevent its use in latency-sensitive applications, such as formation flight. Additionally, the achievable throughput is relatively low in the operational satellite systems, making them incapable of meeting the requirements of high throughput applications, such as online video entertainment via Internet access. In the following discussions, we will briefly consider six existing and emerging satellite systems.

\begin{itemize}[\setlength{\listparindent}{\parindent}]
\item \emph{Iridium}: The Iridium network~\cite{maine1995overview} consists of 66 active satellites used for worldwide voice and data communication from hand-held satellite phones to other transceiver units. Although the Iridium system was primarily designed for supporting personal communications, aeronautical terminals having one to eight channels have been developed for the Iridium system by AlliedSignal Aerospace, which are also used for military transport aircraft~\cite{nicol2000future}.

    As an improved version, Iridium NEXT~\cite{Iridium2015Iridium} began to launch in 2015, maintaining the existing Iridium constellation architecture of 66 cross-linked low-earth orbit satellites covering 100 percent of the globe. It dramatically enhances Iridium's ability to meet the rapidly-expanding demand for truly global mobile communications on land, at sea and in the skies.

\item \emph{Inmarsat}: Inmarsat~\cite{berzins1989inmarsat} is a British satellite telecommunications company, offering global mobile services. It provides telephone and data services for users worldwide, via portable or mobile terminals, which communicate with \gls{GS} through eleven geostationary telecommunications satellites. Inmarsat provides voice/fax/data services for aircraft, employing three levels of terminals: Aero-L (low gain antenna) primarily for packet data including \gls{ACARS} and \gls{ADS}, Aero-H (high gain antenna) for medium-quality voice and fax/data at up to 9600~bit/s, and Aero-I (intermediate gain antenna) for low-quality voice and fax/data at up to 2400~bit/s.
     
    In order to enhance the capacity of the Inmarsat system, Inmarsat signed a contract with Boeing to build a constellation of three Inmarsat-5 satellites, as part of a US \$1.2 billion worldwide wireless broadband network referred to as Inmarsat Global Xpress~\cite{ITU2011Regulation}. The satellites operate in the $Ka$-band in the range of 20-30 GHz. Each Inmarsat-5 carries a payload of 89 small $Ka$-band beams, with the objective that a global $Ka$-band spot coverage can be offered with the aid of all three Inmarsat-5 satellites~\cite{ITU2011Regulation}. There are plans to offer high-speed in-flight broadband on airliners through Inmarsat Global Xpress~\cite{ITU2011Regulation}.

\item \emph{GlobeStar}: The GlobeStar system~\cite{Globalstar2014} is the world's largest provider of mobile satellite voice and data services. The GlobeStar system consists of 52 satellites, of which 48 satellites provide full commercial service for users, while the other four satellites are in-orbit as spares, ready to be activated when needed. GlobeStar uses a version of classic \gls{CDMA} technology based upon the Pan-American IS-95 \gls{CDMA} standard, which has made GlobeStar well known for its crystal clear, ``land-line quality" voice service to commercial and recreational users in more than 120 countries around the world. In 2013, GlobeStar successfully completed launching its constellation of second generation satellites, which support the company's current line-up of voice, as well as duplex and simplex data products and services.

\item \emph{\gls{MTSAT}}: The \gls{MTSAT} system~\cite{MTSAT} relies on a series of weather and aviation-control satellites. They are geostationary satellites owned and operated by the Japanese Ministry of Land, Infrastructure, Transport and Tourism and the Japan Meteorological Agency. Operating in L-band, the \gls{MTSAT} system provides both communications and navigational services for aircraft, and gathers weather data for users throughout the entire Asia-Pacific region. The MTSAT satellite-based augmentation system improves the reliability and accuracy of \gls{GPS} via \gls{MTSAT} for aircraft utilizing the \gls{GPS} position information for their navigation. Unlike the conventional navigational means such as VHF omnidirectional range or distance measuring equipment, it is able to cover a wide range of oceanic and ground areas making it possible to set up flexible flight routes.

\item \emph{\acrfull{COMETS}}: As a collaboration between the \gls{CRL} of the Ministry of Posts and Telecommunications and the \gls{NASDA} in Japan, the \gls{COMETS}~\cite{wakana2000communications} system was developed for future communications and broadcasting. It relies on a two-ton geostationary three-axis stabilized satellite. Particularly, it carries three payloads: advanced mobile communications equipment developed by \gls{CRL} for the $Ka$-band (31/21 GHz) and the millimeter-wave band (47/44 GHz); 21-GHz-band advanced satellite broadcasting equipment developed by \gls{CRL} and \gls{NASDA}; and inter-orbit communication equipment developed by \gls{NASDA}~\cite{miura2002ka}. The broadcasting experiments conducted for high-definition television~\cite{wakana1998comets} showed that it was capable of providing a transmission rate of up to 140 Mbps.

\item \emph{ViaSat Global Network}: ViaSat's airborne satellite communications services offer worldwide access and a range of service levels, providing connectivity and performance options tailored to meeting a variety of needs, including office-in the-sky, real-time broadcasting HD TV, communications on-the-move and air traffic control. The data rate is expected to range from 512~kbps to 10~Mbps~\cite{Viasat2010Government}. The network's performance is optimized for reliable and secure two-way mobile broadband communications. Therefore, airborne operators have the capability of sending full-motion video, making secure phone calls, conducting video conferences, accessing classified networks, and even to perform mission-critical communications in flight.

\item \emph{OneWeb Satellite Constellation}: The OneWeb satellite constellation~\cite{SeldingVu2014asatellite} was started by telecommunications entrepreneur Greg Wyler with the support of Google. It comprises approximately 648 satellites. The OneWeb satellite constellation is aiming for providing global Internet broadband access for ground users, but it can also be exploited for providing global Internet access for airborne users via \gls{A2S} links. The OneWeb satellites operate at approximately 1200~km altitude~\cite{Pasztor2016OneWeb}, and the frequency spectrum used for providing broadband access service is the $Ku$ band~\cite{SeldingVu2014Google} spanning from 11.7 to 12.7~GHz (downlink transmission) and from 14 to 14.5~GHz (uplink transmission).  Each satellite is capable of supporting 6~Gbps throughput, while the user is capable of accessing the Internet at 50 Mbps rate using a phased array based antenna measuring approximately 36 by 16~cm~\cite{SeldingVu2015Virgin}. While the $Ku$ band suffers from rain-induced attenuation proportional to the amount of rainfall, this problem may be solved by appropriate link budget design and power allocation for the satellite network.
\end{itemize}

{\linespread{1.0}
\begin{table*}[!htbp]
\caption{The relationships of aeronautical communication applications, requirements and aeronautical communication systems/techniques.} 
\label{TAB5:relationships-requirements-challenges} 
\centering 
\begin{threeparttable}
\begin{tabular}[t]{C{1.5cm}|L{3.8cm}|C{0.5cm}|C{0.5cm}|C{0.5cm}|C{0.5cm}|C{0.5cm}|C{0.5cm}|C{0.5cm}|C{0.5cm}|C{0.5cm}|} 
\cline{2-11}                        
& In-fliht entertainment~\cite{jahn2003evolution} &$\surd$&$\surd$&$\surd$&$\surd$&$-$      &$\surd$ &$-$      &$-$  &$\surd$\\ [0.5ex] 
\cline{2-11}
& Free flight~\cite{Airbus2015}                     &$-$      &$-$        &$\surd$&$-$      &$-$      &$\surd$&$\surd$&$\surd$&$-$\\ [0.5ex] 
\cline{2-11}
& Formation fly~\cite{Bos2010formation,dijkers2011integrated,Airbus2015}               &$-$      &$\surd$  &$\surd$&$-$       &$\surd$&$-$      &$\surd$&$\surd$&$-$\\ [0.5ex] 
\cline{2-11}
& Aircraft tracking ~\cite{nelson1998flight,medina2008feasibility,medina2012geographic}          &$-$      &$\surd$  &$\surd$&$-$       &$-$      &$\surd$&$\surd$&$\surd$&$-$\\ [0.5ex] 
\cline{2-11}
& Air traffic control~\cite{park2014hybrid,strohmeier2014realities}         &$\surd$&$\surd$  &$\surd$&$-$       &$-$      &$\surd$&$\surd$&$\surd$&$-$\\ [0.5ex] 
\cline{2-11}
& Flight data delivery~\cite{wright2000wireless,wright2001wirelesspread}        &$\surd$&$\surd$  &$\surd$&$\surd$ &$-$      &$\surd$  &$-$     &$-$        &$\surd$\\ [0.5ex] 
\cline{2-11}
& \diagbox[dir=SW,width = 4.4cm,height=2.0cm]{\multirow{5}{*}{Applications}}{\raisebox{-2.0cm}{Requirements}}    &\multirow{2}{*}{\rotatebox{90}{Airport coverage~~~~~~~~~~~}}& \multirow{2}{*}{\rotatebox{90}{Populated coverage~~~~~~~~}} &\multirow{2}{*}{\rotatebox{90}{Unpopulated coverage~~~~~}}&\multirow{2}{*}{\rotatebox{90}{Throughput~~~~~~~~~~~~~~~~~~}} &  \multirow{2}{*}{\rotatebox{90}{Short range latency$^{1}$~~~~~~~}}   &\multirow{2}{*}{\rotatebox{90}{Long range latency$^{2}$~~~~~~~}}  &\multirow{2}{*}{\rotatebox{90}{Security~~~~~~~~~~~~~~~~~~~~~}}& \multirow{2}{*}{\rotatebox{90}{Robustness~~~~~~~~~~~~~~~~~~}}&\multirow{2}{*}{\rotatebox{90}{Low cost~~~~~~~~~~~~~~~~~~~~}}\\ [0.5ex] 
& \diagbox[dir=NW,width = 4.4cm,height=2.0cm]{\vspace{0.3cm}Systems/Techniques}{}             &&  && &     && & &\\ [0.5ex] 
\hline
\multicolumn{1}{|C{1.5cm}|}{A2X}& \multicolumn{1}{L{4.0cm}|}{AANET~\cite{vey2014aeronautical,medina2011airborne}}
&$\sim$&$\sim$  &$\sim$&$\sim$ &$\sim$      &$\sim$&$\sim$      &$\sim$  &$\sim$  \\ [0.5ex] 
\hline
\multicolumn{1}{|C{1.5cm}|}{\multirow{10}{*}{A2G}}& \multicolumn{1}{L{4.0cm}|}{ACARS~\cite{EUROCONTROLACARS}}
&$\uparrow$&$\uparrow$  &$\downarrow$&$\downarrow$ &$\sim$      &/      &$\uparrow$  &$\downarrow$  &$\downarrow$\\ [0.5ex] 
\cline{2-11}
\multicolumn{1}{|C{1.5cm}|}{}& SELCAL~\cite{ASRI2013SELCAL}
&$\uparrow$&$\uparrow$  &$\downarrow$&$\downarrow$ &$\downarrow$      &$\downarrow$   &$\downarrow$  &$\downarrow$  &$\sim$\\ [0.5ex] 
\cline{2-11}
\multicolumn{1}{|C{1.5cm}|}{}& Radar Systems ~\cite{nolan2010fundamentals}
&$\uparrow$&$\sim$  &$\downarrow$&/ &$\uparrow$      &/    &/ &$\uparrow$  &$\downarrow$ \\ [0.5ex] 
\cline{2-11}
\multicolumn{1}{|C{1.5cm}|}{}& ADS~\cite{drouilhet1996automatic,lester2007benefits}
&$\uparrow$&$\uparrow$  &$\sim$&$\downarrow$&$\sim$      &$\downarrow$   &$\sim$ &$\downarrow$  &$\downarrow$ \\ [0.5ex] 
\cline{2-11}
\multicolumn{1}{|C{1.5cm}|}{}& WAM~\cite{james2007multilateration}
&$\uparrow$&$\uparrow$  &$\downarrow$&$\downarrow$&$\sim$      &/   &$\sim$ &$\downarrow$  &$\downarrow$ \\ [0.5ex] 
\cline{2-11}
\multicolumn{1}{|C{1.5cm}|}{}& L-DACS ~\cite{EUROCONTROL2007actionplan,schnell2014ldacs}
&$\downarrow$&$\uparrow$ &$\downarrow$&$\downarrow$&$\sim$      &$\downarrow$  &$\uparrow$&$\downarrow$  &$\downarrow$ \\ [0.5ex] 
\cline{2-11}
\multicolumn{1}{|C{1.5cm}|}{}& EAN ~\cite{Inmarsat2017}
&$\downarrow$&$\uparrow$ &$\downarrow$&$\downarrow$&$\sim$      &$\downarrow$  &$\sim$&$\downarrow$  &$\downarrow$ \\ [0.5ex] 
\cline{2-11}
\multicolumn{1}{|C{1.5cm}|}{}& Gogo ATG ~\cite{gogo2017}
&$\downarrow$&$\uparrow$ &$\downarrow$&$\downarrow$&$\sim$      &$\downarrow$  &$\sim$&$\downarrow$  &$\downarrow$ \\ [0.5ex] 
\hline
\multicolumn{1}{|C{1.5cm}|}{\multirow{5}{*}{A2A}}& \multicolumn{1}{L{4.0cm}|}{ACAS ~\cite{williams2004airborne,kuchar2007traffic}}
&$\uparrow$&$\uparrow$  &$\downarrow$&$\downarrow$ &$\sim$      &$\downarrow$   &$\uparrow$  &$\downarrow$  &$\downarrow$\\ [0.5ex] 
\cline{2-11}
\multicolumn{1}{|C{1.5cm}|}{}& ASAS~\cite{EUROCONTROLASAS}
&$\uparrow$&$\uparrow$  &$\downarrow$&$\downarrow$ &$\sim$      &/ &$\uparrow$ &$\downarrow$  &$\downarrow$\\ [0.5ex] 
\cline{2-11}
\multicolumn{1}{|C{1.5cm}|}{}& L-DACS1 A2A mode  ~\cite{graupl2011dacs1}
&/&$\sim$  &$\downarrow$&$\downarrow$ &$\sim$      &$\downarrow$ &$\uparrow$ &$\downarrow$   &$\downarrow$\\ [0.5ex] 
\cline{2-11}
\multicolumn{1}{|C{1.5cm}|}{}& Free space optical~\cite{chan2006free}
&$\sim$&$\sim$  &$\sim$&$\uparrow$ &$\sim$      &/ &$\uparrow$&$\downarrow$ &$\downarrow$\\ [0.5ex] 
\hline
\multicolumn{1}{|C{1.5cm}|}{\multirow{2}{*}{A2S}}& \multicolumn{1}{L{4.0cm}|}{Satellite systems~\cite{maine1995overview,Iridium2015Iridium,berzins1989inmarsat,Globalstar2014,MTSAT,wakana2000communications,Viasat2010Government}}
&$\uparrow$&$\uparrow$  &$\uparrow$&$\downarrow$ &$\downarrow$   &$\sim$   &$\sim$  &$\uparrow$  &$\downarrow$\\ [0.5ex] 
\hline
\hline
\end{tabular}
\begin{tablenotes}\footnotesize
\item [$\surd$] A crucial requirement.
\item [$-$] Not a crucial requirement.
\item [$\uparrow$] Achieves the requirement to a greater degree than AANET.
\item [$\sim$] Achieves the requirement to the same degree as AANET.
\item [$\downarrow$] Achieves the requirement to a less degree than AANET.
\item [/] Not applicable.
\item ['1'] Short range latency refers to the latency of single hop links.
\item ['2'] Long range latency refers to the latency of multiple-hop links.
\end{tablenotes}
\end{threeparttable}
\end{table*}
}

\subsection{In-Cabin Communications}
Although aircraft constitute the end-points of \glspl{AANET}, passengers on-board the aircraft desire access to the provided throughput. Thus, the existing and potential techniques used for in-cabin communications will be elaborated on as potential extensions.

\begin{itemize}[\setlength{\listparindent}{\parindent}]
\item \emph{\gls{WiFi}}: The popular \gls{WiFi} system is a local area wireless networking technology, which allows an electronic device to exchange data or to connect to the Internet using the 2.4~GHz \gls{UHF} band and the 5~GHz \gls{SHF} radio waves, finding widespread employment in wireless Internet access. This technology can be exploited for the low-cost wireless uploading/downloading of flight information at a high throughput at airports. Furthermore, in response to the passengers' demand, in-flight \gls{WiFi} is now accessible on about $40\%$ of US flights and on international long-haul flights via companies such as American Airline, Lufthansa, Emirates and Qatar Airways.
 
    Since high speed data communications are both desirable and important to society, enhanced \gls{WiFi} techniques are being developed by upgrading the \gls{WiFi} capacity with the aid of  using hybrid technology, such as  GoGo's hybrid ground to orbit technology~\cite{Aircraftinteriors}. The recently developed GoGo@2Ku is capable of providing 70 Mbps peak transmission rate for each aircraft, and its next generation version aims for achieving a 200+~Mbps peak transmission rate~\cite{gogo20162ku}. However, the service is expensive, one-hour pass plan is $\$$ 7.00 or monthly airline plan is $\$$ 49.95 at the time of writing.

\item \emph{Cellular Systems}: There is an increasing interest for the passengers to be able to use their smart phones and laptops on aircraft. More and more firms aim for satisfying the public's urge of mobile communication and surfing the web above the clouds. US and Europe are leading the development of wireless communications onboard of aircraft.
 
    For example, AeroMobile~\cite{AeroMobile2016} enables Virgin Atlantic passengers to use their mobile phones during flights over the UK airspace, while OnAir~\cite{OnAir2016} enables airline passengers to use their personal mobile devices for calls, text messaging, emails and Internet browsing.

    Additionally, passengers are offered the opportunity to use their mobile phones onboard via the \gls{MCA services}~\cite{offcom2008} in Europe. More specifically, mobile phones may connect to a miniature cellular network installed inside the aircraft. The voice and data traffic is then transmitted to a satellite, which routes the traffic to a \gls{GS} for connecting to the conventional telephone network or to the Internet. However, the reliance on satellites results in a low throughput and a high  latency. Furthermore, the \gls{MCA services} are only allowed for use at altitudes above 3000~m in the European airspace, in order to avoid interference with the terrestrial cellular networks.

\item \emph{Power Line Networks}: Power line networks already exist throughout most aircraft. Thus, Jones \textit{et al}.~\cite{jones2006communications} proposed to transmit mobile multimedia signals and to offer Internet access over the aircraft power lines by plugging in devices, in order to reduce both the installation costs and the `tetherless' challenges. The transfer function between the input and the output port on a power line of a transport aircraft was modeled for the sake of optimizing the communication parameters. In~\cite{degardin2010possibility}, preliminary theoretical results were presented for a common-mode configuration, whereas in~\cite{bertuol2011numerical} the emphasis was on the electromagnetic compatibility aspects, such as reducing the effects of coupling between wires within the bundle.
    Degardin {\it{et al}}.~\cite{degardin2012power} modelled the channel propagation and investigated the achievable performance under the tree-shaped architecture of a power line network in an aircraft, and they extended the common-mode preliminary results of~\cite{degardin2010possibility}. They also estimated the expected throughput of different links~\cite{degardin2013theoretical}.  Furthermore, field-programmable gate array-based modems were designed, and the performance of the links was investigated by Degardin {\it{et al}}. in~\cite{degardin2014investigation}, who showed that the BER does not exceed $2 \times 10^{-4}$ at a bit rate below 91~Mbps.

\item \emph{Optical Wireless Networks}: Optical wireless networking offers the potential of exploiting a relatively untapped region of the electromagnetic spectrum for communication, by exploiting \glspl{LED} for information transmission~\cite{gfeller1979wireless}. Optical wireless networking has the advantage of being free from regulation, being untapped, having low-cost front ends and of delivering high data rates. This approach was proposed for intra-cabin communication in~\cite{schmitt2006diffuse}. The wireless path loss distribution within an aircraft cabin was obtained by performing Monte Carlo ray-tracing for the infrared range. Furthermore, the throughput and cell coverage of asymmetrically clipped optical OFDM based and direct-current offset optical OFDM based cellular optical wireless networks were compared in~\cite{dimitrov2011throughput}.
    Zhang {\it{et al}}.~\cite{zhang2013fiber} proposed a Fiber-Wireless (Fi-Wi) network for in-cabin communication by considering the tunnel-shape of the cabin, while Krichene {\it{et al}}.~\cite{krichene2015aeronautical} proposed to exploit the deployment of LEDs in the cabin for designing an in-cabin network architecture. These investigations demonstrated that having an optical wireless network in the cabin is feasible for providing high-data-rate in-flight entertainment services.
\end{itemize}

\begin{figure}[btp!]
\begin{center}
 \includegraphics[width=1.0\columnwidth,angle=0]{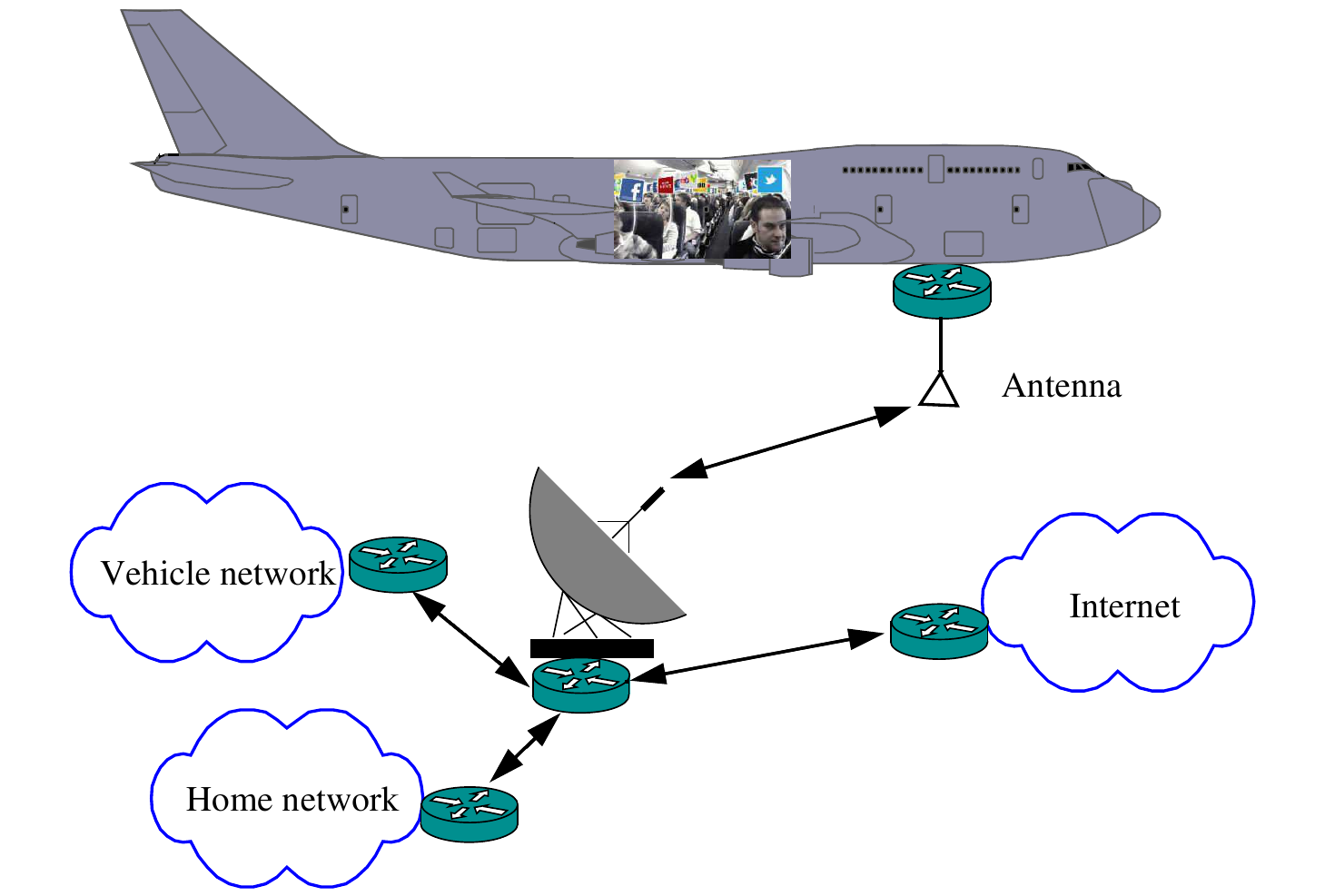}
\end{center}
\vspace{-1mm}
\caption{Multi-hop aeronautical communication via WiMAX.}
\label{FIG12:multihop}
\end{figure}

\subsection{Multi-Hop Communications}
At the time of writing, there is increasing demand for in-flight Wi-Fi connectivity to the Internet, but the existing solutions do not provide value for money. The Institute of Communications and Navigation of the German Aerospace Center (DLR) conceived the concept of networking in the sky for civilian aeronautical communications~\cite{Schnell2006Newsky}, which enables the aircraft to communicate with \glspl{GS} via multi-hop transmission. As studied by Mahmoud etc.~\cite{mahmoud2014aeronautical}, even when only a minority of aircraft are directly connected with \glspl{GS}, the system is capable of connecting most of the remaining aircraft within 3 hops. Explicitly, given the communication range of 150~km for the continental airspace and 300~km for the oceanic airspace, even if $29.9\%$ of aircraft and $41.7\%$ of aircraft were connected via a single hop, links of up to 3 hops are able to connect the remaining $70.1\%$  and $58.3\%$ of aircraft flying within the continental airspace and oceanic airspace, respectively.

As an essential solution for providing additional coverage and/or enhancing the capacity of the aircraft network, multi-hop links are indispensable  for providing Internet access to on-board passengers. Another example of multi-hop networking is illustrated in Fig.~\ref{FIG12:multihop}, where WiMAX operated in the multi-hop scenario has been invoked for providing broadband access to networks on the ground  for on-board passengers~\cite{kamali2012application}. Explicitly, the commercial aircraft is equipped with a WiMAX router, which is capable of communicating with a land-based WiMAX network station. The on-board passengers access the Internet with the aid of the WiMAX router deployed in the aircraft.

\section{Aircraft Networking Challenges}\label{S6}

\begin{figure*}
\begin{center}
 \includegraphics[width=0.8\textwidth,angle=0]{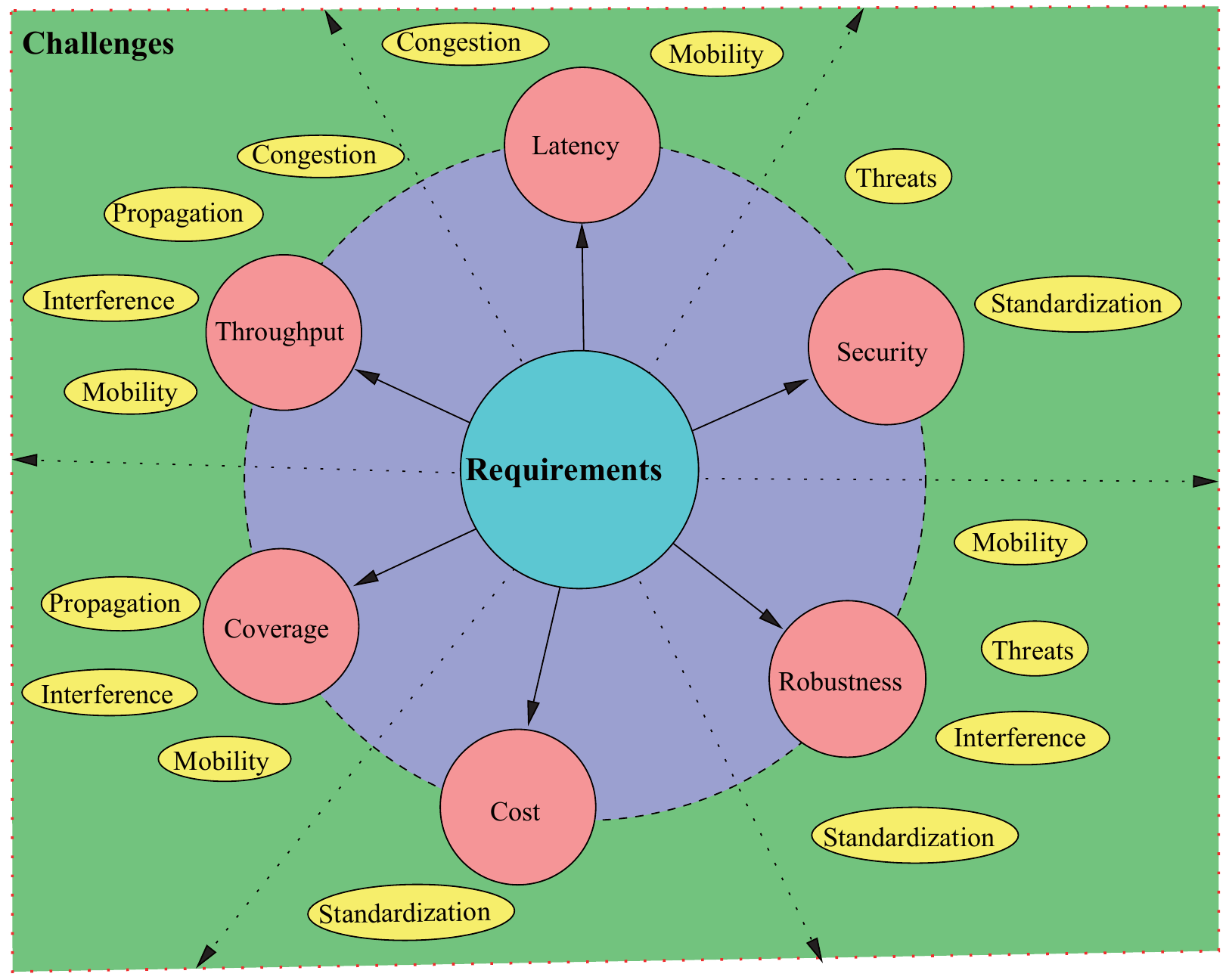}
\end{center}
\vspace{-0mm}
\caption{The challenges of AANET imposed by the requirements.}
\label{FIG11:challenges-imposed-by-requirements}
\end{figure*}

Due to the passengers' desire of establishing an ``Internet above the clouds" and driven by handling the continuously increasing air traffic~\cite{jahn2003evolution,strohmeier2014realities}, both airlines and aeronautical organizations are motivated to develop and establish broadband aeronautical communications among in-flight aircraft. Extensive efforts have been dedicated to building a set of protocols meeting the demanding requirements and applications of future aircraft communications. Different applications of aircraft networks have different requirements, as shown in Table~\ref{TAB5:relationships-requirements-challenges}. Note that the extensional techniques for in-cabin communication are not included in Table~\ref{TAB5:relationships-requirements-challenges}. However, all the existing aeronautical communication systems can only meet part of the requirements of each application. For example, passenger entertainment requires the network to cover unpopulated areas at a low cost. However, none of the existing aeronautical communication systems are able to achieve these requirements at the same time, as also seen in the lower part of Table~\ref{TAB5:relationships-requirements-challenges}. By contrast, \glspl{AANET} are capable of fulfilling all the requirements and support various applications, including those of the emerging aircraft applications, such as free flight and formation flying. However, numerous challenging open issues have to be addressed in \glspl{AANET}, as shown in Fig.~\ref{FIG11:challenges-imposed-by-requirements}.

Particularly, meeting the requirement of global coverage is highly dependent on the frequency band used, the transmit power and network topology management, which will impose challenges in terms of mobility modelling, propagation characterization and interference management in AANETs. The throughput of AANETs is directly affected by the mobility management, interference mitigation, propagation characteristics and network congestion
control. AANETs should accurately model the aircraft mobility for managing routing, for providing connectivity and for avoiding congestion. As illustrated in Fig.~\ref{FIG11:challenges-imposed-by-requirements}, network discovery/healing is also challenging in AANETs, owing to congestion and mobility, resulting in some nodes disappearing from a local network and joining another {\it{ad hoc}} network. The security requirements are very strict for AANETs, since commercial aircraft constitute safety critical systems. Although the highly dynamic nature of AANETs imposes challenges in term of robustness, AANETs must be highly reliable and robust, in order to protect the safety of hundreds of passengers. Thus, AANETs will face great challenges in terms of security threats and robustness. AANETs are a relatively low-cost solution of exchanging information via an {\it{ad hoc}} network established amongst aircraft, which is mainly dependent on the communication protocol design and on software development, without dependence on large-scale infrastructure deployments. However, global standardization is required to ensure global interoperability, but this must also be harmonized with the country-by-country standards and their own interests. In the following discussion, we elaborate on the challenges imposed by the requirements of aeronautical applications, which must be overcome before AANETs can be considered for practical deployment.

\subsection{Mobility}\label{S6-sub1}
There is a strong need for providing connectivity for passengers in aircraft, so that they can continuously communicate with other devices attached to the Internet, at any time and anywhere. However, the connectivity of the network may be frequently interrupted due to the high velocity of aircraft~\cite{bauer2011survey} and occasionally interrupted by weather. This is also an issue for \glspl{FANET}, owing to their time-variant scenarios governed by their high mobility. Hence Cognitive Radio (CR)
technologies based \glspl{UAV} suffer from the  fluctuation of link quality and link outages~\cite{saleem2015integration}. Moreover, owing to the three-dimensional terrain changes, \glspl{UAV} have to cope with highly-dynamic wireless channel fluctuations as well as topology changes~\cite{hayat2016survey}. Hence, the network protocols of FANETs and AANETs have to be more flexible than those of VANETs~\cite{hayat2016survey}. Therefore, the investigation of the mobility characteristics of aircraft is a critical issue for designing and evaluating \glspl{AANET}, so that they can provide robust solutions for connectivity at high velocity. 
  Furthermore, other challenges include the trade-off between the mobility model's accuracy as well as its simplicity for analysis~\cite{kingsbury2009mobile}, and those associated with a high degree of mobility. The inevitable delay problems due to routing over large geographical distances~\cite{bauer2011survey} and the connectivity problems due to the frequent setup and breakup of communication links among aircraft require extremely robust solutions to support high mobility.

\begin{figure*}[htbp!]
\begin{center}
 \includegraphics[width=0.80\textwidth,angle=0]{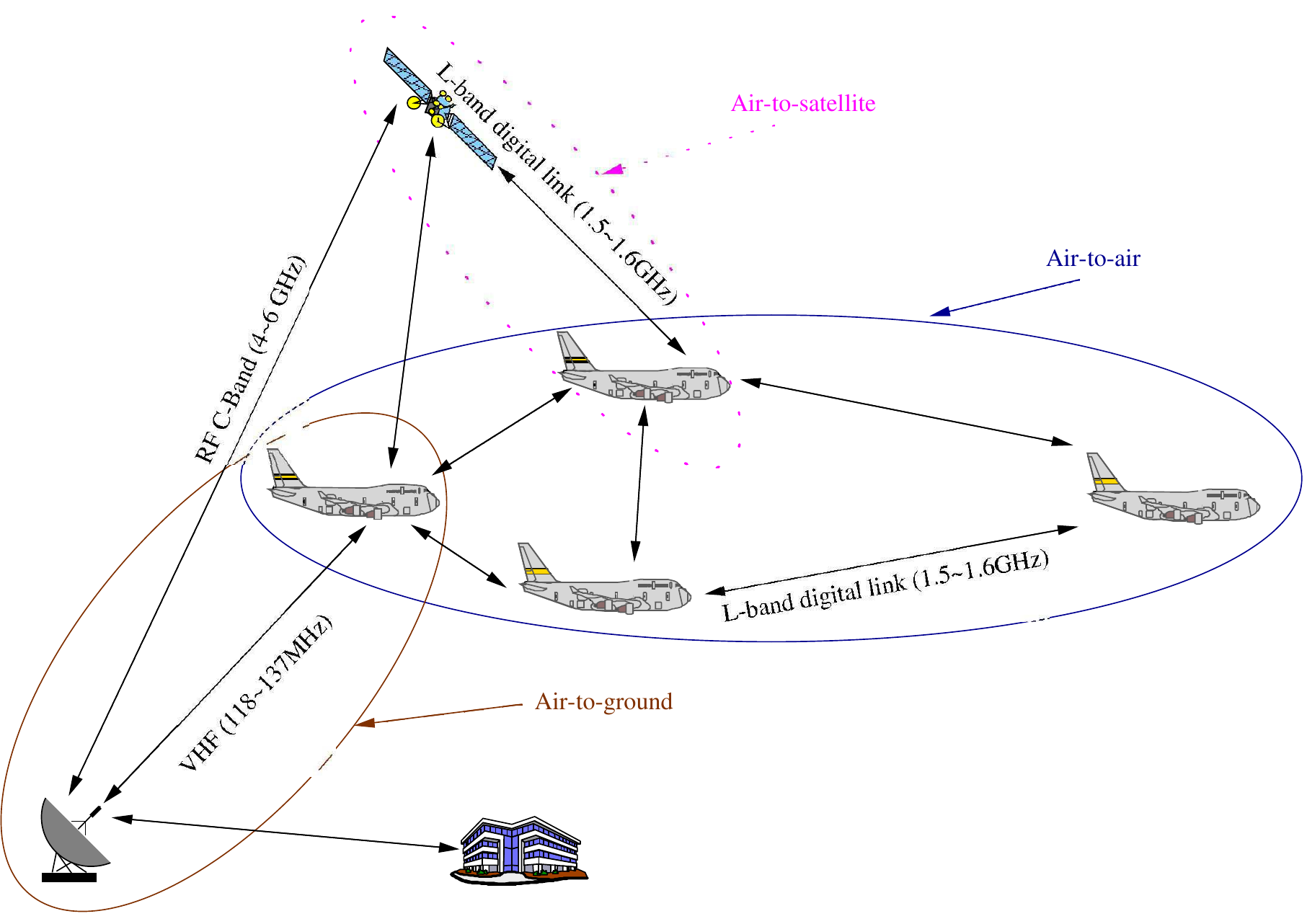}
\end{center}
\vspace{-0mm}
\caption{The spectrum used for various aircraft communication systems.}
\label{FIG12:Electromagnetic_Spectrum}
\end{figure*} 

\subsection{Congestion}
Since \glspl{AANET} are intended for providing Internet access, hence requiring practically all multi-hop traffic to flow through the \glspl{GS}, gateway congestion may be caused at or among the aircraft near these \glspl{GS}. {\color{black}Liu {\it{et al.}}~\cite{liu2018joint,cao2018optimal,cao2018optimalaa2} investigated the impact of gateway placement on the integrated 5G-satellite networks reliability and latency. Similar investigations also have to be applied to aircraft networking for optimizing the gateway selection and/or placement.} Wang {\it{et al}}.~\cite{wang2015throughput} demonstrated that the two-hop model of aeronautical communication networks was capable of achieving the best throughput without long delays. Moreover, by efficiently allocating flows, the traffic may be balanced amongst the gateways to avoid congestion. Furthermore, this problem is strongly coupled with the routing of packets in the network, since the path between an aircraft and a gateway determines the service that the gateway can provide to the aircraft. Additionally, if the wireless channel is shared by all nodes in a wireless network, the transmission of one node may interfere with others, which results in congestion of the network~\cite{hoffmann2013joint}. Therefore, the processes of Internet gateway allocation, routing and scheduling whilst minimizing the average packet delay in the network have to be jointly optimized, which is a challenging non-convex optimization problem.

\subsection{Threats}
It is extremely critical to secure \glspl{AANET} from every conceivable threat, since any threat may result in aviation accidents and incidents involving the lives of hundreds of passengers. Therefore, the precautionary measures should be thoughtful and proactive.
Generally, the security threats to aircraft networks may be categorized into internal and external ones. Internal security threats originate from the in-cabin passenger network, where a malicious user may attempt to gain access to the control network and cause service impairments and/or attempt to take control of the flight. On the other hand, the external security threat is caused by the security vulnerabilities of the communication links~\cite{thanthry2006security}.

{\color{black}\subsection{Propagation}}

\begin{figure}
\begin{center}
 \includegraphics[width=1.0\columnwidth,angle=0]{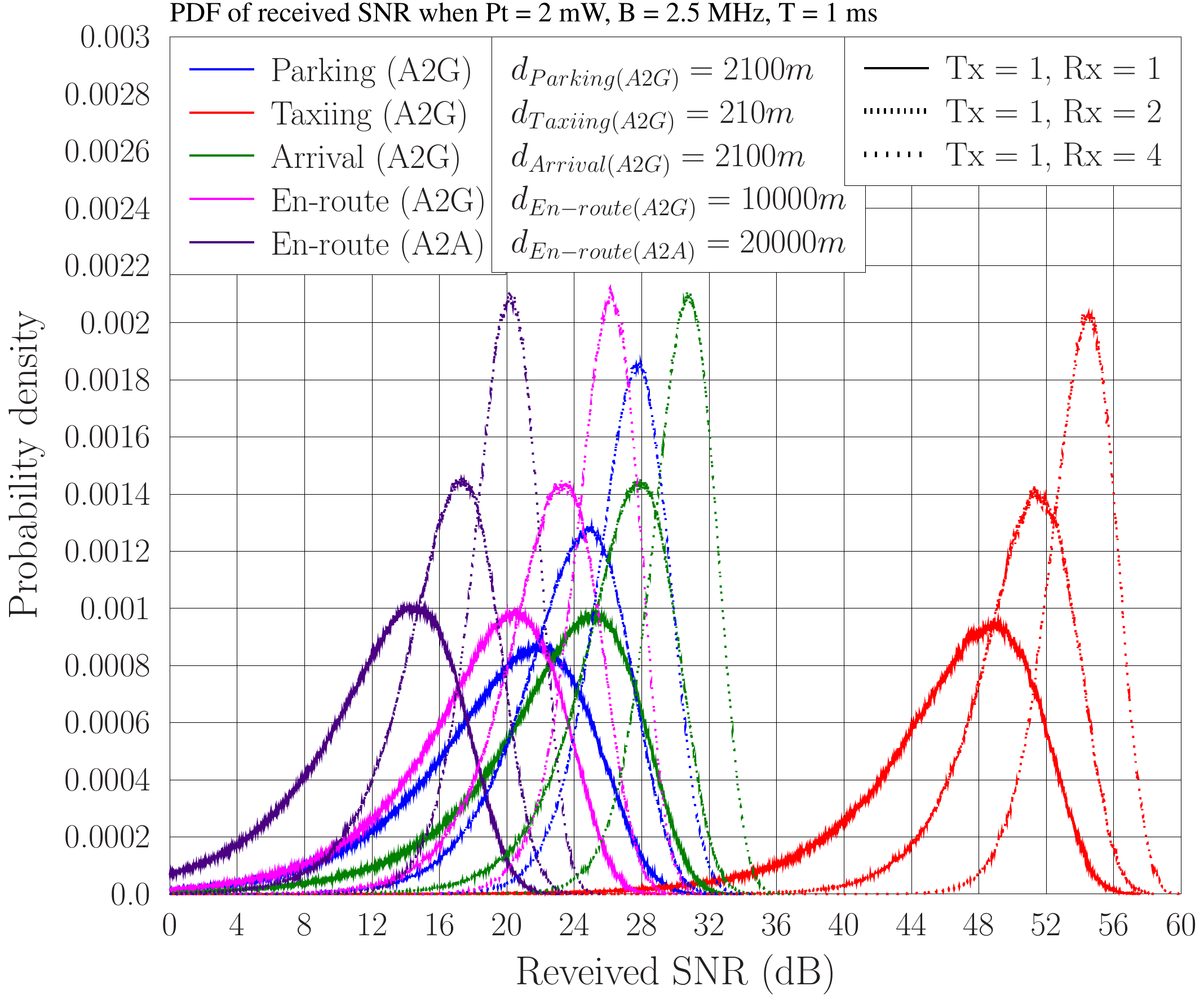}
\end{center}
\vspace{-0mm}
\caption{The probability density of received SNR for different flight phases. Both the large-scale fading and small-scale fading have been considered. The channel parameters are set according to Table.~\ref{TAB1:channel}.}
\label{FIG13:Received_SNR}
\end{figure}

\begin{figure}[tbp!]
\begin{center}
 \includegraphics[width=1.0\columnwidth,angle=0]{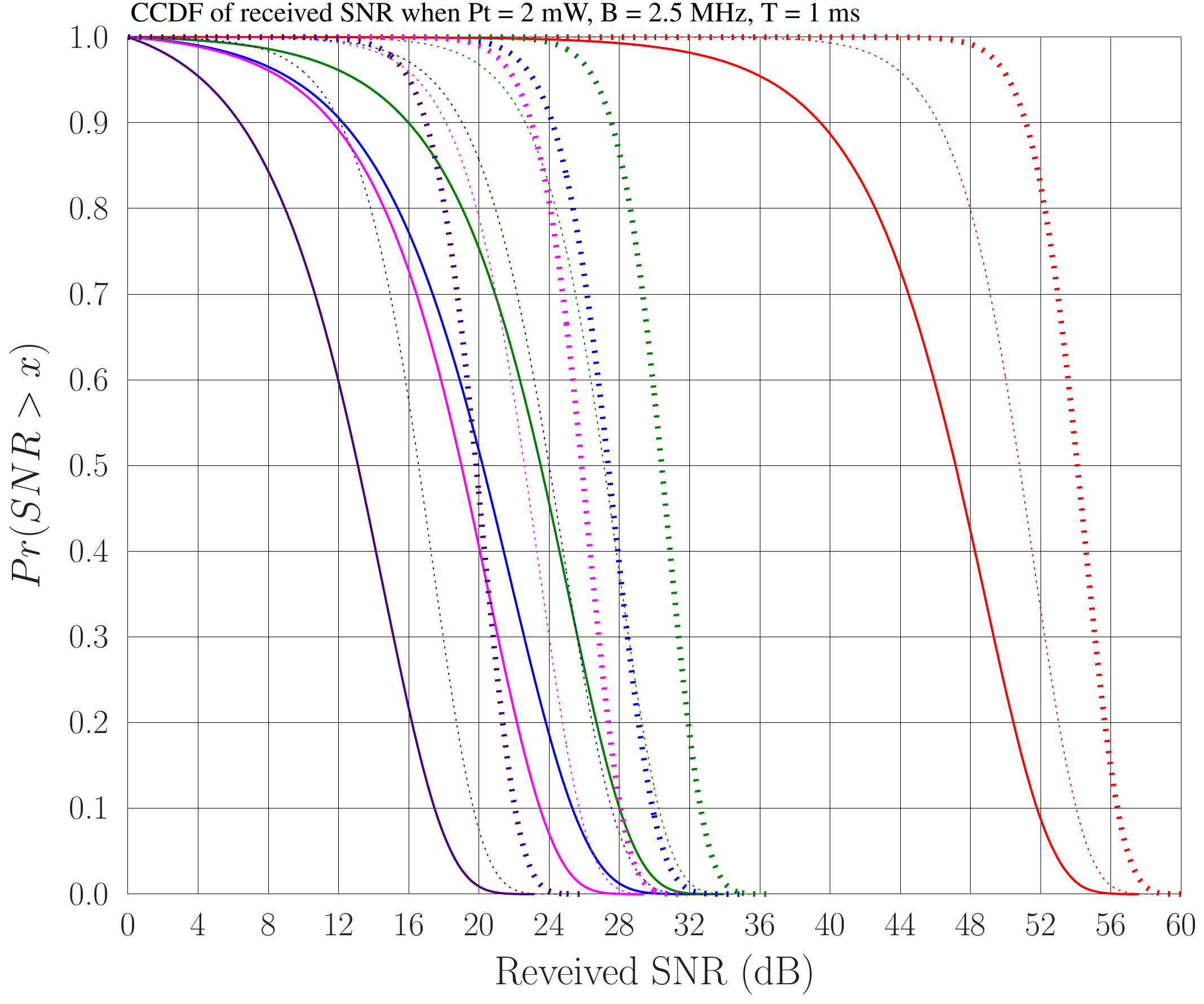}
\end{center}
\vspace{-0mm}
\caption{The CCDF of the received SNR for different flight phases. Both the large-scale fading and the small-scale fading have been considered. The channel parameters are set according to Table.~\ref{TAB1:channel}.}
\label{FIG14:Received_SNR_CCDF}
\end{figure}

In the future, available radio spectrum will become more scarce. However, the signal transmissions in \glspl{AANET} take place over \gls{A2A}, \gls{A2G} and \gls{A2S} across airports, populated and unpopulated areas, each having different bandwidth requirements. The different environments encountered during the flight of an aircraft may lead to different propagation scenarios. As shown in Fig.~\ref{FIG12:Electromagnetic_Spectrum}, an \gls{AANET} may rely on multiple bands, when it exchanges information with a gateway deployed on the ground or a satellite for relaying its information. In order to characterize the channel illustrated in Table~\ref{TAB1:channel}, we present the received \gls{SNR} and the corresponding \gls{CCDF} of different mobility scenarios by considering multiple antennas deployed on aircraft. As illustrated in Fig.~\ref{FIG13:Received_SNR} and Fig.~\ref{FIG14:Received_SNR_CCDF}, the aeronautical channel characteristics are significantly different for different scenarios, which impair the transmission signal quality. It is challenging to design transmission schemes, which can adapt themselves to various propagation modes. Furthermore, it is also a great challenge to accurately model these propagation characteristics, consisting of the signal attenuation and phase variation statistics, the fading rate, the Doppler spread and the delays during wave propagation~\cite{haas2002aeronautical,albano2000bit}. There are prohibitive cost constraints in the way of extensively measuring the aeronautical channel characteristics, especially for multiple access channels that suffer from interference amongst aircraft.

\subsection{Interference}
Another significant challenge is imposed by preventing mutual interference between aeronautical communications and terrestrial wireless communications, when an aircraft is flying over populated areas, especially near an airport~\cite{jain2011analysis}. The potential mutual interference must be mitigated, so that the \glspl{AANET} can adapt to diverse international standards and the specific implementations across the global airports. The detrimental effects of both multipath interference and of co-channel interference tend to increase the noise floor, hence reducing the capacity of the communication system, especially in the high traffic density environments near airports. Additionally, multipath interference may occur due to propagation through frequency selective fading channels, while co-channel interference may arise when signals are mapped to the same carrier frequency.
Therefore, it is desirable to characterize the behavior of the various sources of interference, in order to maintain high-integrity transmission, especially in safety-critical aeronautical applications~\cite{nguyen2007characterization}.

\subsection{Standardization}
Various existing wireless standards defined for aircraft communications~\cite{jain2011analysis} have already existed for decades. However,  there are more disparate requirements for \gls{A2A}, \gls{A2G}, \gls{A2S} communications, as mentioned in Section~\ref{S4}. It is expected that AANETs will require a new global standard for seamless aeronautical  communications~\cite{plass2012seamless}. Therefore, multinational cooperation will be necessary and the system has to be sufficiently flexible. In the light of the associated risks, it is challenging to gain regulatory approval from the various countries for such a safety-critical system and to persuade airlines to install the system on their aircraft. It is also a vital prerequisite that sufficient aircraft join the \glspl{AANET}, so that multi-hop relaying of other aircraft transmissions becomes feasible.
\section{Existing Research in Addressing the AANET Challenges}\label{S7}
In this section, we will review the proposals and techniques devoted to addressing the challenges of aircraft networking. The proposals and techniques discussed may simultaneously address several challenges.

\subsection{Mobility Solutions}
Since \glspl{FANET} support high-mobility nodes and often are mission-oriented, their mobility model has to be flexible to have paths planned in advance or adapted online during the mission, in order to maximize the coverage or avoid collision~\cite{alejo2014collision}.  Similarly, due to the high mobility of the aircraft, their mobility characteristics have a significant effect on the AANETs' mobility model. However, aircraft usually fly at a high velocity, and along a pre-designed route, which specifies the corresponding mobility model~\cite{Jamalipour2009Aeronautical}, as shown in Fig.~\ref{FIG6:scenarios}. Explicitly, the large-scale mobility pattern is randomly distributed for the case of aircraft above continents, since the airports are randomly distributed over the continent. By contrast, aircraft flying over oceans and other unpopulated areas can be identified as pseudo-linear, rapidly moving mobile entities~\cite{sakhaee2007stable,sakhaee2008stable}, since they are all headed for particular populated regions, as seen in the scenario of the North Atlantic from Fig.~\ref{FIG4:Mobility_pattern}. Additionally, a clustering mobility model was established in~\cite{sakhaee2008stable,sakhaee2007stable} by considering aircraft originating from the same source and heading in the same general direction.
Furthermore, Ghosh {\it{et al}}.~\cite{ghosh2015multi} considered the 3D aircraft topology in 3D airspace for mitigating the effects of network disruption.
Taking into the account specific flight phases and the speed of the aircraft, Li   {\it{et al}}.~\cite{li2012improved} proposed a smooth semi-Markov mobility model, which divided the motion of aircraft into flight phases, as shown in Fig.~\ref{FIG6:scenarios}. Petersen {\it{et al}}.~\cite{petersen2016modeling} further developed the Markov  mobility model by taking into account the traffic demand in a certain area, where they modelled both the arrival and departure of an aircraft in any of the states as a Poisson distribution in the developed Markov mobility model.
In order to allow passengers to access the Internet, the architecture of \glspl{AANET} has to support high-velocity mobility. The Internet is based on the \gls{IP} to deliver information, thus the potential solutions conceived for providing Internet connectivity include the so-called network mobility basic support protocol of TCP/IPv6~\cite{petander2006measuring,lucke2013sandra} and the host identity protocol~\cite{moskowitz2008host}. The preferred solution may be host identity protocol,  considering its flexible interoperability, since it could smoothly connect to the Internet and support mobility, as well as provide security and privacy~\cite{nikander2010host}.

\subsection{Scheduling and Routing}

Given the multihop nature of \glspl{AANET}, the packets have to follow multiple wireless paths to arrive at their final destination. As discussed in Section~\ref{S6}, the scheduling and routing strategy has to be investigated for avoiding congestion and for achieving the maximum throughput per aircraft. {\color{black}Luo  {\it{et al}}.~\cite{luo2018frudp} proposed a  reliable user datagram protocol relying on fountain codes. Furthermore, they have also developed a reliable  
multipath routing protocol \cite{luo2018aeromrp} relying on multiple aeronautical networks capable of operating under challenging networking conditions.} 
Furthermore, a number of research projects have investigated scheduling and routing algorithms conceived for \glspl{AANET}. Sakhaee {\it{et al}}.~\cite{sakhaee2007stable,sakhaee2006global} proposed a routing protocol by taking into account the Doppler frequency in the routing procedure. Furthermore, Sakhaee~\cite{sakhaee2007stable} integrated a \gls{QoS} constraint into the cost metric of the routing protocol in his PhD dissertation.
Luo {\it{et al}}.~\cite{luo2017multiple} further developed a QoS-based routing protocol by taking into account the  path
availability period, the residual path capacity and the path latency in their route selection.
Iordanakis {\it{et al}}.~\cite{iordanakis2006ad} proposed an {\it{ad hoc}} routing protocol for aeronautical mobile {\it{ad hoc}} networks by combining the proactive function of {\it{ad hoc}} on-demand distance vector and the reactive function of topology broadcast based on reverse-path forwarding~\cite{ogier2004topology}. Medina {\it{et al}}.~\cite{medina2008feasibility,medina2008topology} proposed to exploit the position information for assisting routing, and they assessed the proposed position-based greedy forwarding algorithm, which demonstrated that all packets were delivered to their destination with a minimum hop count.
A sophisticated routing scheme was proposed by Gankhuyag {\it{et al}}.~\cite{gankhuyag2016novel}, which exploited both location-related and trajectory-related information for establishing their utility function, taking into account both the minimum expected connection duration and the hop count.
Meanwhile, the authors of~\cite{wang2013gr} and~\cite{saifullah2012new} also designed their geographical routing protocols based on the knowledge of location information, which is assumed to be provided by \gls{ADS-B}. Considering the balance between the capacity and traffic load of each \gls{A2G} link, Medina {\it{et al}}.~\cite{medina2010routing,medina2012geographic} also developed a geographic routing strategy by forwarding packets to a set of next-hop candidates and spreading traffic among the set based on queue dynamics. Furthermore, Hoffmann {\it{et al}}.~\cite{hoffmann2013joint} developed a joint routing and scheduling scheme for \glspl{AANET}, which sequentially minimized the weighted hop-count subject to scheduling constraints and minimized the average delay for the previously computed routes. Similar to the solutions of ~\cite{medina2008feasibility,medina2008topology}, the authors of~\cite{tiwari2008mobility} proposed a mobility-aware routing protocol by exploiting the known trajectories of the aircraft to enhance the attainable routing performance. Furthermore, Vey {\it{et al}}.~\cite{vey2016routing}  proposed a node density and trajectory based routing scheme, which exploited the knowledge of the geographic path between the source aircraft and the destination aircraft and considered the actual aircraft density as well as Zhong {\it{et al}}.~\cite{zhong2016aeronautical} also exploited the density of aircraft as well as the geographic path between the source node and the destination node for routing. Peters {\it{et al}}.~\cite{peters2011geographical} developed a geographic routing protocol termed as aeronautical routing protocol for multihop routing in \gls{AANET}, which delivers packets to their destinations in a multi-Mach speed environment using velocity-based heuristics. By contrast, the authors of~\cite{Zhong2014Anew} proposed a topology-based routing mechanism for \glspl{AANET}, which can effectively decrease the probability of routing path breakup, regardless of how high the aircraft density is.

\subsection{Security Mechanisms}

In order to guarantee the security of the aviation network, whilst providing Internet connectivity for the passengers, the separation of the passenger, crew and control networks has been widely recommended~\cite{thanthry2005aviation,thanthry2006security,bilzhause2017datalink}.  However, there is still a high risk of `cyber-physical' security breaches~\cite{sampigethaya2013aviation}, since the safety of air flight depends on data communication, which may suffer from attacks both by remote and onboard devices~\cite{sampigethaya2011future}.

Research efforts have been launched for addressing the security issues in \glspl{AANET}. Sampigethaya {\it{et al}}.~\cite{sampigethaya2008secure} presented some security standards developed for in-aircraft networking~\cite{olive2006commercial}, for electronic distribution of software~\cite{robinson2007electronic}, for onboard health management~\cite{sampigethaya2009secure} and for \gls{ATC}~\cite{strohmeier2014realities}. Mahmoud {\it{et al}}.~\cite{mahmoud2009aeronautical} reviewed security mechanisms designed for aeronautical data link communications. Moreover, an IP-based architecture has been recommended for future aeronautical communications in~\cite{bauer2011survey,plass2012seamless}, and security architectures conceived for the \gls{IP} have received a significant amount of attention by the \gls{SESAR}. IP security~\cite{seo2005security} is one of the most popular solutions, since it operates at the network layer, which makes it suitable for different applications employing different transport layer protocols~\cite{thanthry2006security}. However, the investigation of ~\cite{thanthry2006security} indicated that Secure Sockets Layer (SSL)/Transport Layer Security (TLS)-based security mechanisms are capable of providing a level of security almost equivalent to IP security without reducing the \gls{QoS}. Furthermore, \glspl{AANET} may have to exchange information among aircraft belonging to different airlines for maximizing the aircraft connectivity, which imposes airline confidentiality issues. In order to resolve these security issues, the authors of~\cite{mahmoud2013ads} proposed a secure geographical routing protocol by exploiting the benefits of the greedy perimeter stateless routing of~\cite{karp2000gpsr} and of the \gls{ADS-B} protocol. Nijsure {\it{et al}}.~\cite{nijsure2016adaptive} developed Angle-Of-Arrival (AOA), Time-Difference-Of-Arrival (TDOA) and Frequency-Difference-Of-Arrival (FDOA) techniques in order to provide additional safeguards against \gls{ADS-B} security threats and for aircraft discrimination. Furthermore, Nijsure {\it{et al}}. also implemented a software-defined-radio-based hardware prototype for facilitating AOA/TDOA/FDOA. Since the \gls{ADS-B} messages can be received by any individual \gls{ADS-B} receiver, this results in substantial security concerns for \gls{ADS-B}-based communication systems. Baek {\it{et al}}.~\cite{baek2016protect} proposed an identity-based encryption scheme by modifying the original identity-based encryption of
Boneh and Franklin~\cite{boneh2003identity}. He {\it{et al}}. developed the triple-level hierarchical identity-based signature scheme of~\cite{yang2015new} for practical deployment. A range of further security protocols designed for ADS-B systems may be found in the excellent survey of~\cite{strohmeier2015security}.

\subsection{Aeronautical Channel Characterization}

Aeronautical communications involve \gls{A2G}, \gls{A2A} and \gls{A2S} communications at airports, populated areas and unpopulated areas. The accurate knowledge of the channel characteristics is important for carefully designing and assessing the performance of aeronautical communication protocols, which motivates the research devoted to characterizing the aeronautical channel.

Explicitly, Bello~\cite{bello1973aeronautical} investigated the aeronautical channel between aircraft and satellites, focusing on the effects of indirect paths reflected from and scattered by the surface of the Earth. His work was then further developed by Walter {\it et. al}~\cite{walter2011doppler} by characterizing the \gls{A2A} propagation characteristics, whilst the scattered components of an aeronautical channel were investigated in terms of
its delay and Doppler frequency in~\cite{walter2011doppler}. In contrast to investigating the A2A channel, Haas~\cite{haas2002aeronautical} devoted his efforts to \gls{A2G} aeronautical channel models. As a further development, the \gls{A2G} aeronautical channel at 5.2 GHz radio frequency was measured by Gligorevic~\cite{gligorevic2013airport} at Munich airport. The \gls{A2G} aeronautical channel over the ocean's surface was measured by Lei {\it et. al} \cite{lei2009multipath} at 8.0~GHz and by Meng {\it et. al} \cite{meng2011measurements} at 5.7~GHz, respectively. Facilitated by the development of multiple-antenna technologies in wireless communications, the corresponding radio propagation characteristics were analyzed in~\cite{jensen2007aeronautical} for Alamouti's \gls{STBC} scheme in~\cite{sumiya2011radio}.

Moreover, research efforts have been invested in analyzing the channel capacity of aeronautical channels and the performance of diverse modulation schemes over aeronautical channels~\cite{gong2009performance} as well as in mitigating the effects of
Doppler shifts~\cite{erturk2014doppler}.

\subsection{Interference Mitigation}

A mature aircraft communication system has to be able to mitigate interference, since this limits the achievable capacity, hence also imposing safety risks. In the case of \glspl{FANET}, awareness of the primary and the associated spectrum reuse relying on spectrum sensing is extremely crucial for interference avoidance between the primary users and secondary users~\cite{yucek2009survey}.

\begin{figure*}[tbp!]
\begin{center}
 \includegraphics[width=0.9\textwidth,angle=0]{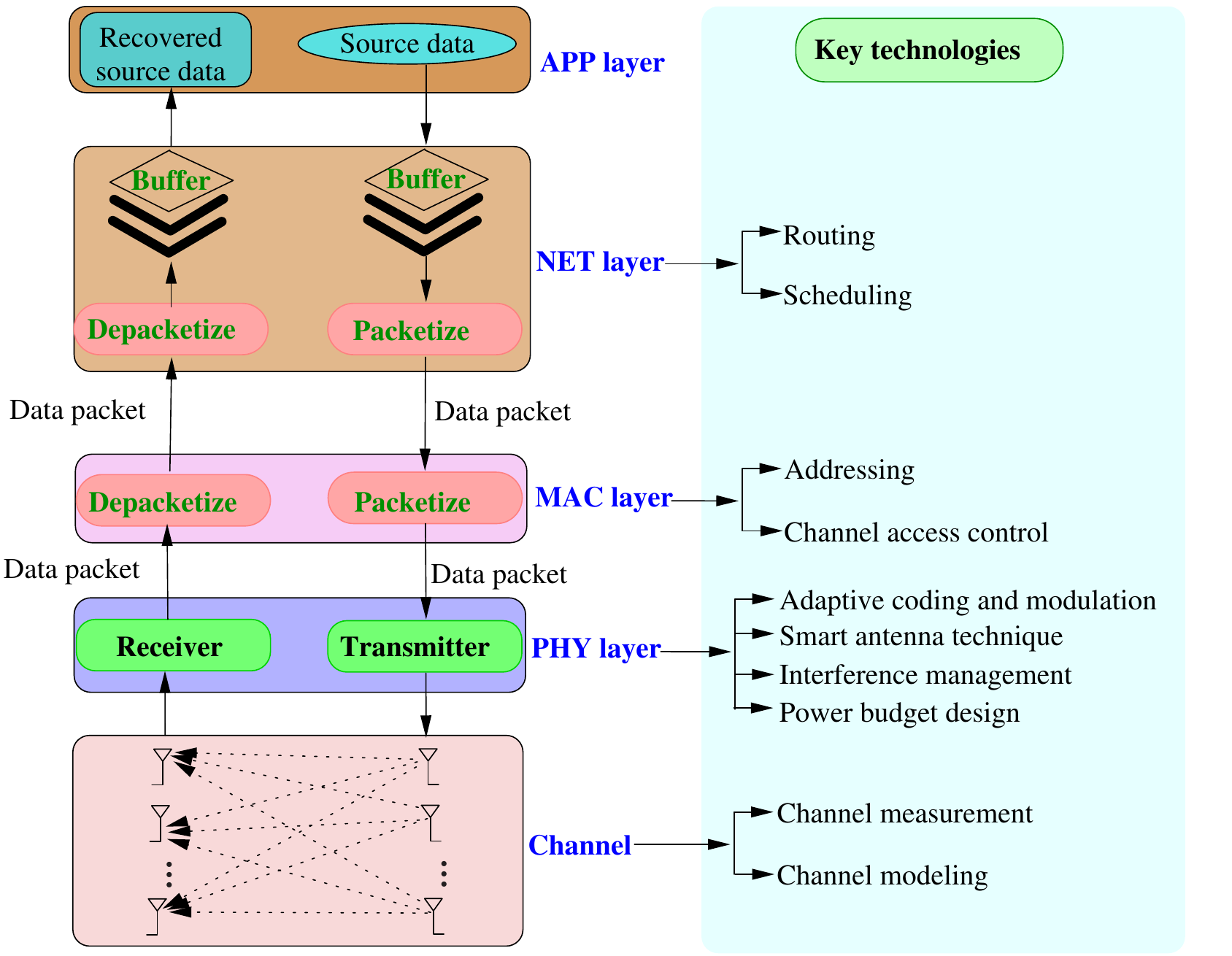}
\end{center}
\vspace{-0mm}
\caption{The design guideline for AANETs.}
\label{FIG122:design_guideline}
\end{figure*}

Extensive efforts have been devoted to mitigating the multipath interference~\cite{popescu2007interference}, the co-channel interference~\cite{nguyen2007characterization,zhang2018adaptive,zhang2018regularized}, the multiple access interference~\cite{medina2010routing,tu2010performance,besse2011interference} and the mutual interference between different wireless communication systems~\cite{tu2009proposal,kamali2011selection,sampigethaya2011future}. More specifically, Popescu {\it{et al}}.~\cite{popescu2007interference} proposed a linear equalizer based on Kalman filtering theory for mitigating the multipath interference and adjacent-channel interference. The co-channel interference characteristics encountered in a high-density traffic environment were analyzed in~\cite{nguyen2007characterization} by using computer simulations, and a multi-user detection scheme was recommended for mitigating co-channel interference. By contrast, Tu {\it{et al}}.~\cite{tu2010performance} characterized the multiple access interference in a sparse air traffic environment, judiciously managing the interference power with the aid of feedback information. Medina {\it{et al}}.~\cite{medina2010routing} recommended sophisticated scheduling for channel access in a \gls{TDMA} regime for mitigating the multiple access interference. {\color{black}As a further development, Fang {\it{et al}}.~\cite{fang2018qos} proposed a hybrid MAC protocol based on pre-allocation of the transmission time slots carefully combined with  random access for mitigating the collision probability.} Besse {\it{et al}}.~\cite{besse2011interference} developed an optimized network engineering tool model to analyze the impact of multiple access interference on the packet delivery probability, concluding that the solution could be either to reduce the transmission rate in order to reduce the interference effects or to use a dedicated channel whilst having  severe co-channel interference. In order to reduce the interference imposed on \gls{HF} communications by other communication systems and that imposed on the communications between \gls{ATC} controllers and pilots, Tu and Shimamoto~\cite{tu2009proposal} proposed a \gls{TDMA}-based multiple access scheme for transmitting an aircraft's own packets and for relaying the neighboring aircraft's packets. Kamali and Kerczewski~\cite{kamali2011selection} investigated {\color{black}the potential interference imposed by the \gls{AeroMACS} into MSS feeder link} in an airport environment, recommending the adaptive allocation of frequencies to different cells using a variable frequency reuse factor. By contrast, the interference imposed both by passenger devices and by intentional jamming were discussed in~\cite{sampigethaya2011future}, with an emphasis on the aviation safety.

\subsection{Efforts for Standardization}

As discussed in \textit{Section~\ref{S6}}, the standardization of aircraft communication is more complex than a pure technical challenge, since it has to balance many other factors, as well such as numerous practical issues, spectrum regulation and national security. Owing to this, it cannot be achieved by a single community or country. At the time of writing, the standards for aeronautical communications are mainly issued by the \gls{ICAO} and \gls{FAA} in the \gls{US}, and by the EUROCONTROL in Europe, given their leading roles in aviation. More specifically, the \gls{FAA} has funded \gls{NextGen}~\cite{planning2007concept} for conceiving the future national airspace system in the \gls{US}. Meanwhile, the European Commission and EUROCONTROL have jointly funded \gls{SESAR}~\cite{Europe2010SESAR} for improving the future air traffic management in Europe. However, the flourishing development of aviation in recent decades has inspired more nations to devote research to aviation. Motivated by this, Neji {\it{et al}}.~\cite{neji2013survey} appealed for multinational cooperation for the sake of establishing international standards. Along these lines, \gls{FAA} and EUROCONTROL initiated a joint study in the framework of Action Plan 17~\cite{EUROCONTROL2007actionplan} to investigate applicable techniques and to provide recommendations for future aircraft communications~\cite{graupl2011dacs1}, which paves the way for an international standard to be accepted by both the \gls{US} and Europe.
{\color{black}
\section{Design guidelines for AANETs}\label{S8aa1}
In this section, we outline the key techniques of AANETs along with our design guidelines for the four-layer protocol stack, namely for the PHY layer, MAC layer, NET layer and APP layer, as shown in Fig.~\ref{FIG122:design_guideline}.

\subsection{Channel and PHY Layer}
Explicitly, the propagation channel of aeronautical communication is intricately linked to the aircraft mobility, which captures the physical movement patterns of aircraft. Since field measurements would be extremely expensive for passenger planes engaged in different maneuvers in different network scenarios, stochastic and/or semi-stochastic mobility modeling may be more realistic solutions. However, random mobility modeling \cite{xie2018comprehensive} typically used in FANETs may fail to accurately characterize AANETs, since the passenger airplanes' routes are typically pre-planned, hence pre-planned semi-stochastic mobility models may be developed for AANETs.

\begin{figure*}[tbp!]
\vspace*{-4mm}
\begin{center}
 \includegraphics[width=1.0\textwidth]{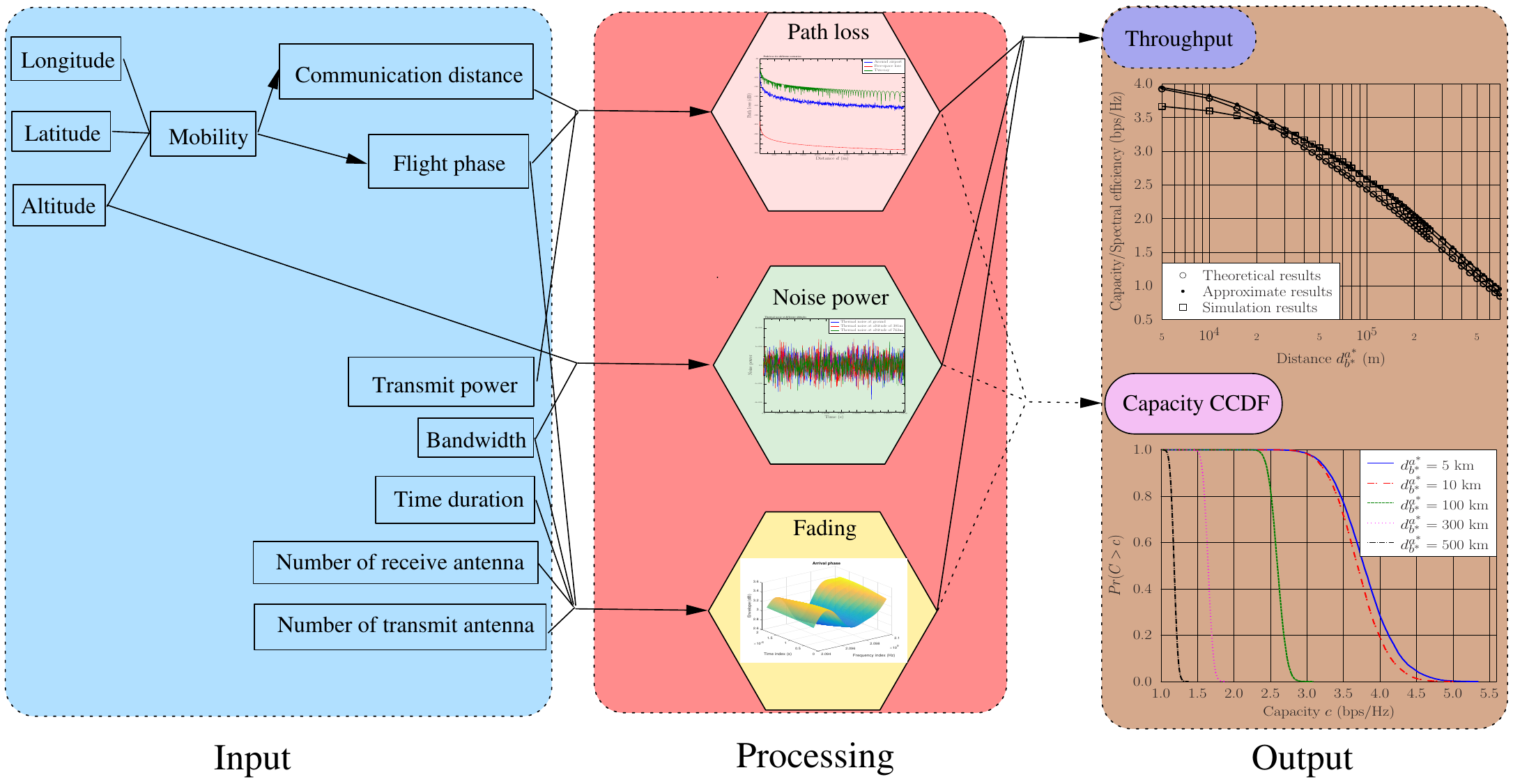}
\end{center}
\vspace*{-5mm}
\caption{Aeronautical channel propagation.}
\label{FIG1CCDF-capacity}
\vspace*{-2mm}
\end{figure*}

\begin{table*}[tp!]
\caption{A design example of RZF-TPC aided and distance-based ACM with $N_t=64$ and $N_r=4$.}
\vspace*{-1mm}
\begin{center}
\begin{tabular}{|C{1.1cm}|C{1.3cm}|C{1.3cm}|C{2.3cm}|C{2.3cm}|C{2.3cm}|C{2.3cm}|C{1.5cm}|}
\hline
 Mode $k$ & Modulation & Code rate & Spectral efficiency (bps/Hz) & Switching threshold $d_k$ (km)
   & Data rate per receive antenna (Mbps) & Total data rate (Mbps) & Routing cost $\mathcal{Q}$ (s{$\cdot$}Hz/bit) \\ \hline
 1 & QPSK   & 0.706 & 1.323 & 500  & 7.974  & 31.895 & 0.76 \\ \hline
 2 & 8-QAM  & 0.642 & 1.813 & 400  & 10.876 & 43.505 & 0.55 \\ \hline
 3 & 8-QAM  & 0.780 & 2.202 & 300  & 13.214 & 52.857 & 0.45 \\ \hline
 4 & 16-QAM & 0.708 & 2.665 & 190  & 15.993 & 63.970 & 0.38 \\ \hline
 5 & 16-QAM & 0.853 & 3.211 & 90   & 19.268 & 77.071 & 0.31 \\ \hline
 6 & 32-QAM & 0.831 & 3.911 & 35   & 23.464 & 93.854 & 0.26 \\ \hline
 7 & 64-QAM & 0.879 & 4.964 & 5.56 & 29.783 & 119.130& 0.20\\ \hline
\end{tabular}
\end{center}
\label{Tab2-RZF-ACM}
\vspace*{-1mm}
\end{table*}

The wireless channel characterized in the middle of Fig.~\ref{FIG1CCDF-capacity} imposes distance-dependent path loss effects  as well as from small-scale fading owing to reflections/scattering and Gaussian-distributed the background noise. Explicitly, apart from the communication distance, the path loss  also depends on the specific flight phase of takeoff/arrival, parking and en-route. Given the transmit power, the position information, bandwidth, the number of transmit antennas and the number of receive antennas and a number of other parameters seen at the left of  Fig.~\ref{FIG1CCDF-capacity}, we can characterize the corresponding aeronautical channel, which directly determines the achievable throughput, as illustrated in the right-hand section of Fig.~\ref{FIG1CCDF-capacity}. 

\begin{figure}[tbp!]
\vspace*{-1mm}
\begin{center}
 \includegraphics[width=0.85\columnwidth]{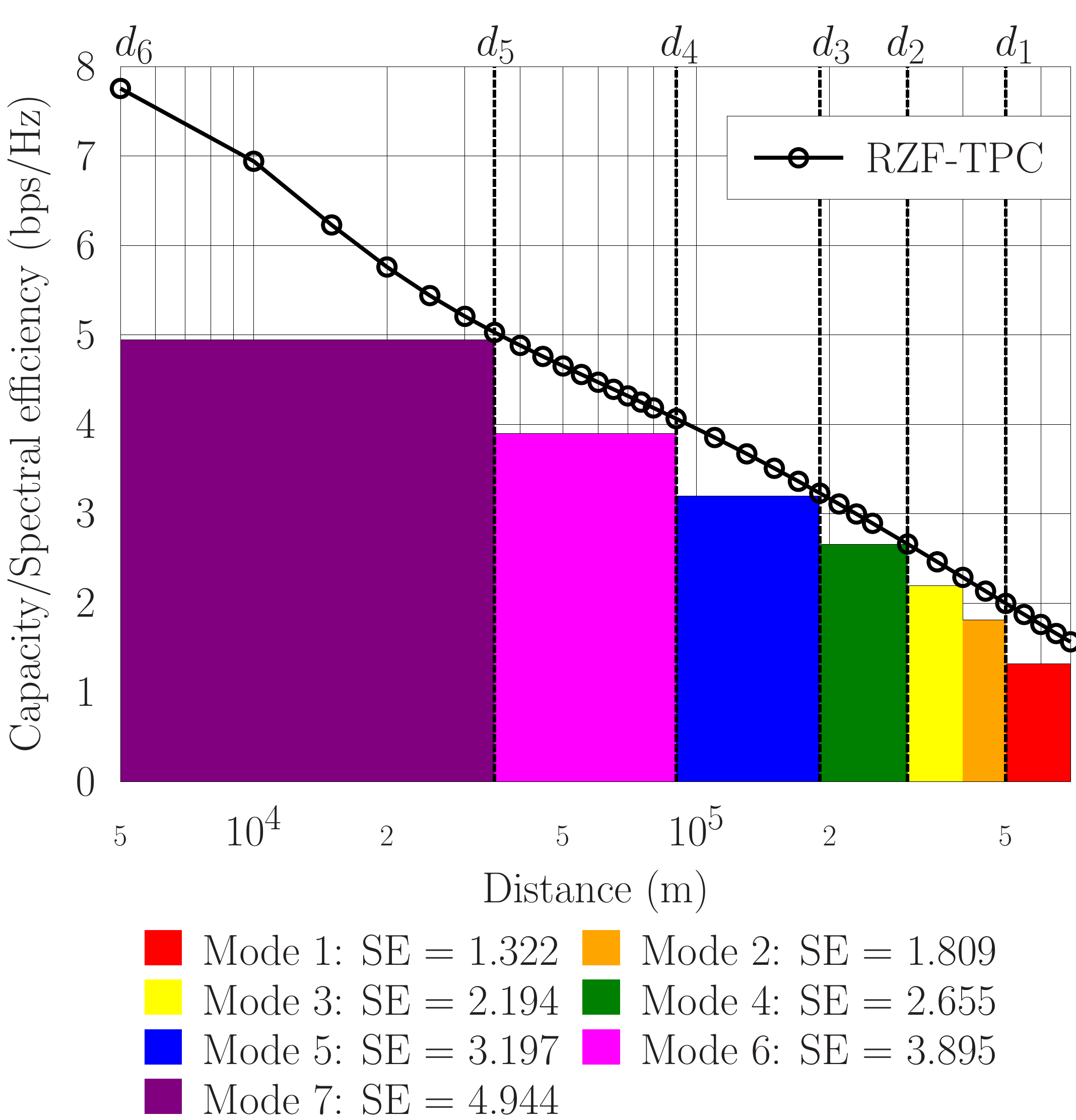} 
\end{center}
\vspace{-1mm}
\caption{The illustration of designing the RZF-TPC aided and distance-based ACM scheme.}
\label{FIG2RZFACM}
\vspace{-1mm}
\end{figure}

To elaborate a little further in technical terms, based on the channel characterization and on our regularized zero-forcing transmit precoding (RZF-TPC) scheme of \cite{zhang2018regularized}, we could design an distance-based adaptive coding and modulation (ACM) \cite{zhang2018adaptive} for A2A aeronautical communications system having seen different-rate ACM modes, as shown in Table~\ref{Tab2-RZF-ACM}. The corresponding system capacity versus distance is shown in Fig.~\ref{FIG2RZFACM}. Explicitly, based on the distance $d_{b^*}^{a^*}$ between aircraft $a^*$ and $b^*$
 measured by its distance measuring equipment, aircraft $a^*$ selects an ACM mode for data transmission according to
\begin{align}
 & \text{If } d_k\le d_{b^*}^{a^*} < d_{k-1}: \text{ choose mode } k , \, k \in \{1,2,\cdots , K\} . \nonumber
\end{align}

Assuming that the maximum communication range is $D_{\max}$, which can be determined by the parameters in the left-hand section of Fig.~\ref{FIG1CCDF-capacity}, no communication is provided for $d_{b^{*}}^{a^{*}}\ge D_{\max}$,
 since the two aircraft are beyond each others' communication range. Moreover, the
 minimum flight-safety based separation must be obeyed, hence the minimum communication distance $D_{\min} $ obeys the minimum separation according to the international civil aviation organization's regulations.

\begin{figure*}[htbp!]
\begin{center}
 \includegraphics[width=0.8\textwidth,angle=0]{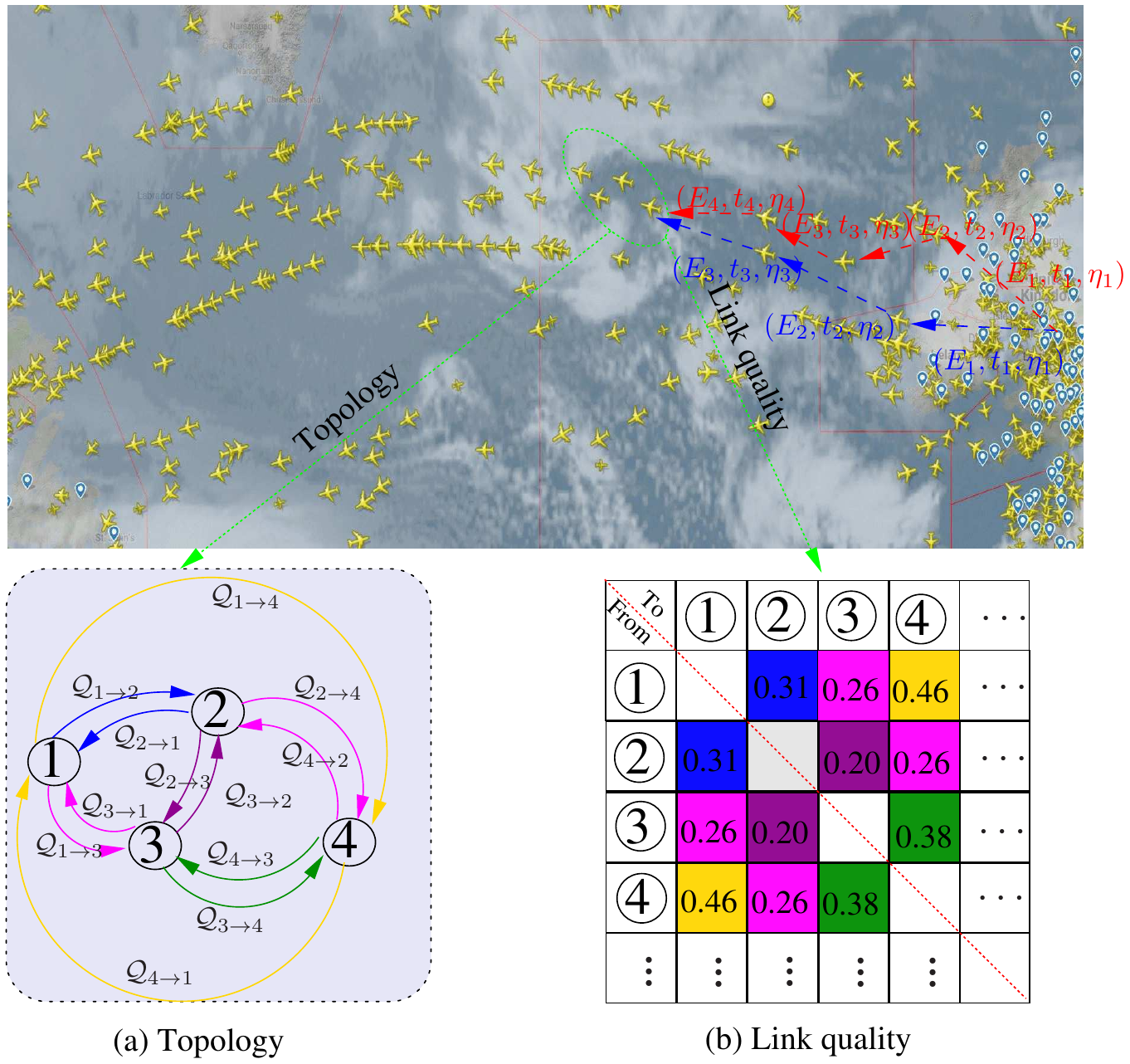}
\end{center}
\vspace{-0mm}
\caption{An example topology of AANET consists of four nodes and their corresponding routing cost table. "$\cdots$" means the routing cost table is expandable according to the number of nodes.}
\label{FIG:topology_example}
\end{figure*}

\subsection{MAC Layer}
The MAC layer takes care of the channel access control mechanisms that make it possible for several nodes to communicate using a shared medium, without suffering from packet collisions transmitted by different nodes.
In AANET scenarios, the combination of having limited wireless spectrum, low latency requirements and high mobility impose significant challenges on the MAC layer. Furthermore, the MAC protocol is expected to support diverse network topologies, that vary dynamically owing to the high velocities of aircraft nodes.
The MAC layer also has to avoid relying on an excessive number of Radio Frequency (RF) chains operating on different frequencies in a FDD manner, which would be unsuitable for aircraft installation. Alternatively, TDD may facilitate multiple nodes to access a shared medium \cite{teymoori2013dt}. Explicitly, only a single node having the token at any instant is allowed to transmit data and then it has to pass the token to another node for avoiding collisions of different nodes transmitting at the same time. This approach maintains reliable and fair access to the network, as well as achieving a high degree of efficiency, flexibility and robustness in the medium access control and topology management.

However, in traditional token-based MAC protocols, when a node joins the ring, it is required to negotiate a position in the ring in order to identify a predecessor and a successor node, which are also required to accordingly update the identity of their successor and predecessor node, respectively. Likewise, when a node leaves the ring, its predecessor node and successor node must update the identity of their successor and predecessor nodes accordingly.
A large amount of coordination and control information must be passed around the network, hence resulting in a high overhead and low efficiency, when the network topology varies rapidly with nodes joining and leaving frequently, which imposes challenges on  AANETs, especially because the aircraft nodes have high velocities, leading to a dynamically evolving topology.

Given the unique features of AANETs, we may advocate a mesh topology-aware token passing management with an associated link quality  table, as shown in Fig.~\ref{FIG:topology_example}, where the color represents the link quality as illustrated in Fig.~\ref{FIG2RZFACM}, whilst the value in the table is the routing cost. To elaborate a little further, the routing cost may be quantified in terms of many different metrics for evaluating a routing protocol, such as the number of hops, delay, reliability  and throughput, just to name a few. In general, this multi-component optimization problem becomes quite complex, especially for networks having many nodes. The best approach is to find the Pareto-front of all optimal solution. More explicitly, the Pareto-front is the collection of all the operating points, which either have the minimum BER, delay, power-consumption etc. None of the Pareto-optimal solutions may be improved, say in terms of the BER without degrading either the delay, or the power-efficiency, or the complexity etc. Nevertheless, here we consider the single-component spectral efficiency optimization for establishing the routing cost table for exemplifying the basic philosophy of our proposed mesh topology-aware token passing management. Consider the four-node mesh network of Fig.~\ref{FIG:topology_example} as an example, where the routing cost table is a $(4 \times 4)$-element table, where the element in the $i$-th row and the $j$-th column identifies the quality of the link spanning from the $i$-th node to the $j$-th node, where the color represents the link's spectral efficiency, as seen in Fig.~\ref{FIG2RZFACM}. The routing cost $\mathcal{Q}$ is defined as the reciprocal of the spectral efficiency. For example, the link leading from node 1 to the node 2 in the routing cost table of Fig.~\ref{FIG:topology_example} has a link quality represented by blue color, which results in a routing cost of $\mathcal{Q} = 1/3.197 = 0.31$. Note that the links between the nodes are bidirectional and may be asymmetric, resulting in different link quality marked by different colors between the elements having the indices $(i,j)$ and $(j,i)$ in the link quality  table. However, in our example we assume simplicity that the link quality of two nodes is symmetric based on the fact that the link quality in A2A aeronautical communication is dominated by the communication distance.  

The link quality table will then be used both by the MAC layer and by the NET layer. In the MAC layer, the link quality table is used in conjunction with the token roll count to select which particular node will pass the token to the next one. When a node's MAC layer has  a token, it will ask he NET layer to provide a set of data packets and to specify the spectral efficiency used.

\subsection{NET Layer}
In the network (NET) layer, scheduling and routing determine the multi-hop paths to be followed by the packets between their source and destination nodes. More specifically, each hop to be taken by the data packets is decided dynamically and opportunistically at each stage of the multi-hop path, rather than being decided by the source node.
In particular, each packet may be received by more than one node and then forwarded by whichever has the first opportunity to transmit.
This dynamic, opportunistic and redundant approach to routing improves the network's robustness to rapidly changing topologies, which is one of the main challenges for routing in AANETs.

The cost of a multi-hop path is given by the sum of the costs of its constituent links.
For example, the path in Fig.~\ref{FIG:topology_example} starting from Node 1 and passing through Node 3 on to Node 4 is denoted as $1\to3\to4$, which has a cost calculated as $\mathcal{Q}_{1\to3\to4}=\mathcal{Q}_{1\to3}+\mathcal{Q}_{3\to4}=0.26+0.38=0.64$. Alternatively, there are also other routing paths from node 1 to node 4, such as $\mathcal{Q}_{1\to2\to4}, \mathcal{Q}_{1\to2\to3\to4}, \mathcal{Q}_{1\to3\to2\to4}, \mathcal{Q}_{1\to4}$. Nevertheless, given the link quality table of Fig.~\ref{FIG:topology_example}, graph theory \cite{evans1992optimization}, relying for example on tree algorithms, shortest-path algorithms, minimum-cost flow algorithms, etc may be exploited for solving the problem of finding the lowest-cost multi-hop path spanning from the source node all the way to the destination node.

Still referring to Fig.~\ref{FIG:topology_example}, we further investigate the routing optimization problems. The mesh network consist of the four airplanes circled by the green dashed ellipse should have a complete connected path all the way to the control tower at London's Hearthrow airport for our example. To elaborate a little further, transmission between a pair of nodes is assumed to incur an energy cost of $E_{i}$, to impose a delay of $t_{i}$ and to have a spectral efficiency of $\eta_{i}$. The cost function associated with a specific routing path contains the aggregate energy consumption $\sum E_{i}$, the aggregate delay $t_{i}$ and the end-to-end spectral efficiency of $\min\{\eta_{i}| i = 1,2,3,\cdots\}$, which is a multi-objective optimization problem determined by diverse  factors. The multi-objective optimization problem can be solved using Pareto optimization techniques, which generate a diverse set of Pareto optimal solutions so that a compelling trade-off might be struck amongst different objectives. An example of a twin-parameter Pareto-optimization problem is shown in Fig.~\ref{FIG:pareto_front}, where all circles represent legitimate operating points and all blue circles represent Pareto optimal points, which are not dominated by any other solutions.

\begin{figure}[htbp!]
\begin{center}
 \includegraphics[width=0.65\columnwidth,angle=-90]{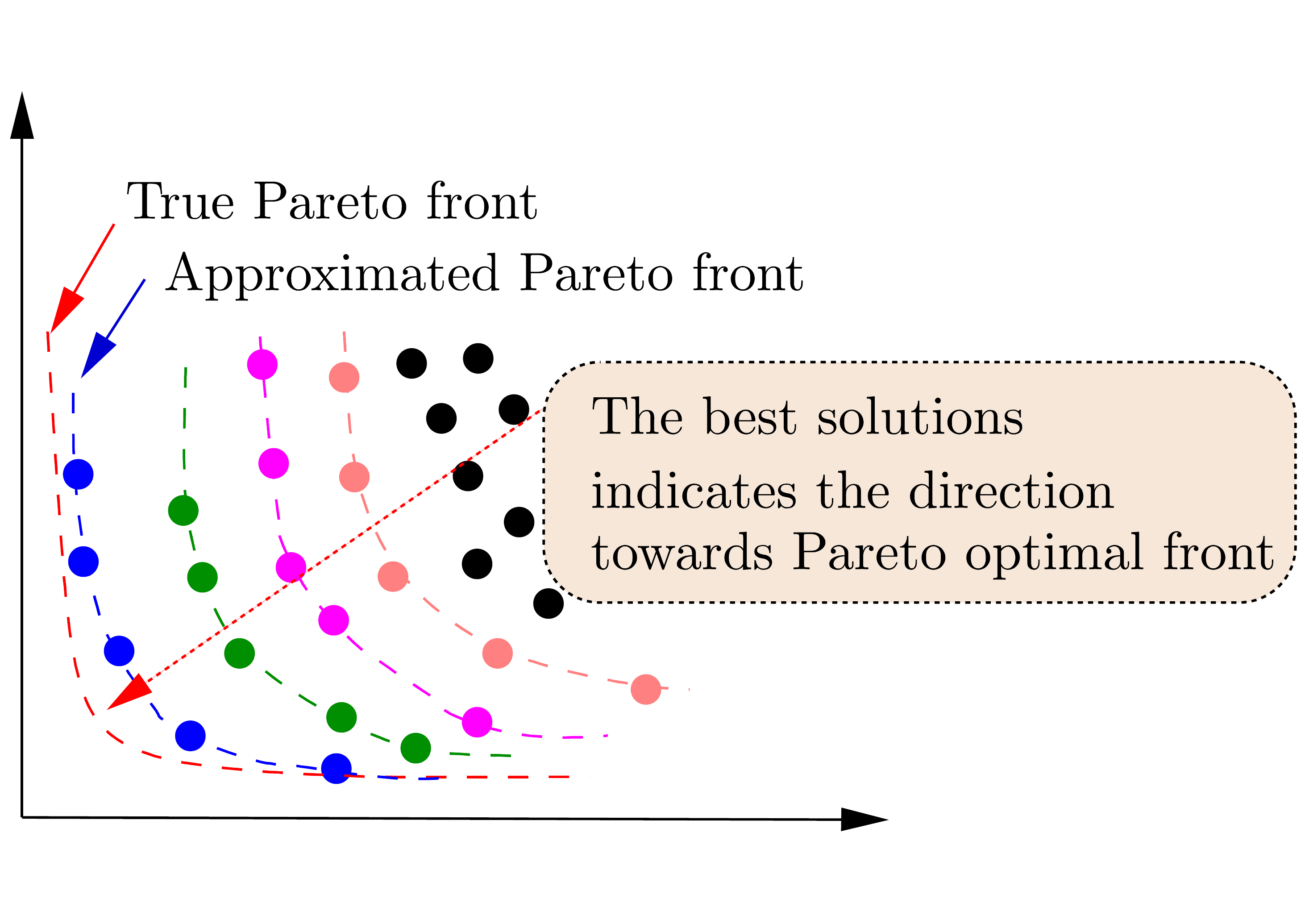}
\end{center}
\vspace{-0mm}
\caption{An example of optimal Pareto front for two objective optimization problems.}
\label{FIG:pareto_front}
\end{figure}

\subsection{Pareto-Optimal Perspective of AANET Optimization}
Again, the design of ANNETs includes that of the PHY layer, MAC layer, NET layer and APP layer, which faces substantial  challenges in terms of meeting diverse objectives. Traditional single-objective may be still used by iteratively optimizing each metric, however, it can only find a local optimum at a potentially excessive computational complexity, signal processing delay and energy consumption. Moreover, the diverse optimization metrics of AANETs are typically not independent of each other, they are mutually linked with each other in terms of influencing the overall system-level performance. It is also a challenge to provide an ultimate comparison among different locally optimal solutions based on different metrics, since the objectives involved ten to conflict with each other, hence requiring a trade-off.

In contrast to the single-objective optimization, multi-objective optimization is capable of finding the global Pareto-optimal solutions by striking a tradeoff amongst conflicting objectives. Explicitly, we summarize a range of popular metrics in Fig.~\ref{FIG:CF_mul_opt}  typically exploited in designing AANETs. Over the past decades, a number of research contributions have focused on addressing one or more objectives as well as jointly addressing a few objectives, as we have discussed in Section~VIII. However, with the rapid improvement of the computational capability of cloud computing \cite{baliga2011green} and quantum computing \cite{hirvensalo2013quantum}, it enables us to systematically conceive cross-layer design and optimization with the aid of multi-objective optimization algorithms, as illustrated in Fig.~\ref{FIG:CF_mul_opt}. More detailed comparison between different multi-objective optimization algorithms could refer to \cite{marler2004survey,fei2017survey} and in the references therein.

\begin{figure*}[htbp!]
\begin{center}
 \includegraphics[width=1.0\textwidth,angle=0]{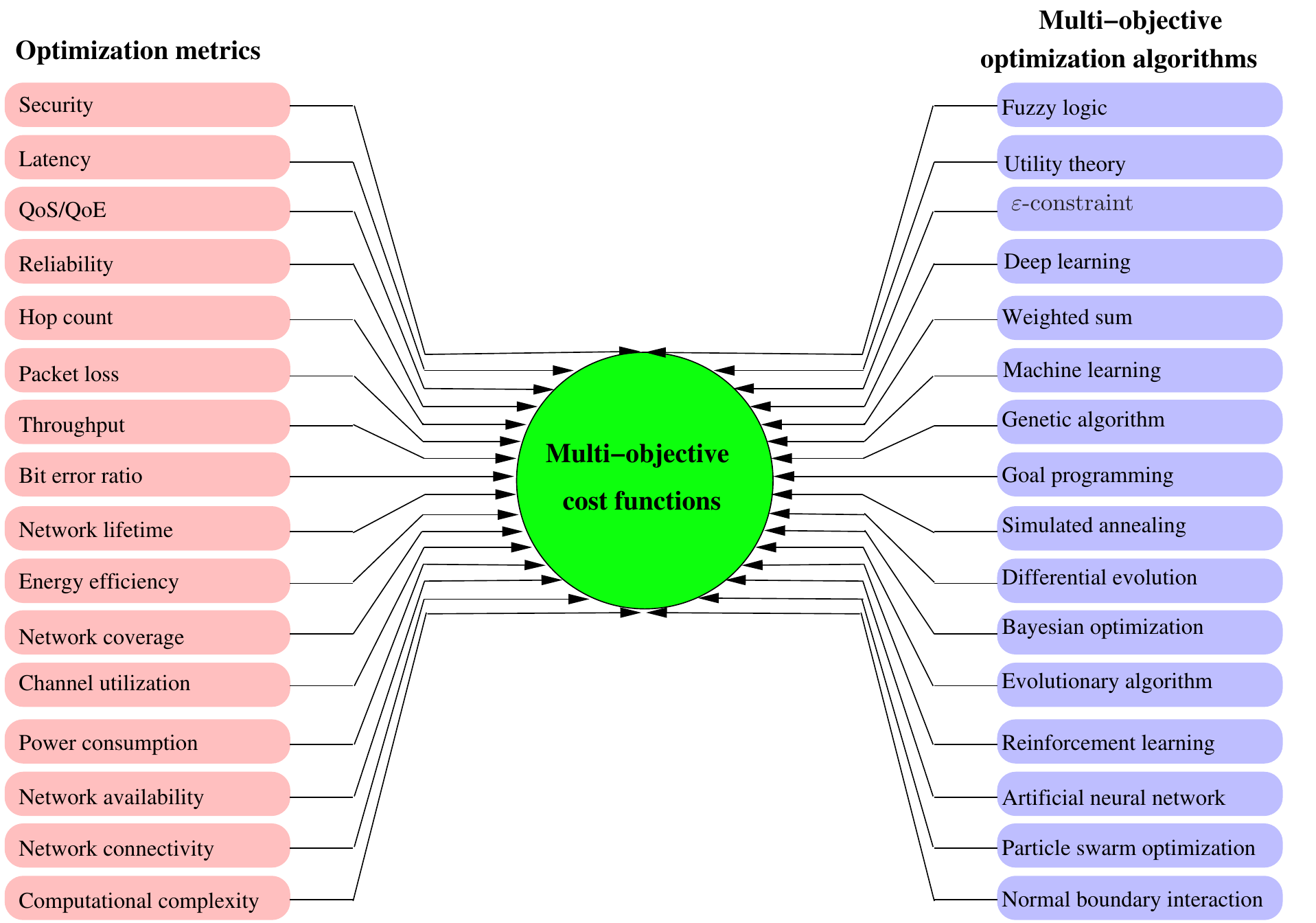}
\end{center}
\vspace{-0mm}
\caption{Potential metrics that may be used for optimizing AANETs and the multi-objective optimization algorithms that may be invoked.}
\label{FIG:CF_mul_opt}
\end{figure*}

}

\section{Prospective Solutions and Open Issues}\label{S8}
\glspl{AANET} aim for building communication links among aircraft. However, they cannot be operated in isolation without satellites and \glspl{GS}, which provide \gls{GPS} signals or backhaul. Thus, the practical \glspl{AANET} rely on multiple layers consisting of satellites, \glspl{GS} and aircraft, while handling information dissemination across the multiple layers in heterogeneous environments, with the objective of meeting the stringent requirements of aeronautical communication in time-sensitive as well as mission-critical applications. The challenges were discussed in Section~\ref{S6}, and the state-of-the-art research contributions devoted to addressing these challenges were discussed in Section~\ref{S7}. Nonetheless, there are many open issues and prospective solutions to be investigated.

\subsection{Prospective Solutions for AANETs}
\begin{itemize}[\setlength{\listparindent}{\parindent}]
\item \emph{Large-Scale Antenna Arrays}: Large-scale MIMO~\cite{larsson2014massive} systems employ hundreds of antennas for serving typically a few dozen terminals, while sharing the same time-frequency resources. This technique achieves a hitherto unprecedented spectrum efficiency, energy efficiency as well as low latency. Hence it is widely accepted as one of the key 5G techniques. Aircraft typically have a large airframe, which may be capable of accommodating dozens of antennas. However, the deployment of large-scale MIMOs is not straightforward due to the form-factor limitation discussed in Section~\ref{S6}. It remains a challenge to fit dozens of antennas on commercial aircraft. The \gls{VHF} band is widely used for existing aeronautical communication systems, but at these wavelengths, the required antenna spacing is high, which will limit the number of antennas that can be installed on the aircraft. Thus, the centimeter-wave carriers having a frequency ranging between 3~GHz and 30~GHz has attracted intense investigations in aeronautical communication~\cite{haas2002aeronautical}. However, the antenna design is crucial due to the limited opportunity for their deployment and fuselage blocking. Motivated by this, conformal antennas\footnotemark\footnotetext{Conformal antennas are flat radio antennas which are designed to conform or follow some prescribed shape upon which they will be mounted.}~\cite{cheng2013millimeter,gao2015conformal} may be considered for the antenna design of aeronautical communications. {\color{black}Furthermore, to accommodate different scenarios, having diverse flight velocities and required throughput, both adaptive coherent/non-coherent and adaptive single/multiple-antenna aided solutions \cite{xu2018adaptive} may also be conceived for aeronautical communications.}

\item \emph{Free Space Optical Communications}: \gls{FSO}~\cite{chan2006free} communications constitute a promising technique which adopts \glspl{LD} as transmitters to communicate, for example, between aircraft as well as between aircraft and a satellite at a high rate. Establishing their applicability to aircraft-ground communications requires further studies owing to its eye-safety concerns. The directional and license-free features of \gls{FSO} are appealing in aeronautical communications, because  conventional radio-frequency communications are fundamentally band-limited. \gls{FSO} communications have also been planned for the provision of connectivity for sub-urban/remote areas in Facebook's forthcoming project~\cite{Cohen2014facebook}, as well as for connectivity between the Moon and Earth in NASA's Lunar laser communication demonstration project~\cite{Boroson2011overview}. Advanced steered laser transceivers~\cite{kaushal2016optical}, which were originally designed for nano-satellites, may also be deployed on aircraft, \glspl{GS} and satellites for providing \gls{FSO} communications between them.

    However, \gls{FSO} communications are vulnerable to mobility, because \gls{LOS} alignment must be maintained for high-integrity communications.  In order to solve the associated pointing and tracking accuracy problems for high-speed aircraft, a feasible solution is to rely on the  built-in \gls{GPS} system of the aircraft, along with the \gls{FSO} system's low transmission latency, which can also assist in formation-flying. Furthermore, since there are no obstacles in the stratosphere, the main disadvantage of terrestrial \gls{FSO} links in terms of requiring a \gls{LOS} channel becomes less of a problem. Thus, the \gls{FSO} communication links among aircraft have a promising potential in terms of constructing an \gls{AANET} for aircraft tracking and collision avoidance. Alternatively, the \glspl{AANET} could also rely on \gls{FSO} for the backhaul with the aid of \glspl{GS} or satellites. Finally, since \gls{FSO} and \gls{RF} links exhibit complementary strengths and weaknesses, a hybrid \gls{FSO}/\gls{RF} link has a substantial promise in large-scale MIMOs.

\item \emph{Heterogeneous Networks}: As seen in Fig.~\ref{FIG1:AANET_topology}, an \gls{AANET} consists of three individual layers, which may be viewed as a Heterogeneous Network (HetNet) that is composed of satellites, aircraft and Internet subnetworks. {\color{black}It is quite a challenge to manage and optimize the whole plethora of metrics across multiple layers.} The architecture of HetNets shifts the design paradigm of the traditional centrally-controlled cellular network to a user-centric distributed network paradigm, which is suitable for emerging \glspl{AANET}. To elaborate a little further, \glspl{AANET} are capable of self-organization. They are autonomous and rely on diverse protocols as well as on potentially-hostile communication links.
    In a HetNet of aircraft, different applications have different QoS requirements, security requirements and user/operator preferences, which may require carefully designed data link selection~\cite{alam2017optimal}, which is cognizant of to the particular transmission characteristics, {\color{black}when constructing the sophisticated Multi-Layered
space-terrestrial integrated network of the future \cite{shi2018across}.} This can provide diverse communication services, such as safety-critical or non-safety-critical communication services, or a combination of both.

    AANETs are also expected to support automatic node discovery and route-repair as well as to exchange cross-layer information amongst aircraft, satellites and \glspl{ATC}, {\color{black}which may be optimized by cross-layer gateway selection, as pioneered by Shi  {\it{et al}}. \cite{shi2018cross} in the integrated satellite-aerial-terrestrial networks. This concept was further optimized by Kato  {\it{et al}}. \cite{kato2019optimizing} using efficient artificial intelligence techniques.} However, the high-mobility-induced dynamics of the aircraft topology impose challenges upon the design of the routing, scheduling, security protocols and on IP management, as well as on cross-layer optimization~\cite{rohrer2011highly} among the physical, MAC, network and transport layers. All these sophisticated, high-flexibility HetNet features should be evaluated in realistic scenarios, including typical airport scenarios as well as both populated and unpopulated areas, which requires substantial efforts from the entire research community~\cite{kwak2012airborne}.

\item \emph{Cooperative Relay Communications}: Relaying messages among aircraft is a pivotal operation in \glspl{AANET}, which is capable of increasing the coverage, throughput and capacity of the \glspl{AANET}. However, the optimization of the multi-hop routing is crucial for maximising the achievable relay performance~\cite{roy2017optimisations,alam2017optimal}. Cooperative relay-aided communication~\cite{bletsas2006simple} is easier to achieve among aircraft than amongst mobile phones, since a mutually beneficial agreement might be easier to strike between airlines. Thus, cooperation constitutes a promising method to offer extra spatial diversity without requiring physical antenna arrays. Moreover, storing packets and retransmitting them when there are favorable communication links is capable of improving the network's resilience, throughput and diversity~\cite{zlatanov2014buffer}, which has motivated the research of buffer-aided relaying~\cite{jamali2015bidirectional01,jamali2015bidirectional02,dong2015energy}. 
    Aircraft, especially those belonging to the same airline or airline alliance, could exploit the buffer-aided relaying technique for improving their connectivity and throughput. As a further development of the buffer-aided idea, `Cache in the air'~\cite{wang2014cache} can cache popular video/audio contents in  intermediate servers, such as local servers, gateways or routers. The concept of caching in the air can also benefit the aircraft network by significantly reducing the associated response latency and by sharing their navigation information as well as their weather conditions, which will enhance the flight safety by the prompt provision of precautionary information for collision avoidance and for storm/airflow warning. Moreover, efficient caching and sharing strategies are capable of supporting the creation of temporary social networks in and around airport lounges, aircraft, etc.

\item \emph{Cognitive Radio Communications}: The existing air traffic systems typically communicate in the \gls{VHF} band (108-137~MHz) and the \gls{HF} band (2.85-23.35~MHz). Apart from surveillance radar and aeronautical navigation systems, the \gls{UHF} band has almost entirely been allocated to television broadcasting and cellular telephony. It is crucial to guarantee interference-free access for these aeronautical communication systems due to safety-of-life. However, it can be foreseen that the wireless tele-traffic of aeronautical communications will rapidly be increasing due to the tremendous growth of the aviation industry, which imposes pressures due to the scarcity of spectrum. Furthermore, unmanned aerial systems also aggravate the spectrum scarcity in the aeronautical domain~\cite{jacob2016cognitive}. Yet, no new aeronautical spectrum assignments can be expected within the immediate future due to the limited availability of wireless spectrum.

    AANETs mainly exchange information via multi-hop \gls{A2A} communication, which may rely on the \gls{SHF} band spanning from 3~GHz to 30~GHz~\cite{zhang2018adaptive}. Historically, separate allocations have also been made for aeronautical surveillance systems and  aeronautical navigation systems. However, it remains quite a challenge to support the ever-increasing demand for wireless access in aeronautical communications without conceiving efficient techniques for spectrum reuse. Thus, there is growing tendency and impetus towards sharing radio spectrum between radio services, provided  that there is no excessive interference. Cognitive Radio (CR)~\cite{haykin2005cognitive,akyildiz2006next} is an emerging paradigm for efficiently exploiting the limited spectral resources.
    CR offers an efficient solution to reuse the existing spectrum without license, which has attracted wide  attention  in aeronautical communications~\cite{wang2010cognitive,fantacci2009performance}.

    However, robust spectrum sensing is required in aeronautical communications, since packet collisions in spectrum usage may lead to catastrophic consequences during landing/takeoff~\cite{jacob2016cognitive}.  Thus the probability of missed detection must tend to zero. Furthermore, the lifetime of the just detected available spectrum should also be  carefully investigated, since the speed of aircraft is high as they can fly at about 16 km per minute. Moreover, integrating and exploiting non-contiguous frequencies  is crucial for providing high throughput broadband Internet access for aircraft. Additionally, a robust handover strategy between frequencies should be designed in order to provide smooth and continuous service, especially for mission-critical communications.
\end{itemize}

\subsection{Open Challenges in the AANET Implementation}

\begin{itemize}[\setlength{\listparindent}{\parindent}]
\item \emph{High Data Rate}: Providing high-rate Internet access for hundreds of passengers in the cabin of a commercial aircraft remains a significant challenge, since it demands extremely high data volume per aircraft. Existing systems mainly use satellite-based solutions, in order to provide global connectivity, although this suffers from a low data rate and high cost. \gls{A2G} stations have been widely deployed both in the USA and in Europe, which have provided faster Internet access and lower cost, but their coverage is limited to the European/North America airspace and the total data volume still remains low. AANETs are capable of extending the coverage of the \gls{A2G} stations designed for aeronautical communications, but the aircraft have to  employ radically improved transceivers for facilitating high data rates. The above-mentioned large-scale MIMO aided adaptive modulation scheme is capable of providing up to 76.7~Mbps \gls{A2A} data rate, using a configuration of 4 receive antennas and 32 transmit antennas~\cite{zhang2018adaptive}. Thus large-scale MIMO schemes constitute a promising solution of providing high data rates for aeronautical communications. Moreover, FSO communication is also a competitive  solution for  providing high data rates, but the laser safety and steering accuracy issues must be addressed~\cite{chan2006free}.

\item \emph{Stable Connectivity}: Maintaining reliable connectivity is fundamental for  AANETs to achieve data delivery. The connectivity amongst aircraft is a function of velocity, position, direction of flight, range of communication and congestion~\cite{ghosh2016acpm}. Due to the highly dynamic nature and larger-scale geographic distribution of high-speed aircraft in contrast to terrestrial wireless communications, AANETs are facing a great challenge in terms of establishing stable multi-hop connectivity amongst aircraft~\cite{sakhaee2008stable}. This challenge is further aggravated by the often unpredictable mobility patterns, the high velocity and the potentially high number of aircraft within a communication range, as discussed in Section~\ref{S6-sub1} and Section~\ref{S3-3}, respectively.
    
    Thus, the routing protocols designed for aeronautical communications should cater for the specific requirements of AANETs and exploit the distinct characteristics of AANETs, as discussed for example by Sakhaee {\it et. al}~\cite{sakhaee2008stable} and Medina {\it et. al}~\cite{medina2011airborne,medina2011geographic}. New strategies, concepts and metrics are required for designing the network protocols, which remains an open research challenge. For example, the probability of an aircraft becoming isolated can be considered for analyzing the connectivity of AANETs.

\item \emph{Testbed Sharing}: Aeronautical communications are safety-related, especially in the context of \gls{ATC}, formation flight and free flight, which requires strict validation of any developed function and technique of AANETs. Thus, creating testbeds representing a proof-of-concept prototype is essential for maturing the technique of AANETs. Having an open testbed would be beneficial for both the academic research community and for the commercial development of AANETs. However, it is challenging to develop an integrated and robust testbed for AANETs, which relies on an aircraft mobility simulator, physical layer, data link layer and network layer emulations.  Moreover, it also faces the challenge of the high cost of developing the testbed. Both the NASA research center~\cite{sheehe2004aviation} and the German aerospace center~\cite{lucke2013sandra} have invested significant efforts in developing their testbeds. However, these testbeds have not been opened for public use, not even for academic research.

\item \emph{Global Harmonization}: Various proposals have been conceived for aeronautical communications by individual countries, which have obtained \gls{ICAO} approval independently of each other. However, none of them have achieved global endorsement. In order to achieve seamless aeronautical communications among aircraft originating from different countries/airlines, an evolutionary approach towards global interoperability has to be developed. For this reason, multi-national cooperation will be necessary for pre-screening, investigation and harmonization of the shortlist of competing technologies.

\item \emph{Compatibility}: An \gls{AANET} is capable of supporting direct \gls{A2A} communication among aircraft without the assistance of \glspl{GS}/satellites and \glspl{ATC}, which reduces the tele-traffic pressure imposed on them and significantly reduces the latency of critical-mission communication as well. Nonetheless, the aircraft should regularly communicate with \glspl{ATC}. Thus, \glspl{AANET} must be capable of operating in the presence of interference, whilst imposing only an acceptable level of interference on the legacy aviation systems to avoid jeopardizing flight safety. Hence, it is necessary to evaluate the
    radio-frequency compatibility of the \glspl{AANET} of the future with the systems already in operation both in \gls{A2G} and in \gls{A2S} communications.

\item \emph{Deployment}: There has to be a certain minimum number of aircraft in the air in order to make the network usable. Thus, a certain minimum number of aircraft has to participate in the \gls{AANET} before its benefits may be quantified. The gestation period of aircraft from a new technology launch to its entry into service is typically 10-15 years, which is significantly longer than that of the 2.5 years typical for cars. Moreover, the aviation industry is more meticulous in critically appraising any new technologies, since its safety issues are under the spotlight right across the globe and they are strictly regulated by governments. Moreover, field tests also prolong the deployment cycle, since it is a challenge to organize dozens of aircraft for  evaluation in a real-word scenario and it is also difficult to get permission to carry out evaluations on passenger flights due to safety of life. Hence, joint efforts are necessary from both the academic and industrial communities for developing and sharing testbeds and for ensuring the security of \glspl{AANET} in order to meet the critical market entry requirements of the aviation industry.
\end{itemize}


\section{Conclusions and Recommendations}\label{S9}
The emerging demands imposed by the ever-increasing air traffic and by the desire to enhance the passengers' in-flight entertainment have stimulated the research efforts of both the academic and of the industrial communities, invested in developing aeronautical communications. \glspl{AANET} may be expected to meet the demands of future aeronautical communications.  However, the specific characteristics, applications, requirements and challenges of \glspl{AANET} have not been comprehensively reviewed in the open literature.

In this paper, we have characterized the scenarios, applications, requirements and challenges of \glspl{AANET}. We have discussed both existing and emerging aircraft communications systems designed for \gls{A2G}, \gls{A2A} and \gls{A2S} communications as well as in-cabin communications. The research community's efforts devoted to developing \glspl{AANET} have been reviewed in this survey. {\color{black}A general design framework for AANETs as well as key technical issues are presented. Moreover, we outline a range of performance metrics as well as a number of representative multi-objective optimization algorithms for designing AANETs. } Finally, some open issues of implementing \glspl{AANET} in practical aeronautical systems have also been discussed.

It can be expected that in the near future the promises of \glspl{AANET} will motivate further research efforts, which will benefit not only aviation, but also the more general area of wireless {\it{ad hoc}} networking. \glspl{AANET} will merge the self-organization of multi-hop {\it{ad hoc}} networks as well as the reliability and robustness of infrastructure-based networks, generating hybrid networking solutions applicable to miscellaneous applications in aircraft communications.

\bibliographystyle{IEEEtran}
\bibliography{AANET_final_accepted_version_Lajos}

\begin{thebibliography}{100}
\providecommand{\url}[1]{#1}
\csname url@samestyle\endcsname
\providecommand{\newblock}{\relax}
\providecommand{\bibinfo}[2]{#2}
\providecommand{\BIBentrySTDinterwordspacing}{\spaceskip=0pt\relax}
\providecommand{\BIBentryALTinterwordstretchfactor}{4}
\providecommand{\BIBentryALTinterwordspacing}{\spaceskip=\fontdimen2\font plus
\BIBentryALTinterwordstretchfactor\fontdimen3\font minus
  \fontdimen4\font\relax}
\providecommand{\BIBforeignlanguage}[2]{{%
\expandafter\ifx\csname l@#1\endcsname\relax
\typeout{** WARNING: IEEEtran.bst: No hyphenation pattern has been}%
\typeout{** loaded for the language `#1'. Using the pattern for}%
\typeout{** the default language instead.}%
\else
\language=\csname l@#1\endcsname
\fi
#2}}
\providecommand{\BIBdecl}{\relax}
\BIBdecl

\bibitem{eurocontrol2013Challenges}
{EUROCONTROL}, ``{Challenges of Growth 2013---Task 4: European Air Traffic in
  2035},''
  \url{https://www.eurocontrol.int/sites/default/files/article/content/documents/official-documents/reports/201306-challenges-of-growth-2013-task-4.pdf},
  2013, [[Online]. Available].

\bibitem{Honeywell2016}
{Honeywell}, ``{Honeywell Survey: Airlines Risk Losing Passengers Due to Poor
  Wi-Fi},''
  \url{https://www.honeywell.com/newsroom/pressreleases/2016/07/honeywell-survey-airlines-risk-losing-passengers-due-to-poor-wifi},
  [[Online]. Available].

\bibitem{erzberger2001method}
H.~Erzberger and B.~D. Mcnally, ``{Method and system for an automated tool for
  en route traffic controllers},'' {US} Patent 6\,314\,362, November 06, 2001.

\bibitem{neji2013survey}
N.~Neji, R.~De~Lacerda, A.~Azoulay, T.~Letertre, and O.~Outtier, ``{Survey on
  the future aeronautical communication system and its development for
  continental communications},'' \emph{IEEE Transactions on Vehicular
  Technology}, vol.~62, no.~1, pp. 182--191, January 2013.

\bibitem{vey2014aeronautical}
Q.~Vey, A.~Pirovano, J.~Radzik, and F.~Garcia, ``{Aeronautical ad hoc network
  for civil aviation},'' in \emph{Proceedings of the 6th International Workshop
  of Nets4Cars/Nets4Trains/Nets4Aircraft}, Offenburg, Germany, May 2014, pp.
  81--93.

\bibitem{medina2011airborne}
D.~Medina and F.~Hoffmann, \emph{{The Airborne Internet}}.\hskip 1em plus 0.5em
  minus 0.4em\relax InTech, September 2011.

\bibitem{medina2011geographic}
D.~Medina, \emph{{Geographic load share routing in the airborne
  Internet}}.\hskip 1em plus 0.5em minus 0.4em\relax Herbert Utz Verlag, 2011.

\bibitem{sakhaee2007stable}
E.~Sakhaee, ``{Stable communication protocol design for aeronautical and
  large-scale pseudo-linear highly mobile ad hoc networks},'' Ph.D.
  dissertation, University of Sydney, 2007.

\bibitem{conti2014Mobile}
M.~Conti and S.~Giordano, ``{Mobile ad hoc networking: milestones, challenges,
  and new research directions},'' \emph{IEEE Communications Magazine}, vol.~52,
  no.~1, pp. 85--96, January 2014.

\bibitem{abid2015survey}
S.~A. Abid, M.~Othman, and N.~Shah, ``{A survey on DHT-based routing for
  large-scale mobile ad hoc networks},'' \emph{ACM Computing Surveys (CSUR)},
  vol.~47, no.~2, p.~20, January 2015.

\bibitem{liu2017information}
X.~Liu, Z.~Li, P.~Yang, and Y.~Dong, ``{Information-centric mobile ad hoc
  networks and content routing: a survey},'' \emph{Ad Hoc Networks}, vol.~58,
  pp. 255--268, April 2017.

\bibitem{dorri2015security}
A.~Dorri and S.~R. Kamel, ``{Security Challenges in Mobile Ad Hoc Networks: A
  Survey},'' \emph{International Journal of Computer Science {\&} Engineering
  Survey}, vol.~6, no.~1, pp. 15--29, February 2015.

\bibitem{sun2014energy}
W.~Sun, Z.~Yang, X.~Zhang, and Y.~Liu, ``{Energy-efficient neighbor discovery
  in mobile ad hoc and wireless sensor networks: A survey},'' \emph{IEEE
  Communications Surveys \& Tutorials}, vol.~16, no.~3, pp. 1448--1459, Third
  Quarter 2014.

\bibitem{nayak2015analysis}
P.~Nayak and P.~Sinha, ``{Analysis of random way point and random walk mobility
  model for reactive routing protocols for MANET using NetSim simulator},'' in
  \emph{IEEE 3rd International Conference on Artificial Intelligence, Modelling
  and Simulation}, Kota Kinabalu, Malaysia, December 2015, pp. 427--432.

\bibitem{raya2007securing}
M.~Raya and J.-P. Hubaux, ``{Securing vehicular ad hoc networks},''
  \emph{Journal of Computer Security}, vol.~15, no.~1, pp. 39--68, January
  2007.

\bibitem{chlamtac2003mobile}
I.~Chlamtac, M.~Conti, and J.~J.-N. Liu, ``{Mobile ad hoc networking:
  imperatives and challenges},'' \emph{Ad hoc networks}, vol.~1, no.~1, pp.
  13--64, July 2003.

\bibitem{wang2018vehicular}
J.~Wang, C.~Jiang, K.~Zhang, T.~Q. Quek, Y.~Ren, and L.~Hanzo, ``{Vehicular
  sensing networks in a smart city: Principles, technologies and
  applications},'' \emph{IEEE Wireless Communications}, vol.~25, no.~1, pp.
  122--132, February 2018.

\bibitem{sou2013modeling}
S.-I. Sou, ``{Modeling emergency messaging for car accident over dichotomized
  headway model in vehicular ad-hoc networks},'' \emph{IEEE Transactions on
  Communications}, vol.~61, no.~2, pp. 802--812, February 2013.

\bibitem{coelingh2012all}
E.~Coelingh and S.~Solyom, ``{All aboard the robotic road train},'' \emph{IEEE
  Spectrum}, vol.~49, no.~11, pp. 34--39, November 2012.

\bibitem{gramaglia2011overhearing}
M.~Gramaglia, I.~Soto, C.~J. Bernardos, and M.~Calderon,
  ``{Overhearing-assisted optimization of address autoconfiguration in
  position-aware VANETs},'' \emph{IEEE Transactions on Vehicular Technology},
  vol.~60, no.~7, pp. 3332--3349, September 2011.

\bibitem{wang2018internet}
J.~Wang, C.~Jiang, Z.~Han, Y.~Ren, and L.~Hanzo, ``{Internet of vehicles:
  Sensing-aided transportation information collection and diffusion},''
  \emph{IEEE Transactions on Vehicular Technology}, vol.~67, no.~5, pp.
  3813--3825, May 2018.

\bibitem{medina2008topology}
D.~Medina, F.~Hoffmann, S.~Ayaz, and C.-H. Rokitansky, ``{Topology
  characterization of high density airspace aeronautical ad hoc networks},'' in
  \emph{Proceedings of 5th IEEE International Conference on the Mobile Ad Hoc
  and Sensor Systems}, Atlanta, USA, September 2008, pp. 295--304.

\bibitem{hsu2009dual}
S.-H. Hsu, Y.-J. Ren, and K.~Chang, ``{A dual-polarized planar-array antenna
  for S-band and X-band airborne applications},'' \emph{IEEE Antennas and
  Propagation Magazine}, vol.~51, no.~4, pp. 70--78, August 2009.

\bibitem{mehltretter2003structural}
L.~Mehltretter, ``{Structural antenna for flight aggregates or aircraft},''
  {US} Patent 6\,636\,182, October 21, 2003.

\bibitem{bauer2011survey}
C.~Bauer and M.~Zitterbart, ``{A survey of protocols to support IP mobility in
  aeronautical communications},'' \emph{IEEE Communications Surveys \&
  Tutorials}, vol.~13, no.~4, pp. 642--657, Fourthquarter 2011.

\bibitem{bekmezci2013flying}
I.~Bekmezci, O.~K. Sahingoz, and {\c{S}}.~Temel, ``{Flying ad-hoc networks
  (FANETs): A survey},'' \emph{Ad Hoc Networks}, vol.~11, no.~3, pp.
  1254--1270, May 2013.

\bibitem{saleem2015integration}
Y.~Saleem, M.~H. Rehmani, and S.~Zeadally, ``{Integration of cognitive radio
  technology with unmanned aerial vehicles: issues, opportunities, and future
  research challenges},'' \emph{Journal of Network and Computer Applications},
  vol.~50, pp. 15--31, April 2015.

\bibitem{gupta2016survey}
L.~Gupta, R.~Jain, and G.~Vaszkun, ``{Survey of important issues in UAV
  communication networks},'' \emph{IEEE Communications Surveys \& Tutorials},
  vol.~18, no.~2, pp. 1123--1152, Secondquarter 2016.

\bibitem{zafar2016flying}
W.~Zafar and B.~M. Khan, ``{Flying ad-hoc networks: technological and social
  implications},'' \emph{IEEE Technology and Society Magazine}, vol.~35, no.~2,
  pp. 67--74, June 2016.

\bibitem{hayat2016survey}
S.~Hayat, E.~Yanmaz, and R.~Muzaffar, ``{Survey on unmanned aerial vehicle
  networks for civil applications: A communications viewpoint},'' \emph{IEEE
  Communications Surveys \& Tutorials}, vol.~18, no.~4, pp. 2624--2661, Fourth
  quarter 2016.

\bibitem{sharma2017cooperative}
V.~Sharma and R.~Kumar, ``{Cooperative frameworks and network models for flying
  ad hoc networks: a survey},'' \emph{Concurrency and Computation: Practice and
  Experience}, vol.~29, no.~4, August 2017.

\bibitem{khuwaja2018survey}
A.~A. Khuwaja, Y.~Chen, N.~Zhao, M.-S. Alouini, and P.~Dobbins, ``A survey of
  channel modeling for uav communications,'' \emph{IEEE Communications Surveys
  \& Tutorials (Early Access)}, 2018.

\bibitem{cao2018airborne}
X.~Cao, P.~Yang, M.~Alzenad, X.~Xi, D.~Wu, and H.~Yanikomeroglu, ``{Airborne
  communication networks: A survey},'' \emph{IEEE Journal on Selected Areas in
  Communications}, vol.~36, no.~10, pp. 1907--1926, September 2018.

\bibitem{liu2018space}
J.~Liu, Y.~Shi, Z.~M. Fadlullah, and N.~Kato, ``{Space-air-ground integrated
  network: A survey},'' \emph{IEEE Communications Surveys \& Tutorials},
  vol.~20, no.~4, pp. 2714--2741, Fourthquarter 2018.

\bibitem{gogo20162ku}
{GoGo}, ``{2Ku High-performance inflight connectivity},''
  \url{https://www.gogoair.com/assets/downloads/gogo-2ku-brochure.pdf}, 2016,
  [[Online]. Available].

\bibitem{wright2001wirelesspread}
T.~H. Wright and R.~Delpak, ``{Wireless spread spectrum ground link-based
  aircraft data communication system for updating flight management files},''
  {US} Patent 6\,173\,159, January 9, 2001.

\bibitem{wright2000wireless}
T.~H. Wright and J.~J. Ziarno, ``{Wireless, frequency-agile spread spectrum
  ground link-based aircraft data communication system},'' {US} Patent
  6\,047\,165, April 4, 2000.

\bibitem{park2014hybrid}
P.~Park, H.~Khadilkar, H.~Balakrishnan, and C.~Tomlin, ``{Hybrid communication
  protocols and control algorithms for NextGen aircraft arrivals},'' \emph{IEEE
  Transactions on Intelligent Transportation Systems}, vol.~15, no.~2, pp.
  615--626, April 2014.

\bibitem{strohmeier2014realities}
M.~Strohmeier, V.~Lenders, I.~Martinovic \emph{et~al.}, ``{Realities and
  challenges of NextGen air traffic management: the case of ADS-B},''
  \emph{IEEE Communications Magazine}, vol.~52, no.~5, pp. 111--118, May 2014.

\bibitem{yu2015aftermath}
Y.~Yu, ``{The aftermath of the missing flight MH370: what can engineers Do?}''
  \emph{Proceedings of the IEEE}, vol. 103, no.~11, pp. 1948--1951, November
  2015.

\bibitem{nelson1998flight}
R.~C. Nelson, \emph{{Flight stability and automatic control}}.\hskip 1em plus
  0.5em minus 0.4em\relax WCB/McGraw Hill, 1998, vol.~2.

\bibitem{kainrath2016communication}
K.~Kainrath, M.~Gruber, H.~Fl{\"u}hr, and E.~Leitgeb, ``{Communication
  techniques for remotely piloted aircraft with integrated modular avionics},''
  in \emph{Proceedings of the International Conference on Broadband
  Communications for Next Generation Networks and Multimedia Applications},
  Graz, Austria, September 2016, pp. 1--6.

\bibitem{medina2012geographic}
D.~Medina, F.~Hoffmann, F.~Rossetto, and C.-H. Rokitansky, ``{A geographic
  routing strategy for north atlantic in-flight Internet access via airborne
  mesh networking},'' \emph{IEEE/ACM Transactions on Networking}, vol.~20,
  no.~4, pp. 1231--1244, August 2012.

\bibitem{rayner1979vortex}
J.~M.~V. Rayner, ``{A vortex theory of animal flight part 2: The forward flight
  of birds},'' \emph{Journal of Fluid Mechanics}, vol.~91, no.~04, pp.
  731--763, April 1979.

\bibitem{Bos2010formation}
D.~A. Bos, H.~P.~A. Dijkers, and T.~L.~M.~e. Gutleb, ``{Formation Flying ---
  Final report (Version 2.0)},'' January 2010.

\bibitem{dijkers2011integrated}
R.~Dijkers, H.~P.~A.~and Van~Nunen, D.~A. Bos, and T.~L.~M.~e. Gutleb,
  ``{Integrated Design of a Long-Haul Commercial Aircraft Optimized for
  Formation Flying},'' in \emph{The 11th AIAA Aviation Technology, Integration,
  and Operations (ATIO) Conference, including the AIA}, Virginia, USA,
  September 2011, pp. 1--8.

\bibitem{Airbus2015}
{Airbus}, ``{Express Skyways},''
  \url{http://www.airbus.com/innovation/future-by-airbus/smarter-skies/aircraft-in-
  free-flight-and- formation-along-express-skyways/}, 2015, [[Online].
  Available].

\bibitem{jahn2003evolution}
A.~Jahn, M.~Holzbock, J.~Muller, R.~Kebel, M.~De~Sanctis, A.~Rogoyski,
  E.~Trachtman, O.~Franzrahe, M.~Werner, and F.~Hu, ``{Evolution of
  aeronautical communications for personal and multimedia services},''
  \emph{IEEE Communications Magazine}, vol.~41, no.~7, pp. 36--43, July 2003.

\bibitem{medina2010routing}
D.~Medina, F.~Hoffmann, F.~Rossetto, and C.-H. Rokitansky, ``{Routing in the
  airborne internet},'' in \emph{IEEE Integrated Communications, Navigation and
  Surveillance Conference}, Herndon, USA, May 2010, pp. A7--1--A7--10.

\bibitem{Aircraft2014}
{Airports Council International}, ``{Aircraft Movements Monthly Ranking, May
  2014},''
  \url{http://www.aci.aero/Data-Centre/Monthly-Traffic-Data/Aircraft-Movements/Monthly},
  2014, [[Online]. Available].

\bibitem{Heathrow}
{Heathrow airport media centre}, ``{No easy fix to airport capacity crisis,
  says Heathrow},''
  \url{http://mediacentre.heathrowairport.com/Press-releases/No-easy-fix-to-airport-capacity-crisis-says-Heathrow-557.aspx},
  [[Online]. Available].

\bibitem{haas2002aeronautical}
E.~Haas, ``{Aeronautical channel modeling},'' \emph{IEEE Transactions on
  Vehicular Technology}, vol.~51, no.~2, pp. 254--264, March 2002.

\bibitem{haque2011ofdm}
J.~Haque, ``{An OFDM based aeronautical communication system},'' Ph.D.
  dissertation, University of South Florida, 2011.

\bibitem{dovis2002small}
F.~Dovis, R.~Fantini, M.~Mondin, and P.~Savi, ``{Small-scale fading for
  high-altitude platform (HAP) propagation channels},'' \emph{IEEE Journal on
  Selected Areas in Communications}, vol.~20, no.~3, pp. 641--647, April 2002.

\bibitem{sakhaee2006global}
E.~Sakhaee and A.~Jamalipour, ``{The global in-flight Internet},'' \emph{IEEE
  Journal on Selected Areas in Communications}, vol.~24, no.~9, pp. 1748--1757,
  September 2006.

\bibitem{tu2009mobile}
H.~D. Tu and S.~Shimamoto, ``{Mobile ad-hoc network based relaying data system
  for oceanic flight routes in aeronautical communications},''
  \emph{International Journal of Computer Networks and Communications}, vol.~1,
  no.~1, pp. 33--44, April 2009.

\bibitem{tu2009proposal}
------, ``{A proposal for high air-traffic oceanic flight routes employing
  ad-hoc networks},'' in \emph{IEEE Wireless Communications and Networking
  Conference}, Budapest, Hungary, April 2009, pp. 1--6.

\bibitem{shirazipourazad2011connectivity}
S.~Shirazipourazad, P.~Ghosh, and A.~Sen, ``{On connectivity of airborne
  networks in presence of region-based faults},'' in \emph{Proceedings of the
  Military Communications Conference}, Baltimore, USA, November 2011, pp.
  1997--2002.

\bibitem{LSTelcom2003Radio}
{W.~Schutz and M.~Schmidt}, ``{Radio Frequency Spectrum Requirement
  Calculations for Future Aeronautical Mobile (Route) System. AM(R)S},''
  \url{https://www.eurocontrol.int/sites/default/files/content/documents/communications/radio-
  frequency-spectrum- requirement- calculations-for- future- aeronautical-
  mobile- system- amrs.pdf}, 2003, [[Online]. Available].

\bibitem{wang2015throughput}
Y.~Wang, M.~C. Ert{\"u}rk, J.~Liu, I.-h. Ra, R.~Sankar, and S.~Morgera,
  ``{Throughput and delay of single-hop and two-hop aeronautical communication
  networks},'' \emph{Journal of Communications and Networks}, vol.~17, no.~1,
  pp. 58--66, February 2015.

\bibitem{xu2012video}
Y.~Xu, C.~Yu, J.~Li, and Y.~Liu, ``{Video telephony for end-consumers:
  measurement study of Google+, iChat, and Skype},'' in \emph{ACM Conference on
  Internet Measurement Conference}, Boston, USA, November 2012, pp. 371--384.

\bibitem{zhang2012profiling}
X.~Zhang, Y.~Xu, H.~Hu, Y.~Liu, Z.~Guo, and Y.~Wang, ``{Profiling skype video
  calls: Rate control and video quality},'' in \emph{The 31st Annual IEEE
  International Conference on Computer Communications}, Orlando, USA, March
  2012, pp. 621--629.

\bibitem{gheorghisor2006preliminary}
I.~Gheorghisor and Y.-S. Hoh, ``{Preliminary analysis of the spectral
  requirements of future ANLE networks},''
  \url{http://www.mitre.org/sites/default/files/pdf/06_0547.pdf}, 2006,
  [[Online]. Available].

\bibitem{SANSInstitute}
{SANS Institute}, ``{Latency and QoS for Voice over IP},''
  \url{http://www.sans.org/reading-room/whitepapers/voip/latency-qos-voice-ip-1349},
  2004, [[Online]. Available].

\bibitem{jansen2011enabling}
J.~Jansen, P.~Cesar, D.~C. Bulterman, T.~Stevens, I.~Kegel, and J.~Issing,
  ``{Enabling composition-based video-conferencing for the home},'' \emph{IEEE
  Transactions on Multimedia}, vol.~13, no.~5, pp. 869--881, October 2011.

\bibitem{stacey2008aeronautical}
D.~Stacey, \emph{{Aeronautical radio communication systems and
  networks}}.\hskip 1em plus 0.5em minus 0.4em\relax John Wiley \& Sons, 2008.

\bibitem{EUROCONTROL2007Future}
{EUROCONTROL}, ``{Future Communications Infrastructure-Technology
  Investigations: Evaluation Scenarios},''
  \url{https://www.eurocontrol.int/sites/default/files/field_tabs/content/documents/communications/fcs-eval-scenarios-v1.0.pdf},
  2007, [[Online]. Available].

\bibitem{thanthry2005aviation}
N.~Thanthry and R.~Pendse, ``{Aviation data networks: security issues and
  network architecture},'' \emph{IEEE Aerospace and Electronic Systems
  Magazine}, vol.~20, no.~6, pp. 3--8, June 2005.

\bibitem{thanthry2006security}
N.~Thanthry, M.~S. Ali, and R.~Pendse, ``{Security, Internet connectivity and
  aircraft data networks},'' \emph{IEEE Aerospace and Electronic Systems
  Magazine}, vol.~21, no.~11, pp. 3--7, November 2006.

\bibitem{medina2008feasibility}
D.~Medina, F.~Hoffmann, S.~Ayaz, and C.-H. Rokitansky, ``{Feasibility of an
  aeronautical mobile ad hoc network over the north atlantic corridor},'' in
  \emph{IEEE Communications Society Conference on Sensor, Mesh and Ad Hoc
  Communications and Networks}, San Francisco, USA, June 2008, pp. 109--116.

\bibitem{ghosh2016acpm}
S.~Ghosh and A.~Nayak, ``{ACPM: An associative connectivity prediction model
  for AANET},'' in \emph{IEEE 8th International Conference on Ubiquitous and
  Future Networks}, Vienna, Austria, July 2016, pp. 605--610.

\bibitem{schnell2014ldacs}
M.~Schnell, U.~Epple, D.~Shutin, and N.~Schneckenburger, ``{LDACS: future
  aeronautical communications for air-traffic management},'' \emph{IEEE
  Communications Magazine}, vol.~52, no.~5, pp. 104--110, May 2014.

\bibitem{sampigethaya2011future}
K.~Sampigethaya, R.~Poovendran, S.~Shetty, T.~Davis, and C.~Royalty, ``{Future
  e-enabled aircraft communications and security: the next 20 years and
  beyond},'' \emph{Proceedings of the IEEE}, vol.~99, no.~11, pp. 2040--2055,
  November 2011.

\bibitem{Aradhana2014Enabling}
A.~Narula-Tam, K.~Namuduri, S.~Chaumette, and D.~Giustiniano, ``{Enabling next
  generation airborne communications [Guest Editorial]},'' \emph{IEEE
  Communications Magazine}, vol.~52, no.~5, pp. 102--103, May 2014.

\bibitem{plass2014flight}
S.~Plass, R.~Hermenier, D.~G. Depoorter, T.~Tordjman, M.~Chatterton,
  M.~Amirfeiz, S.~Scotti, Y.~J. Cheng, P.~Pillai, F.~Durand \emph{et~al.},
  ``{Flight trial demonstration of seamless aeronautical networking},''
  \emph{IEEE Communications Magazine}, vol.~52, no.~5, pp. 119--128, May 2014.

\bibitem{EUROCONTROLACARS}
{EUROCONTROL}, ``{Aircraft Communications, Addressing and Reporting System},''
  \url{http://www.skybrary.aero/index.php/Aircraft_Communications,_Addressing_and_Reporting_System},
  [[Online]. Available].

\bibitem{ASRI2013SELCAL}
{Aviation Spectrum Resources, Inc.}, ``{Selective Calling (SELCAL) Users
  Guide},''
  \url{https://www.asri.aero/wp-content/uploads/2012/07/110914-ASRI-SELCAL-Users-Guide-61742-Rev-C.pdf},
  2013, [[Online]. Available].

\bibitem{wyatt2013aircraft}
D.~Wyatt and M.~Tooley, \emph{{Aircraft communications and navigation
  systems}}.\hskip 1em plus 0.5em minus 0.4em\relax Routledge, 2013.

\bibitem{nolan2010fundamentals}
M.~Nolan, \emph{{Fundamentals of air traffic control}}.\hskip 1em plus 0.5em
  minus 0.4em\relax Cengage Learning, 2010.

\bibitem{howland2005editorial}
P.~Howland, ``{Editorial: Passive radar systems},'' \emph{IEE Proceedings -
  Radar, Sonar and Navigation}, vol. 152, no.~3, pp. 105--106, June 2005.

\bibitem{lester2007benefits}
E.~A. Lester, ``{Benefits and incentives for ADS-B equipage in the national
  airspace system},'' Ph.D. dissertation, Massachusetts Institute of
  Technology, 2007.

\bibitem{richards2010new}
W.~R. Richards, K.~O'Brien, and D.~C. Miller, ``{New air traffic surveillance
  technology},''
  \url{http://www.boeing.com/commercial/aeromagazine/articles/qtr_02_10/pdfs/AERO_Q2-10_article02.pdf},
  2010, [[Online]. Available].

\bibitem{SESAR}
{SESAR}, ``{SESAR Joint Undertaking Program},'' \url{http://www.sesarju.eu/},
  [[Online]. Available].

\bibitem{NextGen}
{NextGen}, ``{NextGen Implementation Plan, FAA 2011.}''
  \url{http://www.faa.gov/nextgen/media/NextGenImplementationPlan2011.pdf},
  2011, [[Online]. Available].

\bibitem{costin2012ghost}
A.~Costin and A.~Francillon, ``{Ghost in the Air (Traffic): On insecurity of
  ADS-B protocol and practical attacks on ADS-B devices},'' \emph{Black Hat
  USA}, pp. 1--12, 2012.

\bibitem{schafer2016opensky}
M.~Schafer, M.~Strohmeier, M.~Smith, M.~Fuchs, R.~Pinheiro, V.~Lenders, and
  I.~Martinovic, ``{OpenSky report 2016: Facts and figures on SSR mode S and
  ADS-B usage},'' in \emph{IEEE/AIAA 35th Digital Avionics Systems Conference},
  Sacramento, USA, December 2016, pp. 1--9.

\bibitem{strohmeier2015security}
M.~Strohmeier, V.~Lenders, and I.~Martinovic, ``{On the security of the
  automatic dependent surveillance-broadcast protocol},'' \emph{IEEE
  Communications Surveys \& Tutorials}, vol.~17, no.~2, pp. 1066--1087,
  Secondquarter 2015.

\bibitem{james2007multilateration}
C.~James, ``{Multilateration: radar's replacement?}'' \emph{Avionics Magazine},
  vol.~31, no.~4, p.~30, April 2007.

\bibitem{EUROCONTROL2007actionplan}
{EUROCONTROL}, ``{Action plan 17: Future communications study --- final
  conclusions and recommendations report},''
  \url{https://www.eurocontrol.int/sites/default/files/field_tabs/content/documents/communications/112007-ap17-final-report.pdf},
  2007, [[Online]. Available].

\bibitem{Eurocontrol2009dacs1s}
------, ``{L-DACS1 system definition proposal: deliverable D2},''
  \url{http://www.eurocontrol.int/sites/default/files/article/content/documents/communications/d2-final-l-dacs1-spec-proposal-v10.pdf},
  2009, [[Online]. Available].

\bibitem{EUROCONTROL2009dacs2t}
------, ``{L-DACS2 Transmitter and Receiver prototype equipment specifications
  : Deliverable D3},''
  \url{http://www.eurocontrol.int/sites/default/files/article/content/documents/communications/18062009-ldacs2-design-d3-v1.2.pdf},
  2009, [[Online]. Available].

\bibitem{TIA2002}
{TIA}, ``{TIA standard family TIA-902, 2002/2003},''
  \url{http://libguides.asu.edu/content.php?pid=259962&sid=2145454}.

\bibitem{rokitansky2007b}
C.-H. Rokitansky, M.~Ehammer, T.~Grdupl, M.~Schnell, S.~Brandes, S.~Gligorevic,
  C.~Rihacek, and M.~Sajatovic, ``{B-AMC A system for future Broadband
  Aeronautical Multi-Carrier communications in the L-Band},'' in
  \emph{IEEE/AIAA 26th Digital Avionics Systems Conference}, Dallas, USA,
  October 2007, pp. 4.D.2--1--4.D.2--13.

\bibitem{etemad2008overview}
K.~Etemad, ``{Overview of mobile WiMAX technology and evolution},'' \emph{IEEE
  Communications magazine}, vol.~46, no.~10, pp. 31--40, October 2008.

\bibitem{neji2012coexistence}
N.~Neji, R.~De~Lacerda, A.~Azoulay, T.~Letertre, and O.~Outtier, ``{Coexistence
  between the future aeronautical system for continental communication L-DACS
  and the Distance Measuring Equipment DME},'' in \emph{IEEE First AESS
  European Conference on Satellite Telecommunications}, Rome, Italy, October
  2012, pp. 1--7.

\bibitem{jain2011analysis}
R.~Jain, F.~Templin, and K.-S. Yin, ``{Analysis of L-Band Digital Aeronautical
  Communication Systems: L-DACS1 and L-DACS2},'' in \emph{IEEE Aerospace
  Conference}, Big Sky, USA, March 2011, pp. 1--10.

\bibitem{Inmarsat2017}
{Inmarsat}, ``{The European Aviation Network (EAN)},''
  \url{https://www.inmarsat.com/aviation/aviation-connectivity-services/european-aviation-network/},
  2017, [[Online]. Available].

\bibitem{Nokia2017}
{Nikia}, ``{European Aviation Network (EAN) is airborne!}''
  \url{http://www.nokia.com/en_int/news/releases/2016/11/28/european-aviation-network-ean-is-airborne},
  2017, [[Online]. Available].

\bibitem{tadayon2016inflight}
N.~Tadayon, G.~Kaddoum, and R.~Noumeir, ``{Inflight broadband connectivity
  using cellular networks},'' \emph{IEEE Access}, vol.~4, pp. 1595--1606, April
  2016.

\bibitem{gogo2017}
{Gogo}, ``{ATG4 coverage},'' \url{https://www.gogoair.com/commercial/atg4},
  [[Online]. Available].

\bibitem{gogo2017AGT4}
------, ``{Gogo ATG-4 – what is it, and how does it work?}''
  \url{http://concourse.gogoair.com/gogo-atg-4-work/}, [[Online]. Available].

\bibitem{williams2004airborne}
{SKYbrary}, ``{Airborne collision avoidance system},''
  \url{http://www.skybrary.aero/index.php/Airborne_Collision_Avoidance_System_(ACAS)},
  2014, [[Online]. Available].

\bibitem{kuchar2007traffic}
J.~Kuchar and A.~C. Drumm, ``{The traffic alert and collision avoidance
  system},'' \emph{Lincoln Laboratory Journal}, vol.~16, no.~2, pp. 277--296,
  2007.

\bibitem{FAA2011TCAS}
{US Department of Transportation Federal Aviation Administration},
  ``{Introduction to TCAS II Version 7.1},''
  \url{http://www.faa.gov/documentLibrary/media/Advisory_Circular/TCAS%20II%20V7.1%20Intro%20booklet.pdf},
  [[Online]. Available].

\bibitem{EUROCONTROLASAS}
{EUROCONTROL}, ``{Airborne Separation Assurance Systems (ASAS)},''
  \url{http://www.skybrary.aero/index.php/Airborne_Separation_Assurance_Systems_(ASAS)},
  [[Online]. Available].

\bibitem{graupl2011dacs1}
T.~Graupl, M.~Ehammer, and S.~Zwettler, ``{L-DACS1 air-to-air data-link
  protocol design and performance},'' in \emph{The 2011 Integrated
  Communications, Navigation and Surveilance Conference}, Westin Dulles
  Herndon, USA, May 2011, pp. 1--14.

\bibitem{chan2006free}
V.~W. Chan, ``Free-space optical communications,'' \emph{Journal of Lightwave
  Technology}, vol.~24, no.~12, pp. 4750--4762, December 2006.

\bibitem{bilgi2010throughput}
M.~Bilgi and M.~Yuksel, ``{Throughput characteristics of free-space-optical
  mobile ad hoc networks},'' in \emph{The 13th ACM International Conference on
  Modeling, Analysis, and Simulation of Wireless and Mobile Systems}, Bodrum,
  Turkey, October 2010, pp. 170--177.

\bibitem{maine1995overview}
K.~Maine, C.~Devieux, and P.~Swan, ``{Overview of Iridium satellite network},''
  in \emph{Proceedings of WESCON}, San Francisco, USA, November 1995, pp.
  483--490.

\bibitem{nicol2000future}
S.~E. Nicol, G.~Walton, L.~D. Westbrook, and D.~A. Wynn, ``{Future satellite
  communications to military aircraft},'' \emph{Electronics \& Communication
  Engineering Journal}, vol.~12, no.~1, pp. 15--26, March 2000.

\bibitem{Iridium2015Iridium}
{Iridium}, ``{About Iridium next},''
  \url{https://www.iridium.com/about/IridiumNEXT.aspx}, 2015, [[Online].
  Available].

\bibitem{berzins1989inmarsat}
G.~Berzins, R.~Phillips, J.~Singh, and P.~Wood, ``{Inmarsat-Worldwide mobile
  satellite services on seas, in air and on land},'' in \emph{The 40th Congress
  of International Astronautical Federation}, Malaga, Spain, October 1989, pp.
  1--11.

\bibitem{ITU2011Regulation}
{ITU}, ``{Regulation of global broadband satellite communications},''
  \url{http://www.itu.int/ITU-D/treg/Events/Seminars/GSR/GSR11/documents/BBReport_BroadbandSatelliteRegulation-E.pdf},
  2011, [[Online]. Available].

\bibitem{Globalstar2014}
{Globalstar}, ``{Globalstar network},''
  \url{http://eu.globalstar.com/en/index.php?cid=3300}, [[Online]. Available].

\bibitem{MTSAT}
{MTSAT}, ``{MTSAT (Multi-functional Transport Satellite) Himawari},''
  \url{http://www.globalsecurity.org/space/world/japan/mtsat.htm}, [[Online].
  Available].

\bibitem{wakana2000communications}
H.~Wakana, S.~Yamamoto, H.~Saito, M.~Ohkawa, E.~Morikawa, and M.~Tanaka,
  ``{Communications and broadcasting engineering test satellite (COMETS) -
  experimental results},'' in \emph{The 18th AIAA International Communications
  Satellite Systems Conference and Exhibit, , Oakland, CA}, Oakland, USA, April
  2000.

\bibitem{miura2002ka}
A.~Miura, S.-i. Yamamoto, H.-B. Li, M.~Tanaka, and H.~Wakana, ``{Ka-band
  aeronautical satellite communications experiments using COMETS},'' \emph{IEEE
  Transactions on Vehicular Technology}, vol.~51, no.~5, pp. 1153--1164,
  September 2002.

\bibitem{wakana1998comets}
H.~Wakana, E.~Morikawa, H.~Saito, M.~Ohkawa, T.~Takahashi, and M.~Tanaka,
  ``{COMETS for Ka-band and millimeter-wave advanced mobile satellite
  communications and 21 GHz advanced satellite broadcasting experiments},'' in
  \emph{IEEE International Conference on Communications}, June 1998, pp.
  79--83.

\bibitem{Viasat2010Government}
{ViaSat}, ``{Government airborne mobile broadband},''
  \url{https://www.viasat.com/files/assets/web/datasheets/airborne_mobile_broadband_brochure_025_web.pdf},
  [[Online]. Available].

\bibitem{SeldingVu2014asatellite}
{Peter B. de Selding}, ``{WorldVu, a satellite startup aiming to provide global
  internet connectivity, continues to grow absent clear Google relationship},''
  Space News, 2014.

\bibitem{Pasztor2016OneWeb}
{Andy Pasztor}, ``{OneWeb satellite startup to set up manufacturing in
  Florida},'' Wall Street Journal, 2016.

\bibitem{SeldingVu2014Google}
{Peter B. de Selding}, ``{Google-backed Global Broadband Venture Secures
  Spectrum for Satellite Network},'' Space News, 2014.

\bibitem{SeldingVu2015Virgin}
------, ``{Virgin, Qualcomm Invest in OneWeb Satellite Internet Venture},''
  Space News, 2015.

\bibitem{drouilhet1996automatic}
P.~R. Drouilhet~Jr, G.~H. Knittel, and V.~A. Orlando, ``{Automatic dependent
  surveillance air navigation system},'' {US} Patent 5\,570\,095, October 29,
  1996.

\bibitem{Aircraftinteriors}
{Aircraft interiors international}, ``{Gogo set to launch GTO 60Mbps internet
  in 2014},''
  \url{http://www.aircraftinteriorsinternational.com/news.php?NewsID=52042},
  [[Online]. Available].

\bibitem{AeroMobile2016}
{AeroMobile}, ``{Airborne Mobile Connectivity},''
  \url{http://www.aeromobile.net/about/}, [[Online]. Available].

\bibitem{OnAir2016}
{OnAir}, ``{On-board device applications and services},''
  \url{http://www.sitaonair.aero/products/on-board-device-applications-and-services/},
  [[Online]. Available].

\bibitem{offcom2008}
Offcom, ``{Mobile Communications on board Aircraft (MCA)},''
  \url{http://stakeholders.ofcom.org.uk/binaries/consultations/mca/statement/mca.pdf},
  2008, [[Online]. Available].

\bibitem{jones2006communications}
C.~H. Jones, ``{Communications over aircraft power lines},'' in \emph{IEEE
  International Symposium on Power Line Communications and Its Applications},
  Orlando, USA, March 2006, pp. 149--154.

\bibitem{degardin2010possibility}
V.~Degardin, E.~Simon, M.~Morelle, M.~Lienard, P.~Degauque, I.~Junqua, and
  S.~Bertuol, ``{On the possibility of using PLC in aircraft},'' in \emph{IEEE
  International Symposium on Power Line Communications and Its Applications},
  Rio de Janeiro, Brazil, March 2010, pp. 337--340.

\bibitem{bertuol2011numerical}
S.~Bertuol, I.~Junqua, V.~Degardin, P.~Degauque, M.~Lienard, M.~Dunand, and
  J.~Genoulaz, ``{Numerical assessment of propagation channel characteristics
  for future application of power line communication in aircraft},'' in
  \emph{The 10th International Symposium on Electromagnetic Compatibility},
  York, UK, September 2011, pp. 506--511.

\bibitem{degardin2012power}
V.~Degardin, M.~Lienard, P.~Degauque, I.~Junqua, and S.~Bertuol, ``{Power line
  communication in aircraft: Channel modelling and performance analysis},'' in
  \emph{The 8th International Caribbean Conference on Devices, Circuits and
  Systems}.\hskip 1em plus 0.5em minus 0.4em\relax Playa del Carmen, Mexico:
  IEEE, March 2012, pp. 1--3.

\bibitem{degardin2013theoretical}
V.~Degardin, I.~Junqua, M.~Lienard, P.~Degauque, and S.~Bertuol, ``{Theoretical
  approach to the feasibility of power-line communication in aircrafts},''
  \emph{IEEE Transactions on Vehicular Technology}, vol.~62, no.~3, pp.
  1362--1366, March 2013.

\bibitem{degardin2014investigation}
V.~Degardin, P.~Laly, M.~Lienard, and P.~Degauque, ``{Investigation on power
  line communication in aircrafts},'' \emph{IET Communications}, vol.~8,
  no.~10, pp. 1868--1874, July 2014.

\bibitem{gfeller1979wireless}
F.~R. Gfeller and U.~Bapst, ``{Wireless in-house data communication via diffuse
  infrared radiation},'' \emph{Proceedings of the IEEE}, vol.~67, no.~11, pp.
  1474--1486, November 1979.

\bibitem{schmitt2006diffuse}
N.~Schmitt, T.~Pistner, C.~Vassilopoulos, D.~Marinos, A.~Boucouvalas,
  M.~Nikolitsa, C.~Aidinis, and G.~Metaxas, ``{Diffuse wireless optical link
  for aircraft intra-cabin passenger communication},'' in \emph{The 5th
  International Symposium on Communication Systems, Networks and Digital Signal
  Processing}, Patras, Greece, July 2006.

\bibitem{dimitrov2011throughput}
S.~Dimitrov, H.~Haas, M.~Cappitelli, and M.~Olbert, ``{On the throughput of an
  OFDM-based cellular optical wireless system for an aircraft cabin},'' in
  \emph{The 5th European Conference on Antennas and Propagation}, Rome, Italy,
  April 2011, pp. 3089--3093.

\bibitem{zhang2013fiber}
C.~Zhang, J.~Yu, and K.~Pang, ``{Fiber-Wireless cabin mobile communications on
  civil Aircraft},'' in \emph{IEEE/AIAA 32nd Digital Avionics Systems
  Conference}, East Syracuse, USA, October 2013, pp. 4B1--4B10.

\bibitem{krichene2015aeronautical}
D.~Krichene, M.~Sliti, W.~Abdallah, and N.~Boudriga, ``{An aeronautical visible
  light communication system to enable in-flight connectivity},'' in \emph{IEEE
  17th International Conference on Transparent Optical Networks}, Budapest,
  Hungary, July 2015, pp. 1--6.

\bibitem{Schnell2006Newsky}
S.~M. and S.~S., ``{Newsky - A Concept for Networking the Sky for Civil
  Aeronautical Communications},'' in \emph{IEEE/AIAA Digital Avionics Systems
  Conference}, Portland, USA, October 2006, pp. 1D4--1--1D46.

\bibitem{mahmoud2014aeronautical}
M.~S.~B. Mahmoud, C.~Guerber, A.~Pirovano, N.~Larrieu \emph{et~al.},
  \emph{Aeronautical Air-Ground Data Link Communications}.\hskip 1em plus 0.5em
  minus 0.4em\relax John Wiley \& Sons, 2014.

\bibitem{kamali2012application}
B.~Kamali, J.~D. Wilson, and R.~J. Kerczewski, ``{Application of multihop relay
  for perfomance enhancement of AeroMACS networks},'' in \emph{IEEE Integrated
  Communications, Navigation and Surveillance Conference}, Herndon, USA, April
  2012, pp. G2--1--G2--11.

\bibitem{kingsbury2009mobile}
R.~W. Kingsbury, ``{Mobile ad hoc networks for oceanic aircraft
  communications},'' Master's thesis, Massachusetts Institute of Technology,
  2009.

\bibitem{liu2018joint}
J.~Liu, Y.~Shi, L.~Zhao, Y.~Cao, W.~Sun, and N.~Kato, ``{Joint placement of
  controllers and gateways in SDN-enabled 5G-satellite integrated network},''
  \emph{IEEE Journal on Selected Areas in Communications}, vol.~36, no.~2, pp.
  221--232, February 2018.

\bibitem{cao2018optimal}
Y.~Cao, H.~Guo, J.~Liu, and N.~Kato, ``{Optimal satellite gateway placement in
  space-ground integrated networks},'' \emph{IEEE Network}, vol.~32, no.~5, pp.
  32--37, October 2018.

\bibitem{cao2018optimalaa2}
Y.~Cao, Y.~Shi, J.~Liu, and N.~Kato, ``{Optimal satellite gateway placement in
  space-ground integrated network for latency minimization with reliability
  guarantee},'' \emph{IEEE Wireless Communications Letters}, vol.~7, no.~2, pp.
  174--177, April 2018.

\bibitem{hoffmann2013joint}
F.~Hoffmann, D.~Medina, and A.~Wolisz, ``{Joint routing and scheduling in
  mobile aeronautical ad hoc networks},'' \emph{IEEE Transactions on Vehicular
  Technology}, vol.~62, no.~6, pp. 2700--2712, July 2013.

\bibitem{albano2000bit}
R.~Albano, L.~Franchina, and S.~A. Kosmopoulos, ``{Bit error performance
  evaluation of double-differential QPSK in faded channels characterized by
  Gaussian plus impulsive noise and Doppler effects},'' \emph{IEEE Transactions
  on Vehicular Technology}, vol.~49, no.~1, pp. 148--158, January 2000.

\bibitem{nguyen2007characterization}
M.~A. Nguyen and A.~I. Zaghloul, ``{On the characterization of cochannel
  interference in an aeronautical mobile environment},'' \emph{IEEE
  Transactions on Vehicular Technology}, vol.~56, no.~2, pp. 837--848, March
  2007.

\bibitem{plass2012seamless}
S.~Plass, ``{Seamless networking for aeronautical communications: one major
  aspect of the SANDRA concept},'' \emph{IEEE Aerospace and Electronic Systems
  Magazine}, vol.~27, no.~9, pp. 21--27, September 2012.

\bibitem{alejo2014collision}
D.~Alejo, J.~A. Cobano, G.~Heredia, and A.~Ollero, ``Collision-free 4d
  trajectory planning in unmanned aerial vehicles for assembly and structure
  construction,'' \emph{Journal of Intelligent \& Robotic Systems}, vol.~73,
  no. 1-4, p. 783, 2014.

\bibitem{Jamalipour2009Aeronautical}
A.~Jamalipour and E.~Sakhaee, ``{Aeronautical Ad-hoc Networks},'' {US} Patent
  2009/0\,092\,074 A1, April 9, 2009.

\bibitem{sakhaee2008stable}
E.~Sakhaee and A.~Jamalipour, ``{Stable clustering and communications in
  pseudolinear highly mobile ad hoc networks},'' \emph{IEEE Transactions on
  Vehicular Technology}, vol.~57, no.~6, pp. 3769--3777, November 2008.

\bibitem{ghosh2015multi}
S.~Ghosh and A.~Nayak, ``{Multi-dimensional clustering and network monitoring
  system for aeronautical ad hoc networks},'' in \emph{The Seventh
  International Conference on Ubiquitous and Future Networks}.\hskip 1em plus
  0.5em minus 0.4em\relax Sapporo, Japan: IEEE, July 2015, pp. 772--777.

\bibitem{li2012improved}
J.~Li, L.~Lei, W.~Liu, Y.~Shen, and G.~Zhu, ``{An improved semi-Markov smooth
  mobility model for aeronautical ad hoc networks},'' in \emph{IEEE 8th
  International Conference on Wireless Communications, Networking and Mobile
  Computing}, Shanghai, China, September 2012, pp. 1--4.

\bibitem{petersen2016modeling}
C.~Petersen, M.~M{\"u}hleisen, and A.~Timm-Giel, ``{Modeling aeronautical data
  traffic demand},'' in \emph{Proceedings of the Wireless Days}, Toulouse,
  France, March 2016, pp. 1--5.

\bibitem{petander2006measuring}
H.~Petander, E.~Perera, K.-C. Lan, and A.~Seneviratne, ``{Measuring and
  improving the performance of network mobility management in IPv6 networks},''
  \emph{IEEE Journal on Selected Areas in Communications}, vol.~24, no.~9, pp.
  1671--1681, September 2006.

\bibitem{lucke2013sandra}
O.~Lucke, D.~G. Depoorter, T.~Tordjman, and F.~Kuhnde, ``{The SANDRA testbed
  for the future aeronautical communication network},'' in \emph{IEEE
  Integrated Communications, Navigation and Surveillance Conference (ICNS)},
  Herndon, USA, April 2013, pp. 1--15.

\bibitem{moskowitz2008host}
R.~Moskowitz, P.~Nikander, P.~Jokela, and T.~Henderson, ``{Host identity
  protocol},'' IETF RFC5201, Tech. Rep., April 2008.

\bibitem{nikander2010host}
P.~Nikander, A.~Gurtov, and T.~R. Henderson, ``{Host identity protocol (HIP):
  Connectivity, mobility, multi-homing, security, and privacy over IPv4 and
  IPv6 networks},'' \emph{IEEE Communications Surveys \& Tutorials}, vol.~12,
  no.~2, pp. 186--204, Secondquarter 2010.

\bibitem{luo2018frudp}
Q.~Luo and J.~Wang, ``{FRUDP: a reliable data transport protocol for
  aeronautical ad hoc networks},'' \emph{IEEE Journal on Selected Areas in
  Communications}, vol.~36, no.~2, pp. 257--267, February 2018.

\bibitem{luo2018aeromrp}
Q.~Luo, J.~Wang, and S.~Liu, ``{AeroMRP: A multipath reliable transport
  protocol for aeronautical ad hoc networks},'' \emph{IEEE Internet of Things
  Journal (Early Access: 10.1109/JIOT.2018.2883736)}, pp. 1--12, November 2018.

\bibitem{luo2017multiple}
Q.~Luo and J.~Wang, ``{Multiple QoS parameters based routing for civil
  aeronautical ad hoc networks},'' \emph{IEEE Internet of Things Journal},
  vol.~4, no.~3, pp. 804--814, June 2017.

\bibitem{iordanakis2006ad}
M.~Iordanakis, D.~Yannis, K.~Karras, G.~Bogdos, G.~Dilintas, M.~Amirfeiz,
  G.~Colangelo, and S.~Baiotti, ``{Ad-hoc routing protocol for aeronautical
  mobile ad-hoc networks},'' in \emph{The Fifth International Symposium on
  Communication Systems, Networks and Digital Signal Processing}.\hskip 1em
  plus 0.5em minus 0.4em\relax Patras, Greece: Citeseer, July 2006.

\bibitem{ogier2004topology}
R.~Ogier, F.~Templin, and M.~Lewis, ``{Topology dissemination based on
  reverse-path forwarding (TBRPF)},'' IETF RFC3684, Tech. Rep., February 2004.

\bibitem{gankhuyag2016novel}
G.~Gankhuyag, A.~P. Shrestha, and S.-J. Yoo, ``{A novel directional routing
  scheme for flying ad-hoc networks},'' in \emph{The International Conference
  on Information and Communication Technology Convergence}.\hskip 1em plus
  0.5em minus 0.4em\relax Jeju, South Korea: IEEE, October 2016, pp. 593--597.

\bibitem{wang2013gr}
S.~Wang, C.~Fan, C.~Deng, W.~Gu, Q.~Sun, and F.~Yang, ``{A-GR: a novel
  geographical routing protocol for AANETs},'' \emph{Journal of Systems
  Architecture}, vol.~59, no.~10, pp. 931--937, November 2013.

\bibitem{saifullah2012new}
K.~Saifullah and K.-I. Kim, ``{A new geographical routing protocol for
  heterogeneous aircraft ad hoc networks},'' in \emph{IEEE/AIAA 31st Digital
  Avionics Systems Conference (DASC)}, Williamsburg, USA, October 2012, pp.
  4B5--1 -- 4B5--9.

\bibitem{tiwari2008mobility}
A.~Tiwari, A.~Ganguli, A.~Sampath, D.~S. Anderson, B.-H. Shen,
  N.~Krishnamurthi, J.~Yadegar, M.~Gerla, and D.~Krzysiak, ``{Mobility aware
  routing for the airborne network backbone},'' in \emph{IEEE Military
  Communications Conference}, San Diego, USA, November 2008, pp. 1--7.

\bibitem{vey2016routing}
Q.~Vey, S.~Puechmorel, A.~Pirovano, and J.~Radzik, ``{Routing in aeronautical
  ad-hoc networks},'' in \emph{IEEE/AIAA 35th Digital Avionics Systems
  Conference}, Sacramento, USA, September 2016, pp. 1--10.

\bibitem{zhong2016aeronautical}
D.~Zhong, Y.~Wang, Y.~Zhu, and T.~You, ``{An Aeronautical Ad Hoc Network
  routing protocol based on air vehicles movement features},'' in \emph{The
  22nd International Conference on Applied Electromagnetics and
  Communications}, Sacramento, USA, December 2016, pp. 1--6.

\bibitem{peters2011geographical}
K.~Peters, A.~Jabbar, E.~K. Etinkaya, and J.~P. Sterbenz, ``{A geographical
  routing protocol for highly-dynamic aeronautical networks},'' in \emph{IEEE
  Wireless Communications and Networking Conference}, Cancun, Mexico, March
  2011, pp. 492--497.

\bibitem{Zhong2014Anew}
Z.~Dong, Z.~Yian, Y.~Tao, D.~Junhua, and K.~Jie, ``{A new data transmission
  mechanism in aeronautical ad hoc network},'' in \emph{Proceedings of the
  International Conference on Big Data and Smart Computing}, Bangkok, Thailand,
  January 2014, pp. 255--260.

\bibitem{bilzhause2017datalink}
A.~Bilzhause, B.~Belgacem, M.~Mostafa, and T.~Graupl, ``Datalink security in
  the l-band digital aeronautical communications system (ldacs) for air traffic
  management,'' \emph{IEEE Aerospace and Electronic Systems Magazine}, vol.~32,
  no.~11, pp. 22--33, November 2017.

\bibitem{sampigethaya2013aviation}
K.~Sampigethaya and R.~Poovendran, ``{Aviation cyber--physical systems:
  foundations for future aircraft and air transport},'' \emph{Proceedings of
  the IEEE}, vol. 101, no.~8, pp. 1834--1855, August 2013.

\bibitem{sampigethaya2008secure}
K.~Sampigethaya, R.~Poovendran, and L.~Bushnell, ``{Secure operation, control,
  and maintenance of future e-enabled airplanes},'' \emph{Proceedings of the
  IEEE}, vol.~96, no.~12, pp. 1992--2007, December 2008.

\bibitem{olive2006commercial}
M.~L. Olive, R.~T. Oishi, and S.~Arentz, ``{Commercial aircraft information
  security-an overview of ARINC report 811},'' in \emph{IEEE/AIAA 25th Digital
  Avionics Systems Conference,}.\hskip 1em plus 0.5em minus 0.4em\relax
  Portland, USA: IEEE, October 2006, pp. 1--12.

\bibitem{robinson2007electronic}
R.~Robinson, M.~Li, S.~Lintelman, K.~Sampigethaya, R.~Poovendran,
  D.~Von~Oheimb, J.-U. Bu{\ss}er, and J.~Cuellar, ``{Electronic distribution of
  airplane software and the impact of information security on airplane
  safety},'' in \emph{The 26th International Conference on Computer Safety,
  Reliability and Security}, Nuremberg, Germany, September 2007, pp. 28--39.

\bibitem{sampigethaya2009secure}
K.~Sampigethaya, R.~Poovendran, L.~Bushnell, M.~Li, R.~Robinson, and
  S.~Lintelman, ``{Secure wireless collection and distribution of commercial
  airplane health data},'' \emph{IEEE Aerospace and Electronic Systems
  Magazine}, vol.~24, no.~7, pp. 14--20, July 2009.

\bibitem{mahmoud2009aeronautical}
B.~Mahmoud, M.~Slim, N.~Larrieu, and A.~Pirovano, ``{An aeronautical data link
  security overview},'' in \emph{IEEE/AIAA 28th Digital Avionics Systems
  Conference}, Orlando, USA, October 2009, pp. 4.A.4--1--4.A.4--14.

\bibitem{seo2005security}
K.~Seo and S.~Kent, ``{Security architecture for the internet protocol},''
  RFC4301, Tech. Rep., December 2005.

\bibitem{mahmoud2013ads}
B.~Mahmoud, M.~Slim, and N.~Larrieu, ``{An ADS-B based secure geographical
  routing protocol for aeronautical ad hoc networks},'' in \emph{IEEE 37th
  Annual Computer Software and Applications Conference Workshops (COMPSACW)},
  Kyoto, Japan, July 2013, pp. 556--562.

\bibitem{karp2000gpsr}
B.~Karp and H.-T. Kung, ``{GPSR: Greedy perimeter stateless routing for
  wireless networks},'' in \emph{The 6th Annual International Conference on
  Mobile Computing and Networking}, Boston, USA, August 2000, pp. 243--254.

\bibitem{nijsure2016adaptive}
Y.~A. Nijsure, G.~Kaddoum, G.~Gagnon, F.~Gagnon, C.~Yuen, and R.~Mahapatra,
  ``{Adaptive air-to-ground secure communication system based on ADS-B and
  wide-area multilateration},'' \emph{IEEE Transactions on Vehicular
  Technology}, vol.~65, no.~5, pp. 3150--3165, June 2016.

\bibitem{baek2016protect}
J.~Baek, E.~Hableel, Y.-J. Byon, D.~S. Wong, K.~Jang, and H.~Yeo, ``{How to
  protect ADS-B: confidentiality framework and efficient realization based on
  staged identity-based encryption},'' \emph{IEEE Transactions on Intelligent
  Transportation Systems}, vol.~PP, no.~99, pp. 1--11, August 2016.

\bibitem{boneh2003identity}
D.~Boneh and M.~Franklin, ``{Identity-based encryption from the Weil
  pairing},'' \emph{SIAM journal on Computing}, vol.~32, no.~3, pp. 586--615,
  March 2003.

\bibitem{yang2015new}
A.~Yang, X.~Tan, J.~Baek, and D.~S. Wong, ``{A new ADS-B authentication
  framework based on efficient hierarchical identity-based signature with batch
  verification},'' \emph{IEEE Transactions on Services Computing}, vol.~10,
  no.~2, pp. 165--175, 2017.

\bibitem{bello1973aeronautical}
P.~A. Bello, ``{Aeronautical channel characterization},'' \emph{IEEE
  Transactions on Communications}, vol.~21, no.~5, pp. 548--563, May 1973.

\bibitem{walter2011doppler}
M.~Walter and M.~Schnell, ``{The Doppler-delay characteristic of the
  aeronautical scatter channel},'' in \emph{IEEE Vehicular Technology
  Conference}, San Francisco, USA, September 2011, pp. 1--5.

\bibitem{gligorevic2013airport}
S.~Gligorevic, ``{Airport surface propagation channel in the C-Band:
  measurements and modeling},'' \emph{IEEE Transactions on Antennas and
  Propagation}, vol.~61, no.~9, pp. 4792--4802, September 2013.

\bibitem{lei2009multipath}
Q.~Lei and M.~Rice, ``{Multipath channel model for over-water aeronautical
  telemetry},'' \emph{IEEE Transactions on Aerospace and Electronic Systems},
  vol.~45, no.~2, pp. 735--742, April 2009.

\bibitem{meng2011measurements}
Y.~S. Meng and Y.~H. Lee, ``{Measurements and characterizations of
  air-to-ground channel over sea surface at C-band with low airborne
  altitudes},'' \emph{IEEE Transactions on Vehicular Technology}, vol.~60,
  no.~4, pp. 1943--1948, May 2011.

\bibitem{jensen2007aeronautical}
M.~Jensen, M.~D. Rice, A.~L. Anderson \emph{et~al.}, ``{Aeronautical telemetry
  using multiple-antenna transmitters},'' \emph{IEEE Transactions on Aerospace
  and Electronic Systems}, vol.~43, no.~1, pp. 262--272, January 2007.

\bibitem{sumiya2011radio}
Y.~Sumiya and Y.~Ogawa, ``{Radio propagation analysis using an aircraft model
  for MIMO antenna system in an anechoic chamber},'' in \emph{IEEE Aerospace
  Conference}, Big Sky, USA, March 2011, pp. 1--9.

\bibitem{gong2009performance}
M.~Gong, C.~Zhang, H.~Han, and X.~Lin, ``{Performance Analysis of Different
  Phase Shift Keying Modulation Schemes in Aeronautical Channels},''
  \emph{Tsinghua Science \& Technology}, vol.~14, pp. 1--6, December 2009.

\bibitem{erturk2014doppler}
M.~C. Erturk, J.~Haque, W.~Moreno, H.~Arslan \emph{et~al.}, ``{Doppler
  mitigation in OFDM-based aeronautical communications},'' \emph{IEEE
  Transactions on Aerospace and Electronic Systems}, vol.~50, no.~1, pp.
  120--129, January 2014.

\bibitem{yucek2009survey}
T.~Yucek and H.~Arslan, ``{A survey of spectrum sensing algorithms for
  cognitive radio applications},'' \emph{IEEE communications surveys \&
  tutorials}, vol.~11, no.~1, pp. 116--130, Firstquarter 2009.

\bibitem{popescu2007interference}
O.~Popescu, M.~Saquib, D.~C. Popescu, and M.~D. Rice, ``{Interference
  mitigation in aeronautical telemetry systems using Kalman filter},''
  \emph{IEEE Transactions on Aerospace and Electronic Systems}, vol.~43, no.~4,
  pp. 1624--1630, October 2007.

\bibitem{zhang2018adaptive}
J.~Zhang, S.~Chen, R.~G. Maunder, R.~Zhang, and L.~Hanzo, ``{Adaptive coding
  and modulation for large-scale antenna array based aeronautical
  communications in the presence of co-channel interference},'' \emph{IEEE
  Transactions on Wireless Communications}, vol.~17, no.~2, pp. 1343--1357,
  February 2018.

\bibitem{zhang2018regularized}
------, ``{Regularized zero-forcing precoding aided adaptive coding and
  modulation for large-scale antenna array based air-to-air communications},''
  \emph{IEEE Journal on Selected Areas in Communications}, vol.~36, no.~10, pp.
  2087--2103, September 2018.

\bibitem{tu2010performance}
H.~D. Tu, J.~Park, S.~Shimamoto, and J.~Kitaori, ``{Performance evaluation of
  communication system proposed for oceanic air traffic control},'' in
  \emph{IEEE Wireless Communications and Networking Conference}, Sydney,
  Australia, April 2010, pp. 1--6.

\bibitem{besse2011interference}
F.~Besse, A.~Pirovano, F.~Garcia, and J.~Radzik, ``{Interference estimation in
  an aeronautical ad hoc network},'' in \emph{IEEE/AIAA 30th Digital Avionics
  Systems Conference}, Seattle, USA, October 2011, pp. 4C6--1.

\bibitem{kamali2011selection}
B.~Kamali and R.~J. Kerczewski, ``{On selection of proper IEEE 802.16-based
  standard for Aeronautical Mobile Airport Surface Communications (AeroMACS)
  application},'' in \emph{IEEE Integrated Communications, Navigation and
  Surveilance Conference}, Herndon, USA, May 2011, pp. G3--1--G3--8.

\bibitem{fang2018qos}
Z.~Fang, Q.~Qiu, Y.~Ding, and L.~Ding, ``{A QoS guarantee based hybrid media
  access control protocol of aeronautical ad hoc network},'' \emph{IEEE
  Access}, vol.~6, pp. 5954--5961, March 2018.

\bibitem{planning2007concept}
{Joint Planning and Development Office}, ``{Concept of operations for the next
  generation air transportation system},''
  \url{http://citeseerx.ist.psu.edu/viewdoc/download?doi=10.1.1.112.922&rep=rep1&type=pdf},
  2007, [[Online]. Available].

\bibitem{Europe2010SESAR}
{SESAR Undertaking}, ``{SESAR --- The future of flying},''
  \url{http://ec.europa.eu/transport/modes/air/sesar/doc/2010_the_future_of_flying_en.pdf},
  2010, [[Online]. Available].

\bibitem{xie2018comprehensive}
J.~Xie, Y.~Wan, B.~Wang, S.~Fu, K.~Lu, and J.~H. Kim, ``{A comprehensive
  3-dimensional random mobility modeling framework for airborne networks},''
  \emph{IEEE Access}, vol.~6, pp. 22\,849--22\,862, May 2018.

\bibitem{teymoori2013dt}
P.~Teymoori, N.~Yazdani, and A.~Khonsari, ``{DT-MAC: An efficient and scalable
  medium access control protocol for wireless networks},'' \emph{IEEE
  Transactions on Wireless Communications}, vol.~12, no.~3, pp. 1268--1278,
  March 2013.

\bibitem{evans1992optimization}
J.~Evans, \emph{{Optimization algorithms for networks and graphs}}.\hskip 1em
  plus 0.5em minus 0.4em\relax Routledge, 1992.

\bibitem{baliga2011green}
J.~Baliga, R.~W. Ayre, K.~Hinton, and R.~S. Tucker, ``{Green cloud computing:
  Balancing energy in processing, storage, and transport},'' \emph{Proceedings
  of the IEEE}, vol.~99, no.~1, pp. 149--167, January 2011.

\bibitem{hirvensalo2013quantum}
M.~Hirvensalo, \emph{Quantum computing}.\hskip 1em plus 0.5em minus 0.4em\relax
  Springer, 2013.

\bibitem{marler2004survey}
R.~T. Marler and J.~S. Arora, ``{Survey of multi-objective optimization methods
  for engineering},'' \emph{Structural and multidisciplinary optimization},
  vol.~26, no.~6, pp. 369--395, April 2004.

\bibitem{fei2017survey}
Z.~Fei, B.~Li, S.~Yang, C.~Xing, H.~Chen, and L.~Hanzo, ``{A survey of
  multi-objective optimization in wireless sensor networks: Metrics,
  algorithms, and open problems},'' \emph{IEEE Communications Surveys \&
  Tutorials}, vol.~19, no.~1, pp. 550--586, Firstquarter 2017.

\bibitem{larsson2014massive}
E.~G. Larsson, O.~Edfors, F.~Tufvesson, and T.~L. Marzetta, ``{Massive MIMO for
  next generation wireless systems},'' \emph{IEEE Communications Magazine},
  vol.~52, no.~2, pp. 186--195, February 2014.

\bibitem{cheng2013millimeter}
Y.~J. Cheng, H.~Xu, D.~Ma, J.~Wu, L.~Wang, and Y.~Fan, ``{Millimeter-wave
  shaped-beam substrate integrated conformal array antenna},'' \emph{IEEE
  Transactions on Antennas and Propagation}, vol.~61, no.~9, pp. 4558--4566,
  September 2013.

\bibitem{gao2015conformal}
X.~Gao, Z.~Shen, and C.~Hua, ``Conformal vhf log-periodic balloon antenna,''
  \emph{IEEE Transactions on Antennas and Propagation}, vol.~63, no.~6, pp.
  2756--2761, June 2015.

\bibitem{xu2018adaptive}
C.~Xu, J.~Zhang, T.~Bai, S.~P. Botsinis, R.~G. Maunder, R.~Zhang, and L.~Hanzo,
  ``{Adaptive coherent/non-Coherent single/multiple-antenna aided channel coded
  ground-to-air aeronautical communication},'' \emph{IEEE Transactions on
  Communications}, vol.~67, no.~2, pp. 1099--1116, February 2019.

\bibitem{Cohen2014facebook}
{Cohen~D.}, ``{Facebook's connectivity lab: drones, plans, satellites, lasers
  to further internet.org mission of bringing connectivity to the whole
  world},'' 2014.

\bibitem{Boroson2011overview}
D.~M. Boroson, ``{Overview of the lunar laser communications demonstration},''
  in \emph{The International Conference on Space Optical Systems and
  Applications}, Kobe, Japan, May 2014, pp. 1--7.

\bibitem{kaushal2016optical}
H.~Kaushal and G.~Kaddoum, ``{Optical communication in space: challenges and
  mitigation techniques},'' \emph{IEEE Communications Surveys \& Tutorials},
  vol.~19, no.~1, pp. 57--96, Firstquarter 2017.

\bibitem{alam2017optimal}
A.~S. Alam, Y.-F. Hu, P.~Pillai, K.~Xu, and A.~Smith, ``{Optimal Datalink
  Selection for Future Aeronautical Telecommunication Networks},'' \emph{IEEE
  Transactions on Aerospace and Electronic Systems}, vol.~53, no.~5, pp. 2502
  -- 2515, May 2017.

\bibitem{shi2018across}
Y.~Shi, Y.~Cao, J.~Liu, and N.~Kato, ``{A cross-domain SDN architecture for
  multi-layered space-terrestrial integrated networks},'' \emph{IEEE Network},
  vol.~33, no.~1, pp. 29--35, February 2018.

\bibitem{shi2018cross}
Y.~Shi, J.~Liu, Z.~M. Fadlullah, and N.~Kato, ``{Cross-layer data delivery in
  satellite-aerial-terrestrial communication},'' \emph{IEEE Wireless
  Communications}, vol.~25, no.~3, pp. 138--143, June 2018.

\bibitem{kato2019optimizing}
N.~Kato, Z.~M. Fadlullah, n.~F. Tang, B.~Mao, S.~Tani, A.~Okamura, and J.~Liu,
  ``{Optimizing space-air-ground integrated networks by artificial
  intelligence},'' \emph{IEEE Wireless Communications: DOI:
  10.1109/MWC.2018.1800365}, 2019.

\bibitem{rohrer2011highly}
J.~P. Rohrer, A.~Jabbar, E.~K. Cetinkaya, E.~Perrins, and J.~P. Sterbenz,
  ``{Highly-dynamic cross-layered aeronautical network architecture},''
  \emph{IEEE Transactions on Aerospace and Electronic Systems}, vol.~47, no.~4,
  pp. 2742--2765, October 2011.

\bibitem{kwak2012airborne}
K.~J. Kwak, Y.~Sagduyu, J.~Deng, J.~Yackoski, and J.~Li, ``{Airborne Network
  Evaluation: Challenges and High Fidelity Emulation Solution},'' \emph{IEEE
  Communications Magazine}, vol.~52, no.~10, pp. 30--36, October 2014.

\bibitem{roy2017optimisations}
S.~B. Roy, A.~Ambede, A.~Vinod, and A.~Madhukumar, ``{Optimisations in
  aeronautical communications using aircrafts as relays},'' in \emph{IEEE
  Integrated Communications, Navigation and Surveillance Conference}, Herndon,
  USA, April 2017, pp. 5C2--1 -- 5C2--7.

\bibitem{bletsas2006simple}
A.~Bletsas, A.~Khisti, D.~P. Reed, and A.~Lippman, ``{A simple cooperative
  diversity method based on network path selection},'' \emph{IEEE Journal on
  Selected Areas in Communications}, vol.~24, no.~3, pp. 659--672, March 2006.

\bibitem{zlatanov2014buffer}
N.~Zlatanov, A.~Ikhlef, T.~Islam, and R.~Schober, ``{Buffer-aided cooperative
  communications: opportunities and challenges},'' \emph{IEEE Communications
  Magazine}, vol.~52, no.~4, pp. 146--153, April 2014.

\bibitem{jamali2015bidirectional01}
V.~Jamali, N.~Zlatanov, and R.~Schober, ``{Bidirectional buffer-aided relay
  networks with fixed rate transmission—Part I: Delay-unconstrained case},''
  \emph{IEEE Transactions on Wireless Communications}, vol.~14, no.~3, pp.
  1323--1338, March 2015.

\bibitem{jamali2015bidirectional02}
------, ``{Bidirectional buffer-aided relay networks with fixed rate
  transmission—Part II: Delay-constrained case},'' \emph{IEEE Transactions on
  Wireless Communications}, vol.~14, no.~3, pp. 1339--1355, March 2015.

\bibitem{dong2015energy}
C.~Dong, L.-L. Yang, J.~Zuo, S.~X. Ng, and L.~Hanzo, ``{Energy, delay, and
  outage analysis of a buffer-aided three-node network relying on opportunistic
  routing},'' \emph{IEEE Transactions on Communications}, vol.~63, no.~3, pp.
  667--682, March 2015.

\bibitem{wang2014cache}
X.~Wang, M.~Chen, T.~Taleb, A.~Ksentini, and V.~Leung, ``{Cache in the air:
  exploiting content caching and delivery techniques for 5G systems},''
  \emph{IEEE Communications Magazine}, vol.~52, no.~2, pp. 131--139, February
  2014.

\bibitem{jacob2016cognitive}
P.~Jacob, R.~P. Sirigina, A.~Madhukumar, and V.~A. Prasad, ``{Cognitive radio
  for aeronautical communications: a survey},'' \emph{IEEE Access}, vol.~4, pp.
  3417--3443, July 2016.

\bibitem{haykin2005cognitive}
S.~Haykin, ``{Cognitive radio: brain-empowered wireless communications},''
  \emph{IEEE Journal on Selected Areas in Communications}, vol.~23, no.~2, pp.
  201--220, February 2005.

\bibitem{akyildiz2006next}
I.~F. Akyildiz, W.-Y. Lee, M.~C. Vuran, and S.~Mohanty, ``{Next
  generation/dynamic spectrum access/cognitive radio wireless networks: A
  survey},'' \emph{Computer networks}, vol.~50, no.~13, pp. 2127--2159,
  September 2006.

\bibitem{wang2010cognitive}
Y.~Wang, ``{Cognitive radio for aeronautical air-ground communications},''
  \emph{IEEE Aerospace and Electronic Systems Magazine}, vol.~25, no.~5, pp.
  18--23, May 2010.

\bibitem{fantacci2009performance}
R.~Fantacci and A.~Tani, ``{Performance evaluation of a spectrum-sensing
  technique for cognitive radio applications in B-VHF communication systems},''
  \emph{IEEE Transactions on Vehicular Technology}, vol.~58, no.~4, pp.
  1722--1730, May 2009.

\bibitem{sheehe2004aviation}
C.~Sheehe and T.~Mulkerin, ``{Aviation communications emulation testbed},'' in
  \emph{The 23rd Digital Avionics Systems Conference}, vol.~2, Salt Lake City,
  USA, October 2004, pp. 11--A.

\end{thebibliography}
\section*{List of abbreviations and acronyms}

\begin{table}[h!]
\begin{tabular}{L{2cm}L{5cm}}
A2A & Air-to-Air\\
A2G & Air-to-Ground\\
A2S & Air-to-Satellite\\
AANET & Aeronautical {\it{ad hoc}} Network\\
ACARS & Aircraft Communication Addressing and
Reporting System \\
ACAS & Airborne Collision Avoidance System\\
ACM & Adaptive Coding and Modulation \\
ADS & Automatic Dependent Surveillance\\
ADS-B & Automatic Dependent Surveillance-Broadcast\\
ADS-C & Automatic Dependent Surveillance-Contract\\
AeroMACS & Aeronautical Mobile Airport Communication System \\
AMACS & All-purpose Multi-channel Aviation Communication System \\
AOA & Angle-Of-Arrival\\
ASAS & Airborne Separation Assurance System\\
ATC & Air Traffic Control\\
ATCT & Air Traffic Control Tower\\
ATM & Air Traffic Management\\
B-AMC & Broadband Aeronautical Multi-carrier Communications\\
BER & Bit Error Ratio\\
BS & Base Station\\
CCDF & Complementary Cumulative Distribution Function\\
CDMA & Code Division Multiple Access\\
COMETS & Communications and Broadcasting Engineering Test Satellites\\
CPFSK & Continuous-Phase Frequency-Shift Keying\\
CR & Cognitive Radio\\
CRL & Communications Research Laboratory\\
CRT & Cognitive Radio Technology\\
DL & DownLink\\
EAN & European Aviation Network\\
FAA & Federal Aviation Administration\\
FANET & Flying {\it{ad hoc}} Network\\
FCI & Future Communications Infrastructure\\
FDD & Frequency-Duplex Division\\
FDOA & Frequency-Difference-Of-Arrival\\
FM & Frequency Modulation\\
FSO & Free-Space Optical\\
GMSK &Gaussian Minimum-Shift Keying\\
GPS & Global Positioning System\\
GS &Ground Station\\
GSM & Global System for Mobile communications\\
HetNet & Heterogeneous Network\\
HAP  & High-Altitude Platform\\
HF &  High Frequency\\
ICAO  & International Civil Aviation Organization\\
IP  & Internet Protocol\\
km/h  & kilometers per hour
\end{tabular}
\end{table}

\begin{table}[h!]
\begin{tabular}{L{2cm}L{5cm}}

LAP  & Low-Altitude Platforms\\
L-DACS  & L-band Digital Aeronautical Communication
System\\
LED  & Light Emitting Diode\\
LOS  & Line-Of-Sight\\
LTE  & Long-Term Evolution\\
MAC  & Media Access Control\\
MANET  & Mobile {\it{ad hoc}} Network\\
MCA services &  Mobile Communication services on Aircraft\\
MHz  & MegaHertz\\
MIMO  & Multiple-Input Multiple-Output\\
MTSAT  & Multi-functional Transport SATellite\\
NASDA  & National Space Development Agency\\
NextGen  & Next Generation air transportation\\
NET  & Network\\
NM  & Nautical Mile\\
OFDM  & Orthogonal Frequency-Division Multiplexing\\
PHY  & PHYsical\\
PPM  & Pulse-Position Modulation\\
PSR  & Primary Surveillance Radar\\
QoE  & Quality-of-Experience\\
QoS  & Quality-of-Service\\
RA  & Resolution Advisorie\\
SELCAL  & SELective CALling\\
SESAR  & Single European Sky Air Traffic Management
Research\\
SHF  & Super High Frequency\\
SNR  & Signal-to-Noise Ratio\\
SSL & Secure Sockets Layer\\
SSR  & Secondary Surveillance Radar\\
STBC  & Space-Time Block Coding\\
TA  & Traffic Advisorie\\
TCAS  & Traffic Collision Avoidance System\\
TDD  & Time-Division Duplex\\
TDMA  & Time Division Multiple Access\\
TDOA & Time-Difference-Of-Arrival \\
TLS & Transport Layer Security \\
TRACON  & Terminal Radar Approach CONtrol\\
UAT & Universal Access Transceiver\\
UAV  & Unmanned Aerial Vehicles\\
UHF &  Ultra High Frequency\\
UL &  UpLink\\
US  & United States\\
VANET  & Vehicular {\it{ad hoc}} Network\\
VDL  & VHF Data Link\\
VHF  & Very High Frequency\\
VoIP  & Voice over IP\\
WAM  & Wide-Area Multilateration\\
WiFi  & Wireless Fidelity\\
WiMAX  & Worldwide interoperability for
Microwave Acces\\
\end{tabular}
\end{table}

\end{document}